\documentclass[a4paper,11pt]{article}
\usepackage{jcappub} 
\usepackage{lineno} 
\usepackage[titletoc,title]{appendix}
\usepackage{orcidlink}
\usepackage{hyperref}
\usepackage{mathtools}
\usepackage{bm}
\usepackage{soul}
\usepackage[normalem]{ulem}

\newcommand{\GeV}{\text{GeV}}
\newcommand{\SZero}{S_{E,0}}
\usepackage{comment}

\usepackage{cleveref}

\newcommand{\beq}{\begin{equation}}
\newcommand{\eeq}{\end{equation}}

\newcommand{\beqn}{\begin{eqnarray}}
\newcommand{\eeqn}{\end{eqnarray}}

\usepackage{footnote}

\begin{document}

\title{
 Non-Minimally Coupled Chain Inflation at High Scales
}
\date{}

 \author[a,b]{Miguel Barroso Varela\orcidlink{0009-0006-9844-7661},}
 \author[a,b]{Orfeu Bertolami \orcidlink{0000-0002-7672-0560},}
  \author[c,d]{Katherine Freese\orcidlink{0000-0001-9725-7395}}

 \author[c,e]{and Evangelos I. Sfakianakis\orcidlink{0000-0002-1525-2433}}

 \affiliation[a]{{Departamento de Física e Astronomia, Faculdade de Ciências, Universidade do Porto, Rua do Campo Alegre s/n, 4169-007 Porto, Portugal}}
 \affiliation[b]{{Centro de Física das Universidades do Minho e do Porto, Rua do Campo Alegre s/n, 4169-007 Porto, Portugal}}
 \affiliation[c]{Texas Center for Cosmology and Astroparticle Physics, Weinberg Institute for Theoretical Physics, Department of Physics, The University of Texas at Austin, Austin, TX 78712, USA}
 \affiliation[d]{The Oskar Klein Centre, Department of Astronomy, Stockholm University, 106 91 Stockholm, Sweden}
 \affiliation[e]{Department of Physics, Harvard University, Cambridge, MA, 02131, USA}

\emailAdd{up201907272@fc.up.pt}
\emailAdd{orfeu.bertolami@fc.up.pt}
\emailAdd{ktfreese@utexas.edu}
\emailAdd{evangelos.sfakianakis@austin.utexas.edu}

\abstract
{  Chain inflation offers an alternative to standard slow-roll dynamics, with accelerated expansion proceeding through a sequence of rapid quantum tunneling events between metastable vacua. At the high energy scales relevant for the early Universe, scalar fields are generically expected to couple non-minimally to gravity via operators like $\xi R\phi^2$, allowed by symmetry and required as counterterms for interacting theories in curved spacetime. We study the dynamical and observational consequences of this coupling for chain inflation. We find the modifications to the model for arbitrary $\xi$ and focus on interesting phenomenology for $\xi ={\cal O}( 10)$. We show that, in the Einstein frame, the non-minimal coupling induces a field-dependent amplification of the Euclidean bounce action, thus modifying the tunneling rate across the chain. We develop an analytic framework connecting this modified tunneling dynamics to the scalar spectral index, its running, the primordial curvature power spectrum, and the stochastic gravitational wave background from bubble collisions. As one consequence, the non-minimal coupling breaks the rigid relation between the scalar tilt and inflationary scale that drives the minimally coupled pure tilted cosine model to very low energies ($V_*^{1/4}\lesssim 3\,\rm{GeV}$, where $V_*$ is the value of the inflationary potential when the CMB-relevant modes exit the horizon), allowing for viable high-scale chain inflation with $V_*^{1/4}\sim 10^{11}\,\rm{GeV}$. Furthermore, non-minimally coupled chain inflation at high scales produces a peaked stochastic gravitational wave signal in the dHz-kHz bands, accessible to upcoming interferometers such as the Einstein Telescope and Cosmic Explorer. Finally, the model predicts a distinct running of the spectral index that will be testable by the Simons Observatory, making it a prime target for multi-messenger cosmology.
}

\maketitle

\section{Introduction}\label{sec:Introduction}
\begin{figure}
    \centering
    \includegraphics[width=\linewidth]{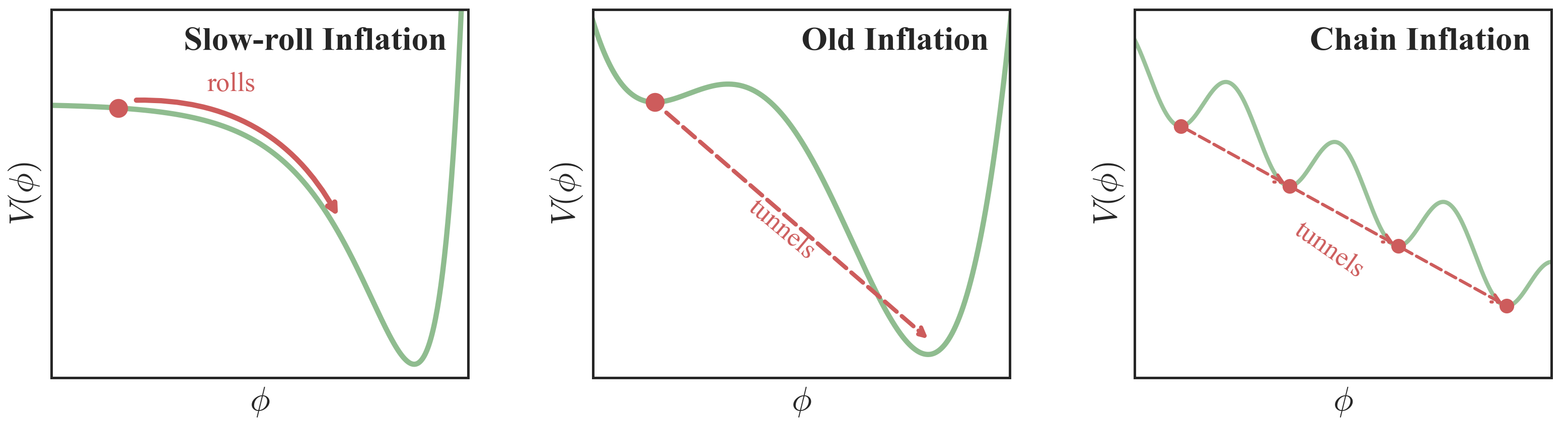}
    \caption{Comparison of slow-roll inflation (left), old inflation (center) and chain inflation (right) potentials. Unlike old inflation, in chain inflation the inflaton spends a short amount of time in each false vacuum, thus avoiding the empty Universe problem.}
    \label{fig:SlowRoll_vs_Chain}
\end{figure}

Despite the remarkable phenomenological success of slow-roll inflation, chain inflation offers a compelling alternative cosmological mechanism with distinct observational signatures \cite{Freese:2004vs,Freese:2005kt}. In this scenario, the Universe undergoes a series of rapid first-order phase transitions, consecutively tunneling through a cascade of local minima instead of slowly rolling down a flat potential. This is a natural extension to the original model of Guth's ``old inflation" \cite{Guth:1980zm}, in which the inflaton resides in a metastable vacuum, during which the Universe expands by a tremendous amount, before tunneling to the true vacuum. However, the requirements of sufficient expansion and percolation of the true vacuum bubbles are impossible to satisfy simultaneously in ``old" inflation \cite{Guth:1982pn}, which became known as the ``empty Universe" problem\footnote{An alternative possible solution is double field inflation \cite{Linde:1990gz,Adams:1990ds}, in which the field rolls in one dimension while requiring tunneling in the other dimension to reach the true vacuum. The rolling of the field diminishes the barrier, thus switching the tunneling rate from very slow to very fast and allowing the Universe to reheat uniformly.}.
Chain inflation solves this issue allowing for the inflaton to stay in each metastable vacuum for only a short period (a fraction of an $e$-fold), thereby letting each phase transition complete before proceeding to the next one; sufficient inflation results due to the fact that the field tunnels through thousands of vacua in the chain before reaching the minimum.  Chain inflation thus ensures a graceful exit from inflation while providing sufficient expansion to solve the flatness and horizon problems \cite{Freese:2004vs} (see Figure~\ref{fig:SlowRoll_vs_Chain}).

From an effective-field-theory perspective, a scalar field $\phi$ evolving in the {high energy density environment
of the early Universe (with correspondingly large space-time Ricci scalar $R$)} need not be minimally coupled to gravity. The operator $R\phi^2$ 
has mass dimension four and is compatible with the usual symmetries of a scalar field theory. Moreover, even if absent at tree level, the non-minimal coupling $\xi R\phi^2$ is generally generated by renormalization of an interacting scalar theory in curved spacetime (such as our expanding Universe) and should be included as a counterterm \cite{Birrell:1982ix,Parker:2009uva,Buchbinder:1992rb}. The value of $\xi$ is therefore a physical parameter of the low-energy theory, and the minimally coupled limit $\xi=0$ is thus not generically protected against radiative corrections.  Even though $\xi$ can be small and thus negligible, the regime $\xi\gtrsim {\cal O}(1)$ should be examined as it provides interesting phenomenology.  This makes the non-minimally coupled theory a natural setting in which to revisit the dynamics and observational predictions of chain inflation, especially in high-scale regimes where curvature effects can have significant dynamical consequences.

The generic presence of such a coupling is particularly relevant for chain inflation because its predictions are exponentially sensitive to the tunneling rate.
 In the minimally coupled ideal tilted cosine realization, the local tunneling parameters are constant along the chain, leading to a rigid relation between the scalar spectral index, the duration of inflation, and the inflationary energy scale. As we will show, this drives the simplest minimally coupled model to very low scales, $V_*^{1/4}\lesssim 3\,{\rm GeV}$ (where $V_*$ is the height of the inflationary potential  when the CMB-relevant modes exit the horizon), once the observed range of the scalar spectral index, $n_s\simeq 0.965 - 0.975$, is imposed \cite{Planck:2018jri,AtacamaCosmologyTelescope:2025blo,AtacamaCosmologyTelescope:2025nti}. This result should not be viewed merely as a peculiarity to be repaired, but rather as evidence that the tunneling rate must evolve non-trivially across the chain if chain inflation is to be realized at high energies.
 If chain inflation is achieved by a series of tunneling events of varying potential differences in the context of the string landscape, higher-scale inflation is allowed.  As shown in this paper, even in the context of a pure tilted cosine potential, a non-minimal coupling provides a gravitationally motivated way to generate precisely such an evolution of the tunneling rate that allows high-scale inflation.

In this work, we study chain inflation in the presence of a non-minimal coupling of the form $\xi R\phi^2$. In the Einstein frame, this coupling flattens the effective potential and induces a field-dependent amplification of the Euclidean bounce action. As a result, the tunneling rate evolves dynamically along the chain, dilating the lifetime of the false vacua either at earlier or later stages of inflation, depending on the field range. This breaks the rigid connection between $n_s$ and the inflationary energy scale found in the minimally coupled tilted cosine model and opens a viable high-scale region with $V_*^{1/4}\sim 10^{11}\,{\rm GeV}$.

We also explore the distinct observational signatures of this high-scale realization. While low-scale chain inflation predicts gravitational waves (GWs) in the nHz regime, placing it within the tentative signals reported by stochastic gravitational wave background (SGWB) experiments \cite{Freese:2022qrl}, high-scale chain inflation, like that allowed by our NMC framework, shifts the peak of the spectrum into the dHz-kHz bands. This places the model directly within the sensitivities of upcoming interferometers such as the Einstein Telescope \cite{Punturo:2010zz} and Cosmic Explorer \cite{Reitze:2019iox}.
At the same time, the curvature perturbations can exhibit a distinctive running of the spectral index, making the model testable by future Cosmic Microwave Background (CMB) experiments such as the Simons Observatory \cite{SimonsObservatory:2018koc}.

This paper is organized as follows. In Section~\ref{sec:StandardChainInflation}, we introduce the fundamental properties of chain inflation, including the vacuum-radiation dynamics and the mechanisms for the termination of the chain, after which we detail the simplest chain inflation potential - the tilted cosine - as both a useful local approximation for generic potentials and a good representation of potentials with nearly constant parameters. 
This section culminates in an analytic derivation of the low-scale bound on this class of models.
In Section~\ref{sec:NMC}, we introduce the non-minimally coupled chain inflation model and derive its effects on the bounce action and tunneling rate in the sub-Planckian field regime.
In Sections~\ref{sec:Results} and~\ref{subsec:GWSpectrum}, we study the resulting cosmological observables, including the scalar spectral index, the shape of the primordial curvature power spectrum, and the gravitational wave background from bubble collisions.
We present our conclusions in Section~\ref{sec:Conclusions}, along with promising avenues for future work. We work in units for which $c=\hbar=1$ and the reduced Planck mass $M_{\rm Pl}=(8\pi G)^{-1/2}=1$, although this quantity will sometimes be restored in equations for which we require a clear comparison with units of GeV.

\section{Standard Chain Inflation \texorpdfstring{$(\xi = 0)$}{(xi=0)}: Analytics and CMB Constraints}
\label{sec:StandardChainInflation}

Throughout the current Section~\ref{sec:StandardChainInflation}, we restrict our discussion to standard chain inflation, in which there is no explicit coupling to gravity (corresponding to $\xi=0$).
We first review the standard chain inflation scenario. 
Secondly,  we use this simple example as the test-bed to develop our analytical framework. Using it, we provide for the first time an approximate -- but extremely accurate -- analytical derivation for  the spectral tilt of density fluctuations from chain inflation $n_s$ as a function of the energy scale of chain inflation (at the pivot scale) $V_*$. 

We begin by reviewing the stochastic density perturbations in chain inflation,  the comparison to Cosmic Microwave Background (CMB) data,
and the resulting 
 bound $V_*^{1/4}\lesssim10^{12}$ GeV~\cite{Winkler:2020ape,Freese:2021noj}  for generic chain inflation potentials. Then
we turn to  the case of a pure tilted cosine potential and find a more restrictive bound:
 the monotonically increasing shape of $n_s(V_*)$ restricts the model to low scales $V_*^{1/4}\lesssim {\cal O}(1)\,{\rm GeV} $ in order to satisfy CMB bounds.

\subsection{Stochastic Density Perturbations}

Unlike standard slow-roll inflation, where primordial density perturbations arise from the quantum fluctuations of a slowly rolling scalar field, perturbations in chain inflation are generated by the probabilistic nature of the tunneling process itself and were derived in Ref.~\cite{Winkler:2020ape}. As the Universe expands, different Hubble patches undergo phase transitions at slightly different times. These local variations in the tunneling time translate into fluctuations in the energy density, which subsequently freeze out as curvature perturbations.

The rate of bubble nucleation per unit spacetime volume is given by 
\begin{equation}
\label{eq:tunnelingrate}
\Gamma = \mathcal{A} e^{-S_E} \, , 
\end{equation}
where $S_E$ is the $O(4)$-symmetric Euclidean bounce action extrapolating between adjacent vacua, and $\mathcal{A}$ is a mass-dimension four pre-factor. By simple dimensional analysis, the typical lifetime of a false vacuum in the chain must be inversely proportional to the transition rate $\Delta t \propto \Gamma^{-1/4}$.
For chain inflation to successfully percolate and avoid the ``empty Universe" problem of old inflation~\cite{Guth:1982pn, Turner:1992tz}, the transition rate must be  fast, $\Gamma/H^4 >9/4\pi$, such that the Universe expands by only a small fraction of an $e$-fold in each individual minimum.

As derived in detail in Ref.~\cite{Winkler:2020ape}, the amplitude of the primordial curvature power spectrum, $\Delta_{\mathcal{R}}^2$, depends solely on two parameters: the transition rate and the Hubble parameter $H$, both evaluated at the moment the pivot scale $k_*$ 
crosses the horizon. Since the pivot scale corresponds to the comoving wavenumber of the CMB-relevant modes, all quantities denoted by ``$*$" refer to the time during inflation when CMB modes exited the horizon.
Using a numerically fitted pre-factor, the power spectrum can be written as \cite{Winkler:2020ape}
\begin{equation}\label{eq:ScalarPowerSpectrumConstraint}
    A_s = \Delta_{\mathcal{R}}^2\big\lvert_{k=k_*} \approx 0.06\left(\frac{\Gamma_*^{1/4}}{H_*}\right)^{-5/3} \,.
\end{equation}
Requiring that the above relation reproduces  the observed amplitude from the {\it Planck} satellite, $A_s \simeq 2.1 \times 10^{-9}$ \cite{Planck:2018jri},  fixes the  ratio of the bubble nucleation rate to  the Hubble rate at horizon crossing:
\begin{equation}\label{eq:GammaH_Ratio}
    \frac{\Gamma_*^{1/4}}{H_*} = \left(\frac{A_s}{0.06}\right)^{-3/5} \approx 3 \times 10^4 \,.
\end{equation}

Having used the amplitude of scalar fluctuations, we move to the second piece of information that CMB experiments provide: the scalar spectral index $n_s$. This 
quantifies the scale dependence of scalar perturbations and in standard slow-roll inflation can be expressed as a linear function of the slow-roll parameters. Using the general form of $\Delta_{\mathcal{R}}^2(k) \propto (\Gamma^{1/4}/H)^{-5/3}$ and the horizon crossing condition $d\ln k \approx H dt$, which holds as long as the Hubble scale does not change much (commonly defined by the first slow-roll parameter $\epsilon\equiv-\dot H/H^2$ obeying $\epsilon\ll 1$), we can evaluate the logarithmic derivative of the power spectrum as
\begin{equation}
    \frac{d \ln \Delta^2_\mathcal{R}}{H dt} = \frac{1}{H} \frac{d}{ dt} \left[\frac{5}{3}\ln H -\frac{5}{12}\ln\Gamma   \right] = \frac{5}{12} \left( 4\frac{\dot{H}}{H^2} - \frac{\dot{\Gamma}}{H\Gamma} \right) \,.
\end{equation}
This yields the scalar spectral index
\begin{equation}\label{eq:ScalarSpectralIndex_Raw}
    n_s  =1+ \frac{d\ln\Delta^2_\mathcal{R}}{d \ln k}\biggr\rvert_{k=k_*} \approx 1+\frac{5}{12}\left( \frac{4\dot{H}}{H^2} - \frac{\dot{\Gamma}}{H\Gamma} \right)\Biggr\rvert_{t=t_*} \,.
\end{equation}
To express $n_s$ in terms of the potential parameters, we must explicitly evaluate the background cosmological evolution $\dot{H}/H^2$ and the tunneling dynamics $\dot{\Gamma}/(H\Gamma)$. Similarly, the running of the spectral index can be written in terms of the derivative of this expression evaluated at the pivot scale, thus giving
\begin{equation}\label{eq:running_general}
    \alpha_s\equiv\frac{d n_s}{d \log k} = \left. \frac{d^2 \log \Delta_{\mathcal{R}}^2}{d \log k^2} \right|_{k=k_*} = \frac{5}{12} \left( - \frac{8\dot{H}^2}{H^4} + \frac{\dot{H}\dot{\Gamma}}{H^3\Gamma} + \frac{\dot{\Gamma}^2}{H^2\Gamma^2} + \frac{4\ddot{H}}{H^3} - \frac{\ddot{\Gamma}}{H^2\Gamma} \right)\Bigg|_{t=t_*} \, ,
\end{equation}
as originally derived in Ref. \cite{Freese:2023szd}.

By combining the CMB amplitude constraint \eqref{eq:ScalarPowerSpectrumConstraint} with the prediction for the scalar spectral index in Eq.~\eqref{eq:ScalarSpectralIndex_Raw} while imposing observationally compatible values of $n_s\sim0.965-0.975$ and bounding the amount of fine tuning in the theory, the maximum scale allowed by generic models of chain inflation has been found to be $V_*^{1/4}\lesssim10^{12}$ GeV~\cite{Winkler:2020ape,Freese:2021noj}. Additionally, for the model to be compatible with BBN bounds \cite{Kawasaki:2000en}, one must also ensure that $V_*^{1/4}\gtrsim10$ MeV. These set the goal posts between which any plausible model of chain inflation must be constructed, namely
\begin{equation}
    10\,{\rm MeV}\lesssim V_*^{1/4}\lesssim 10^{12}\, {\rm GeV} \, .
\end{equation}

\subsection{Effective Action Expansion}
\label{sec:SE}

To determine the CMB observables of the theory, we must evaluate the variation of the transition rate, $\dot{\Gamma}/\Gamma$. Because $\Gamma$ is exponentially sensitive to the Euclidean bounce action, its evolution is dominated by variations in $S_E$ along the chain. Let $\Delta\phi_{\text{total}} = \phi_c - \phi_*$ be the total field excursion from horizon crossing (where the field value is $\phi_*$) to the end of inflation (where the field value is $\phi_c$). We can parameterize the deviation from a constant transition rate by Taylor expanding $S_E(\phi)$ in terms of the fractional field displacement $\tilde{\phi} = (\phi - \phi_*)/\Delta\phi_{\text{total}}$:
\begin{equation}\label{eq:S_expansion}
    S_E(\phi) \approx S_E(\phi_*) + \sum_{j=1}^{4} S_j \tilde{\phi}^j \,.
\end{equation}
The coefficients $S_j$ uniquely quantify the higher-order deformations of the chain. The transition rate then scales as $\Gamma(\phi) \approx \Gamma_* \exp(- \sum S_j \tilde{\phi}^j)$. 

Taking the time derivative and evaluating it at horizon crossing ($t=t_*$ for $\tilde{\phi}=0$), only the linear term survives:
\begin{equation}
    \frac{\dot{\Gamma}_*}{\Gamma_*} = -S_1 \dot{\tilde{\phi}}_* \,.
\end{equation}
The rate of field traversal is simply $\dot{\tilde{\phi}} = \dot{\phi}/\Delta\phi_{\text{total}}$. Approximating $\dot{\phi} \approx \Delta\phi/\Delta t$, where $\Delta \phi$ is the spacing between adjacent minima in field-space, and substituting the total number of transitions $N_* = \Delta\phi_{\text{total}}/\Delta\phi \approx V_*/\Delta V$, where $\Delta V$ denotes the energy difference between adjacent minima in the chain, we find:
\begin{equation}\label{eq:GammaDot_Over_Gamma}
    \frac{\dot{\Gamma}_*}{H_*\Gamma_*} = -S_1 \frac{\Delta\phi}{H_* \Delta t \Delta\phi_{\text{total}}} = -S_1 \frac{1}{H_* N_* \Delta t} = -S_1 \frac{\Delta V}{V_*}\frac{1.4\Gamma_*^{1/4}}{H_*} \,.
\end{equation}

The change in the Hubble scale can be computed using the discrete change after each transition
\begin{equation}
\frac{d (H^2)}{dt} \approx \frac{\Delta (H^2)}{\Delta t}
= -\frac{\Delta V}{3} 1.4 \Gamma^{1/4}\,,
\end{equation}
where the minus sign in the above equation complies with our convention of $\Delta V>0$ and the fact that the potential (and therefore the Hubble scale) gets reduced after each transition. Using $d_t H^2 = 2 H \dot H$ and $V_* \approx 3M_{\rm Pl}^2 H_*^2$, we can immediately compute the first slow-roll parameter
\begin{equation}\label{eq:HDot_Over_H2}
\epsilon\equiv - \frac{\dot H}{H^2} =\frac{\Delta V}{3 } \frac{1.4 \Gamma^{1/4}}{2H^3}
= \frac{1}{2}\frac{\Delta V}{V_*}\frac{1.4\Gamma^{1/4}}{H} \, .
\end{equation}
Finally, we substitute our derived expressions for $\dot{H}/H^2$ (Eq.~\eqref{eq:HDot_Over_H2}) and $\dot{\Gamma}/(H\Gamma)$ (Eq.~\eqref{eq:GammaDot_Over_Gamma}) back into the master formula for the spectral index (Eq.~\eqref{eq:ScalarSpectralIndex_Raw}):
\begin{equation}\label{eq:ns_Effective}
\begin{aligned}
    n_s  &\approx 1+ \frac{5}{12} \left[ 4\left( -\frac{1}{2}\frac{\Delta V}{V_*}\frac{1.4\Gamma_*^{1/4}}{H_*} \right) - \left( -S_1 \frac{\Delta V}{V_*}\frac{1.4\Gamma_*^{1/4}}{H_*} \right) \right] \\
    &= 1-\frac{5}{12} \frac{\Delta V}{V_*}\frac{1.4\Gamma_*^{1/4}}{H_*} \left( 2 - S_1 \right) \,.
\end{aligned}
\end{equation}

Analogous expressions for the running ($d n_s/d \log k$) and higher derivatives of $n_s$ depend on the corresponding higher-order time derivatives of $H$ and $\Gamma$, as shown in Eq.~\eqref{eq:running_general}. In particular, the running of the spectral index can be written in terms of $S_1$ and $S_2$, as it is obtained from a derivative of $n_s$, which is itself dependent on first-order variations of the Euclidean action throughout the chain. We can thus develop Eq.~\eqref{eq:running_general} into
\begin{equation}\label{eq:running_effective}
    \alpha_s\equiv\frac{dn_s}{d\log k}= \frac{5}{12} \left( \frac{\Delta V}{V_*} \frac{1.4 \Gamma_*^{1/4}}{H_*}\right)^2 \left( \frac{-S_1^2}{4} + 2S_2 + S_1 - 3 \right) \, ,
\end{equation}
which can be further simplified in terms of Eq.~\eqref{eq:ns_Effective} as
\begin{equation}\label{eq:Running_simplified}
    \alpha_s = \frac{3}{5} \frac{(1 - n_s)^2}{(2 - S_1)^2} \left( -S_1^2 + 8S_2 + 4S_1 - 12 \right) \, ,
\end{equation}
such that the running of the spectral index is fully fixed by $n_s$ itself in combination with the action expansion coefficients $S_1$ and $S_2$. This will be particularly useful, as it will allow us to fix a stringent bound on the running of any chain inflationary theory.

The running-of-running ($d^2 n_s/d\log k^2$) and higher-order derivatives of the scalar spectral index ($d^pn_s/d^p\log k$) will depend on all  $S_i$ coefficients with $i\le p+1$. Schematically
\begin{equation}
\label{eq:d_plogns}
    \frac{d^p n_s }{d^p\log k} = \left (\frac{\Delta V}{V_*} \frac{1.4 \Gamma_*^{1/4}}{H_*}\right)^{p+1}  \times {\cal F}_p(S_i) \, ,
\end{equation}
where ${\cal F}_p(S_i)$ is a function of the coefficients $S_i$.
Using Eq.~\eqref{eq:ns_Effective}, the first factor on the right-hand side of the above equation scales as $\left (\frac{\Delta V}{V_*} \frac{1.4 \Gamma_*^{1/4}}{H_*}\right)^{p+1}\propto (1-n_s)^{p+1}\sim 0.03^{\,p+1}$ for the $p$th-order running. This becomes smaller for larger values of $p$. We will thus restrict our attention to $p\le 1$. Therefore, unless the effective action expansion contains coefficients of $\mathcal{O}(30)$ or greater, one expects a generic theory of chain inflation to be characterized by the spectral index $n_s$ and its running $\alpha_s$. 

\subsection{Vacuum-radiation Dynamics 
}\label{subsec:Radiation}

We now focus on radiation density during chain inflation, its connection to the first slow-roll parameter $\epsilon$, shown in Eq.~\eqref{eq:HDot_Over_H2}, and a direct analogy to $\epsilon$ in standard slow-roll inflation, providing an intuitive connection between the two paradigms.

As the inflaton cascades down the potential, the vacuum energy $\rho_{\text{v}} = V(\phi)$ decreases discretely with each tunneling event. The energy difference between adjacent vacua, $\Delta V$, is converted into a  bath of radiation of density $\rho_r$ via bubble collisions. The radiation bath increases its energy immediately after each transition, while simultaneously diluting due to the expansion of the Universe between two consecutive transitions. The energy balance of the radiation component is written as
\begin{equation}\label{eq:rhordot}
    \dot{\rho}_r = \frac{\Delta V}{\Delta t} - 4H\rho_r \,.
\end{equation}
We can rewrite this equation in terms of derivatives with respect to $\phi$ as
\begin{equation}
\label{eq:rhordot2}
    \frac{d\rho_r}{d\phi} = \frac{d\rho}{dt}\frac{dt}{d\phi}\approx\dot \rho_r \frac{\Delta t}{\Delta\phi}=\frac{\Delta V}{\Delta\phi}-\frac{4H\Delta t}{\Delta\phi}\rho_r=\frac{\Delta V}{\Delta\phi}-\frac{4H(\phi)}{1.4\Gamma^{1/4}(\phi)\Delta\phi}\rho_r \, ,
\end{equation}
where we note that both the Hubble rate $H^2=(V(\phi)+\rho_r(\phi))/3$ and the transition rate $\Gamma(\phi)$ evolve with the field value, while we neglected the variation of $\Delta V$ since its effect will be dominantly present through $\Gamma$, which is exponentially sensitive to this quantity, as will be discussed in more detail later in this work. Although this equation can in general lead to intricate dynamics, if the timescale for the variation of the radiation density is slower than the Hubble time we may assume that it follows a tracking solution 
that arises from setting $\dot\rho_r=0$ in Eq.~\eqref{eq:rhordot2}
\begin{equation}\label{eq:trackingSolution}
    \rho_{r,tracking}(\phi)\approx \frac{1.4 \Gamma^{1/4}(\phi) \Delta V}{4H(\phi)}\, ,
\end{equation}
such that the radiation density smoothly evolves to match the inflaton potential's evolution as it progresses through the chain\footnote{It is worth noting that if $\Gamma$ is constant and $H$ necessarily decays, then the radiation density grows, but if $\Gamma$ changes, it is possible to have a decreasing radiation density. To guarantee that inflation indeed terminates, all we need to require is that radiation overcomes the vacuum at some point, which is guaranteed as long as there is a solution to Eq.~\eqref{eq:Vac_Rad_Equality} with $\tilde\phi_{eq}\leq1$, where $\tilde\phi_{eq}$ denotes the value of $\tilde\phi$ at the moment of vacuum-radiation equality. This is always guaranteed by any chain-like potential that terminates around $\rho_{\rm v}\approx0$, as required by the low vacuum energy present in the Universe at present.}.

Using the standard Friedmann acceleration equation, $\dot{H} = -\frac{1}{2M_{\rm Pl}^2}(\rho + p)$, and noting that the vacuum exerts a pressure $p_{\text{v}} = -\rho_{\text{v}}$ while radiation exerts a pressure $p_r = \rho_r/3$, we find $\rho+p = \frac{4}{3}\rho_r$. The evolution of the Hubble scale can now be computed as 
 \begin{equation}
     \dot{H} = -\frac{2\rho_r}{3M_{\rm Pl}^2} =- \frac{\Delta V}{ 6 M_{\rm Pl}^2 H \Delta t}  \,.
 \end{equation}
 Dividing by $H^2 \approx V_* / 3M_{\rm Pl}^2$ and substituting the attractor solution Eq.~\eqref{eq:trackingSolution}, we obtain the first slow-roll parameter $\epsilon$ given in Eq.~\eqref{eq:HDot_Over_H2}. We thus derived a simple but interesting relation $\rho_r/V_*=\epsilon/2$, which encapsulates that as inflation proceeds, part of the vacuum energy  is slowly depleted, going into radiation.

This  has a direct analogue in ordinary slow-roll inflation, where
potential energy is slowly transferred to kinetic energy of the inflaton, keeping the total energy as $\rho_{\rm SR,tot}=V+\dot\phi^2/2$, whereas in chain inflation $\rho_{\rm chain,tot}=V+\rho_{r}$. One would be tempted to draw the analogy  $\dot\phi^2/2 \leftrightarrow \rho_{r}$. However the correspondence should be made at the level of $\rho+p$, not at the level of the energy density alone. In slow roll, during inflation, the vacuum energy is slowly converted into scalar kinetic energy, for which $w_{\rm kin}=1$, so that 
\begin{equation}
\rho_{\rm SR}+p_{\rm SR}=\dot\phi^2=2\rho_{\rm kin}.
\end{equation}
In chain inflation, the vacuum energy released at each tunneling event is instead converted into radiation, for which $w_r=1/3$, giving
\begin{equation}
\rho_{\rm chain}+p_{\rm chain}=\frac43\rho_r.
\end{equation}
Thus both mechanisms describe a vacuum-dominated background whose energy is gradually transferred into a subdominant component with $w>-1$, thereby sourcing $\dot H<0$ and a nonzero $\epsilon=-\dot H/H^2$. The difference is that the receiving component has a different equation of state in the two cases. Consequently,
\begin{equation}
\epsilon_{\rm SR}\simeq 3\frac{\rho_{\rm kin}}{V},
\qquad
\epsilon_{\rm chain}\simeq 2\frac{\rho_r}{V},
\end{equation}
so that the chain inflation tracking solution implies
$
{\rho_r}/{V}\simeq {\epsilon_{\rm chain}}/{2}
$.
This is the sense in which radiation in chain inflation plays the role of the non-vacuum component that controls the departure from exact de Sitter expansion, similarly to the inflaton velocity in slow-roll inflation.

The  value  of the radiation density given by the tracking solution of Eq.~\eqref{eq:trackingSolution} is approximately of the order of the amount of energy released in one $e$-fold, as we release $\Delta V$ energy per transition and have roughly $\Gamma^{1/4}/H$ transitions per $e$-fold. Note that the tracking solution at the pivot scale is particularly simple due to the CMB power spectrum amplitude constraint
\begin{equation}\label{eq:RadiationAttractorSolution}
    \rho_{r,*}\equiv \rho_{r,tracking}(\phi_*)=\frac{1.4 \Gamma_*^{1/4} \Delta V}{4H_*}\approx10^4\Delta V \, .
\end{equation}

 Even though the radiation attractor is a useful approximation during inflation, as shown in Appendix \ref{sec:RadiationAppendix}, one should not rely on it at the end of chain inflation.  The gravitational wave spectrum of the theory, given its sensitivity on the amount of radiation present during post-inflationary phase transitions, should be obtained using the correct value of $\rho_{r}(\tilde\phi_{eq})$ at the end of chain inflation (i.e.~at vacuum-radiation equality), in lieu of using $\rho_{r,*}$, particularly in models with large negative values of $S_1$. This will be discussed in more detail in Section~\ref{subsec:GWSpectrum}.

\subsection{Graceful Exit to Radiation Dominated Universe}
\label{sec:graceful_exit}
Inflationary cosmology  requires a graceful exit from vacuum domination into a radiation-dominated
Universe. We take inflation to end 
once the inflaton reaches a critical field value $\phi_c$, at the point of vacuum/radiation equality; henceforth subscript $c$ refers to the end of inflation.
We will remain agnostic about the microscopic mechanism responsible for
ending the tunneling in the chain, and will consider two scenarios for reheating in chain inflation, as previously discussed in Ref.~\cite{Freese:2023szd}.

In the first scenario for ending inflation, a \emph{chain-to-stop}
exit, the potential barriers at the end of the chain grow (e.g.~triggered by an auxiliary field which receives a vacuum expectation value),
increasing the Euclidean bounce action
and suppressing the tunneling rate.  A few ever slower transitions may occur until the inflaton effectively becomes trapped in a minimum with a lifetime larger than the age of the Universe; this trapping point defines the end of inflation with field value $\phi_c$.
  Depending on the microscopic realization, there may be
one or a few late transitions after $\phi_c$ is reached. In this case there may be a brief period of vacuum domination after the field has reached $\phi_c$,
but its duration must be less than a fraction of an $e$-fold in order to allow for thermalization.
The final transition
differs from intermediate transitions during inflation: once accelerated expansion has
ended, particles produced near the bubble walls are no longer separated by the
quasi-de Sitter expansion and can reheat the bubble interior on a timescale of order
one Hubble time.  The cosmological observables and gravitational wave signal are completely dominated by the last phase transition, whether it be before or after the field has reached the critical value $\phi_c$.

In the second scenario, a \emph{chain-to-roll}
exit,  the barriers decrease near the end of the chain to the point where the field rolls to the absolute minimum, or become sufficiently shallow that the field rapidly tunnels to the bottom
of the potential. The final stage is then modeled by an effectively instantaneous
conversion of the remaining inflaton energy into radiation at $\phi_c$. These two types of chain-ending processes are described in more detail in Appendix \ref{sec:GracefulExitAppendix}.

For the purposes of the gravitational-wave signal, the detailed microphysics of the exit enters primarily through the amount of vacuum energy released in the last relevant transition. We therefore model two limiting cases as follows. For the case of models with a post-inflationary transition, we take \(\rho_{\rm v}=\Delta V\), thereby capturing the burst of energy released in the final transition between neighboring vacua. 
For models with no post-inflationary transition, we take \(\rho_{\rm v}\simeq 0\), modeling an effectively instantaneous reheating of the Universe.

Regardless of the details of the final transition (pre- or post-inflationary), 
we capture the properties of the graceful exit by the strength of the final transition, i.e.~the ratio of
vacuum energy to radiation energy right before the final tunneling event. This is defined as
\begin{equation}\label{eq:alpha}
    \alpha\equiv\frac{\rho_{\rm v}}{\rho_r}\Big\lvert_{t_{\rm last}} \, .
\end{equation}
If there is a second period of vacuum domination, it is possible to obtain $\alpha>1$. However, as was shown in {Section II of} Ref.~\cite{Freese:2022qrl},  there is an upper bound $\alpha \lesssim 20$ to ensure that the vacuum-dominated expansion does not outpace bubble nucleation and violate the percolation condition. 

Considering that the radiation density at the end of inflation is given by Eq.~\eqref{eq:Rad_AtEquality}, this establishes a minimum value for $\alpha$. This is obtained by taking the lowest possible non-zero vacuum energy for the last transition ($\rho_{\rm v}=\Delta V$) and the largest final radiation density $\rho_{r}(\tilde\phi_{\rm eq})$, i.e. its value if the final transition in the chain happens right at vacuum-radiation equality ($\tilde\phi=\tilde\phi_{eq}$), equivalent to the end of the inflationary expansion period. This gives
\begin{equation}\label{eq:alpha_min}
    \alpha_{min} =\frac{\Delta V}{\rho_{r}(\tilde\phi_{\rm eq})}= \frac{4H_*}{1.4 \Gamma_*^{1/4}}\frac{0.018}{(2-S_1)(1-\tilde\phi_{eq})} \approx \frac{1.71 \times 10^{-6}}{(2-S_1)(1-\tilde\phi_{eq})}\, ,
\end{equation}
where we have used the CMB amplitude constraint on the final step. 
{Solving the vacuum-radiation equality condition (Eq.~\eqref{eq:Vac_Rad_Equality} in  Appendix~\ref{sec:RadiationAppendix}), we derive a conservative lower bound for the entire parameter range considered in this work, $1-\tilde\phi_{\rm eq}\gtrsim10^{-3}$.
 Eq.~\eqref{eq:alpha_min} then implies $\alpha_{\rm min}\lesssim 10^{-3}$ for  $2-S_1=\mathcal{O}(1)$.
}

\subsection{The Pure Tilted Cosine and the Low-scale Constraint}

We now move to compute the spectral index of the simplest chain inflation model, governed by a tilted cosine potential
\begin{equation}
    V(\phi) = -\mu^3 \phi + \Lambda^4 \cos\left(\frac{\phi}{f}\right) + C\, ,
    \label{eq:tilted_cosine}
\end{equation}
describing a  decreasing potential from $V(\phi_*)=V_*$ (at CMB scales during inflation) to $V(\phi_c)\approx0$ at the end of inflation through a series of subsequent metastable minima. In the above definition, $\mu^3$ is the
tilt, $\Lambda^4$ sets the local barrier height, and $f$ is the
period in field-space. The potential in Eq.~\eqref{eq:tilted_cosine} is a particularly useful example, since it can be seen as an effective path in a more generic multidimensional potential, as one would expect to find in the high-energy string landscape. However, the tilted cosine potential describes a simplified version of such a landscape, where   each tunneling event is governed by the same parameters.
The tilted cosine chain inflation model was previously studied \cite{Winkler:2020ape,Freese:2022qrl}, and we begin by reiterating some of the results from those papers.  We further develop analytic studies of relevant quantities for chain inflation with this potential.

A new result of the current paper, which we will derive in this section, is the fact that the tilted cosine chain inflation model  is restricted to low scale inflation with overall potential height 
$V_*^{1/4} \lesssim  3 \ \GeV\,$ at the $2\sigma$ level.

In order for the field to undergo tunneling rather than behave as in the standard slow-roll scenario, the potential must exhibit distinct ordered local minima. This imposes the condition that the barrier height $\sim \Lambda^4$  must dominate over the energy difference between successive vacua $\Delta V = \mu^3 f$, or equivalently 
\begin{equation}\label{eq:x_definition}
    x\equiv\frac{\mu^3f}{\Lambda^4} < 1 \, ,
\end{equation}
where we have introduced the dimensionless tunneling parameter $x$.

The tunneling rate $\Gamma = \mathcal{A}e^{-S_E}$ is exponentially dependent on the Euclidean bounce action $S_E$.
Since the parameters $\{\mu, \Lambda, f\}$ are  constant, the barrier shapes are identical for every transition. Consequently, the Euclidean bounce action $S_E$ is invariant, meaning the expansion coefficients defined in Eq.~\eqref{eq:S_expansion} vanish ($S_{i\geq 1} = 0$). 

Considering the symmetry of the tilted cosine potential given in Eq.~\eqref{eq:tilted_cosine} applied to the Euclidean equation of motion, the bounce action can be calculated as
\begin{equation}\label{eq:ActionNumericalApproximation}
    S_E=\frac{f^4}{\Lambda^4}\mathcal{S}(x) \, ,
\end{equation}
where $\mathcal{S}(x)$ is the rescaled bounce action, which only depends on the dimensionless parameter $x$ defined in Eq.~\eqref{eq:x_definition}. Appendix \ref{sec:ActionSymmetryAppendix} contains the proof that $S_E$ can be written in terms of the ratio of $f/\Lambda$ multiplied by some function of $x$, which can be used for any set of consecutive minima in any potential under this local parametrization.
The numerical solution to the bounce equation for the tilted cosine potential was found in Ref. \cite{Winkler:2020ape}, where it was also shown that a consistent analytical approximation is given by 
\begin{equation}
\label{eq:calSofx}
    \mathcal{S}(x)=\sqrt{(1-x^2)(1-0.86x^2)}\mathcal{S}_{\rm thin-wall} \quad \text{where} \quad \mathcal{S}_{\rm thin-wall}=\frac{4}{\pi}\left(\frac{12}{x}\right)^3 \,.
\end{equation}
Here $\mathcal{S}_{\rm thin-wall}$ denotes the rescaled bounce action in the thin-wall approximation, also known as the ``envelope approximation," which holds under the assumption of a small energy difference between the two vacua. However, as found in Ref. \cite{Winkler:2020ape}, the thin-wall approximation is only applicable for $x\lesssim0.4$, beyond which the full expression above must be considered. In the case of fast tunneling ($x\gtrsim0.8$), naturally preferred by chain inflationary models with many phase transitions per $e$-fold, the full expression is crucial in obtaining accurate results, as shown in Ref.~\cite{Winkler:2020ape}. There it was found that the thin-wall approximation consistently overestimates the Euclidean action, such that the transition rate would be severely underestimated due to its exponential dependence on $S_E$ and thus the lifetime of each vacuum would be predicted to be much longer than it would be in reality. This would, for example, lead to the incorrect prediction for the $e$-folds of expansion in the chain, therefore having a direct effect on the associated predicted CMB observables.
Thus in this paper we work in the regime of fast tunneling.

{As $x$ increases, the speed of tunneling increases. As $x$ approaches unity, the minima become sufficiently shallow that ripples in the field generated behind an expanding bubble wall can carry the field over the subsequent barrier and classically induce the next transition, thereby initiating a cascade rather than a controlled sequence of tunneling events between neighboring vacua. 
The real-time $1+1$ dimensional simulations of Ref.~\cite{Cline:2011fi} place the onset of this runaway behavior at
$x\simeq 0.9581\approx 0.96$. We therefore take $x=0.96$ as the approximate upper
limit of the allowed parameter space, while  a physical realization should lie somewhat below this value  to avoid triggering  a so-called ``tunneling catastrophe".}

The pre-factor $\mathcal{A}$ in the tunneling rate, also found numerically in Ref. \cite{Winkler:2020ape}, can be written in terms of the effective mass of the scalar-field at each minimum of the potential, $m^2=V''(\phi_{\rm min})$, as
\begin{equation}
    \mathcal{A}=m^4\frac{S_E^2}{4\pi^2}\exp\left(13.15-\frac{15.8}{x^{2.9}}\right) =\frac{\Lambda^8}{f^4}(1-x^2)\frac{S_E^2}{4\pi^2}\exp\left(13.15-\frac{15.8}{x^{2.9}}\right) \, ,
\end{equation}
where we have used the fact that the mass term can be written as $m^2=V''(\phi_{\rm min})=\frac{\Lambda^4}{f^2}\sqrt{1-x^2}$, since at each minimum the tilted cosine potential obeys $V'(\phi_{\rm min})=0$ leading to $\sin(\phi_{\min}/f)=x$. Thus the transition rate can be analytically approximated by
\begin{equation}\label{eq:GeneralGamma}
    \Gamma=\frac{\Lambda^8}{f^4}(1-x^2)\frac{S_E^2}{4\pi^2}e^{-S_E}\exp \left(13.15-\frac{15.8}{x^{2.9}}\right)\,.
\end{equation}
One can see that  the transition rate (and hence the lifetime of each vacuum) scales with the Euclidean action primarily as 
$e^{-S_E}$. Thus even slight variations in the parametrized tilted cosine parameters $\{\mu^3,\Lambda^4,f\}$ can have considerable effects on the transition rate, thus speeding up or slowing down the tunneling chain. In order for the tilted cosine potential of Eq.~\eqref{eq:tilted_cosine} to be a valid approximation,  any variation of the potential parameters  must remain small  over several transitions.

We define
\begin{itemize}
    \item $N_*\equiv$ total number of phase transitions between CMB window and the end of inflation;
    \item $\mathcal{N}_*\equiv$ total number of $e$-folds of expansion between CMB window and the end of inflation.
\end{itemize}
As a reminder, the subscript ``$*$" refers to the horizon crossing of the pivot scale during inflation.
Setting $S_1 = 0$ in Eq.~\eqref{eq:ns_Effective}, the spectral index is completely determined by the evolution of the Hubble scale (equivalently the first slow-roll parameter $\epsilon$)
\begin{equation}\label{eq:ns_PureTiltedCosine}
    1-n_s = \frac{5}{6}\frac{1.4\Gamma^{1/4}_*}{H_*}\frac{\Delta V}{V_*} = \frac{5}{6}(4.17 \times 10^4) \frac{1}{N_*} \approx \frac{3.5\times10^4}{N_*}\,,
\end{equation}
where we have substituted the CMB amplitude constraint from Eq.~\eqref{eq:GammaH_Ratio}. This reveals a remarkable rigidity: the total number of phase transitions between horizon crossing and the end of inflation, $N_*$, is  fixed by the measured value of the spectral index and increases if $n_s$ approaches the scale-invariant spectrum.

With $N_*$ locked, the total number of $e$-folds of expansion between the pivot scale and the end of inflation, $\mathcal{N}_*$, is similarly constrained. Given the translational invariance of the chain, we fix the  field value at horizon crossing to $\phi_* = 0$ for simplicity. We then assume the chain is terminated through a generic chain-to-stop mechanism at a critical field value $\phi_c$, as described in Appendix \ref{sec:GracefulExitAppendix}. We remain fully agnostic about the detailed nature of this mechanism, as in its simplest form it simply determines the point at which the tunneling chain terminates, with no other consequence on the preceding dynamics.

The number of $e$-folds is given by integrating the expansion rate over the field excursion, starting from the definition of the $e$-folding number $d{\cal N}= H dt$
\begin{equation}
        \mathcal{N}_*= \int^{t_c}_{t_*} H dt = \int^{\phi_c}_{\phi_*} H \frac{dt}{d\phi}d\phi = \int^{\phi_c}_{\phi_*} \frac{H}{\dot\phi}d\phi
\end{equation}
We can now use the typical duration of each transition $\Delta t(\phi)=({1.4\Gamma^{1/4}(\phi)})^{-1}$, in order to approximate the field velocity as $\dot\phi \simeq \Delta\phi/\Delta t = \Delta\phi\times 1.4\Gamma^{1/4}(\phi)$, where $\Delta\phi$ is the ``jump" of the field between two consecutive minima and is thus entirely dictated by the periodicity of the tilted cosine. The above relation for the $e$-folding number thus becomes
\begin{equation}\label{eq:LowScale_Efolds}
\begin{aligned}
    \mathcal{N}_* &= \int_{0}^{\phi_c} d\phi\ \frac{H(\phi)}{1.4\Gamma_*^{1/4}\Delta\phi} 
    = \frac{1}{1.4\Gamma_*^{1/4}\Delta\phi}\int_0^{\phi_c} d\phi\ \sqrt{\frac{\mu^3(\phi_c-\phi)+\rho_{r}}{3M_{\rm Pl}^2}} \\
    &
    = \frac{H_*}{1.4\Gamma_*^{1/4}\Delta\phi}\int_0^{\phi_c} d\phi\ \sqrt{1-\frac{\phi}{\phi_c} + \frac{\rho_r}{V_*}} 
    = \frac{H_*}{1.4\Gamma_*^{1/4}}\frac{\phi_c}{\Delta\phi}\int^1_0 du\ \sqrt{1 - u + \frac{\rho_r}{V_*}} \\
    &\simeq \frac{N_*H_*}{1.4\Gamma_*^{1/4}} \left [ \frac{2}{3}+ \frac{\rho_r}{V_*} +
    \mathcal{O}\left (
\frac{\rho_r^{3/2}}{V_*^{3/2}} 
    \right )
    \right ]\simeq \frac{0.6}{1-n_s} \,,
\end{aligned}
\end{equation}
where we defined the dimensionless variable $u = \phi/\phi_c$, and applied the identity $\phi_c/\Delta\phi = N_*$.
In the second-to-last step we Taylor expanded the resulting  integral in powers of $\rho_r/V_* = \epsilon/2\ll 1$. The fraction of the radiation to total energy density can be also be expressed in terms of the spectral index through Eq.~\eqref{eq:ns_PureTiltedCosine} as $\rho_r/V_* = 0.3 (1-n_s)$. For $n_s\simeq 0.965$ the radiation fraction is $\rho_r/V_*\simeq 0.01$ and thus can safely be ignored in the calculation of ${\cal N}_*$.
It is worth noting that in the numerical results the term $\rho_r/V_*$ is retained, despite its small effect. A more detailed discussion of the radiation density in the tilted cosine model is shown in Appendix \ref{sec:RadiationAppendix}.

For observationally viable values of the spectral index, $n_s \approx 0.965$, Eq.~\eqref{eq:LowScale_Efolds} dictates that the Universe expands by only $15\lesssim \mathcal{N}_* \lesssim 30$ $e$-folds between the pivot scale and the end of inflation. This low number of $e$-folds naturally forces the model into the extreme lower end of allowed inflationary energy scales. To determine the physical consequences of the limited  expansion ($15\lesssim \mathcal{N}_* \lesssim  30$), we need to  match the inflationary expansion history to the subsequent thermal history of the Universe.

We can write the total number of $e$-folds from horizon crossing to the end of inflation as
\begin{equation}
    \mathcal{N}_* = \ln\left( \frac{a_c}{a_*} \right) \,,
\end{equation}
where $a_*$ and $a_c$ are the scale factors at pivot scale crossing and chain termination, respectively. The scale factor at horizon crossing is fixed by the kinematic condition $k_* = a_* H_*$. To determine $a_c$, we invoke the conservation of comoving entropy from the end of inflation to the present day, $S_c = S_0$. Assuming no significant entropy production post-inflation\footnote{In the case of a post-inflationary vacuum domination period ($\alpha>1$), there may be entropy production, in which case we must correct the equation of conservation of comoving entropy with a factor $\Delta_S>1$ such that $S_0=\Delta_SS_c$.}, this yields  $g_{*S}(T_c) T_c^3 a_c^3 = g_{*S}(T_0) T_0^3 a_0^3 $, leading to
\begin{equation}
 \frac{a_c}{a_0} = \left( \frac{g_{*S}(T_0)}{g_{*S}(T_c)} \right)^{1/3} \frac{T_0}{T_c} \,,
\end{equation}
where $T_0$ is the present CMB temperature, and $g_{*S}$ denotes the effective number of relativistic degrees of freedom for entropy. 

Setting the present scale factor to $a_0 = 1$ and combining these relations, we obtain an exact analytical expression for the required number of $e$-folds:
\begin{equation}\label{eq:N_efolds_exact}
    \mathcal{N}_* = \ln\left[ \frac{a_c}{a_0} \frac{a_0 H_*}{k_*} \right] = \ln\left[ \left( \frac{g_{*S}(T_0)}{g_{*S}(T_c)} \right)^{1/3} \frac{T_0}{k_*} \frac{H_*}{T_c} \right] \,.
\end{equation}
We can evaluate the constant term in this expression using standard cosmological parameters. For the \textit{Planck} pivot scale $k_* = 0.05 \text{ Mpc}^{-1} \approx 3.19 \times 10^{-40}$ GeV, the present CMB temperature $T_0 = 2.725 \text{ K} \approx 2.35 \times 10^{-13}$ GeV, and the degrees of freedom today $g_{*S}(T_0) \simeq 3.91$ and at high energies $g_{*S}(T_c) \sim 106.75$, the logarithmic pre-factor evaluates to
\begin{equation}
    \ln\left[ \left( \frac{3.91}{106.75} \right)^{1/3} \frac{2.35 \times 10^{-13} \text{ GeV}}{3.19 \times 10^{-40} \text{ GeV}} \right] \approx 62 \,.
\end{equation}
We may thus write the $e$-folds of expansion in terms of the inflationary energy scale as
 \begin{equation}\label{eq:N_efolds_62}
    \mathcal{N}_* \approx 62 + \ln\left( \frac{H_*}{T_c} \right) \,.
\end{equation}
We can use the relation between the scale of inflation, the reheating temperature and the number of $e$-folds to translate the bounds on ${\cal N}_*$ onto a bound for $\ln (H_*/T_c)$ as
\begin{equation}
-47\lesssim \ln \left (\frac{H_*}{T_c}\right )\lesssim -32 \, .
\end{equation}
Both $H_*$ and $T_*$ depend on the potential scale $V_*$. The former is given through the Friedmann equation, $H_* = \sqrt{V_*/(3M_{\rm Pl}^2)}$.
To determine the exact relation between the reheating temperature $T_c$ and the potential step $\Delta V$, we start by equating the radiation attractor solution \eqref{eq:RadiationAttractorSolution} at the end of inflation to the relativistic energy density of the thermal bath. Using the Stefan-Boltzmann law, we write
\begin{equation}
    \rho_{r,c} = \frac{\pi^2}{30} g_*(T_c) T_c^4 = \frac{1.4}{4} \frac{\Gamma_*^{1/4}}{H_*}\Delta V \,,
\end{equation}
where $g_*(T_c)$ is the effective number of relativistic degrees of freedom for the energy density. Because the kinematic ratio $\Gamma_*^{1/4}/H_*$ is strictly fixed by the CMB scalar power spectrum amplitude $A_s$ via Eq.~\eqref{eq:ScalarPowerSpectrumConstraint}, we can substitute it directly to obtain:
\begin{equation}
    \frac{\pi^2}{30} g_*(T_c) T_c^4 = \frac{1.4}{4} \left( \frac{A_s}{0.06} \right)^{-3/5} \Delta V \,.
\end{equation}
Solving for the reheating temperature yields:
\begin{equation}
\label{eq:Tclowscale}
    T_c = \left[ \frac{10.5}{\pi^2 g_*(T_c)} \left( \frac{A_s}{0.06} \right)^{-3/5} \right]^{1/4} \Delta V^{1/4} \,.
\end{equation}
Using the observed \textit{Planck} amplitude $A_s \simeq 2.10 \times 10^{-9}$ and assuming the Standard Model particle content at high energies ($g_* \simeq 106.75$), the dimensionless pre-factor evaluates to $\approx 4.15$. We therefore find the  relation:
\begin{equation}\label{eq:Tc_exact}
    T_c \approx 4.15 \, \Delta V^{1/4} \,.
\end{equation}
We can also connect the reheat temperature to the tensor to scalar ratio, by noting that the total number of transitions is $N_* = V_*/\Delta V$ and through Eq.~\eqref{eq:ns_PureTiltedCosine} we get
\begin{equation}
T_c = 4.15 \left [\frac{ (1-n_s)   V_* }{3.5\times 10^4 }\right ]^{1/4} \simeq 0.3 {V_*^{1/4} (1-n_s)^{1/4}} \, .
\end{equation}
Exponentiating Eq.~\eqref{eq:N_efolds_62} and using the formulae for $T_c$ and $H_*$, we get 
\begin{equation}\label{eq:ns_InflationScale_Relation}
    V_*^{1/4}  \approx \exp\left(\frac{0.6}{1-n_s}\right)(1-n_s)^{1/4}\times 1.50\times10^{-9} \ \GeV \,.
\end{equation}

Equation~\eqref{eq:ns_InflationScale_Relation} explicitly demonstrates the central limitation of the tilted cosine potential: the inflationary scale and the CMB observables are  tied to each other. Given the tight observational constraints on the spectral index, $1-n_s=0.037\pm0.005$ at the $68\%$ confidence level (\textit{Planck}+BICEP/Keck \cite{Planck:2018jri}), the largest possible inflationary scale for this potential at the $2\sigma$ level is severely bounded:
\begin{equation}
     V_*^{1/4} \lesssim  3 \ \GeV\,.
\end{equation}

This is still cosmologically viable, as it leads to a temperature $T_c\lesssim300$ MeV, which is only slightly above the ${\cal O}({\rm MeV})$ scale required by BBN, even when considering the largest possible temperature allowed at the $2\sigma$ level by \textit{Planck} for this model. However the proximity of  reheating  to BBN leaves little room for accommodating early Universe phenomenology, such as leptogenesis or baryogenesis models. 
For instance, standard Electroweak Baryogenesis requires the Universe to reach the critical temperature of the electroweak phase transition to prevent sphaleron freeze-out, imposing $T_c \gtrsim 100$ GeV \cite{Buchmuller:2012tv}.
As another example, in the seesaw model, the baryon asymmetry of the Universe can be generated by the decay of the lightest right-handed neutrino, and if these are produced thermally, then leptogenesis requires a minimum reheating temperature of the Universe $T_c\gtrsim10^8-10^9$ GeV \cite{Davidson:2002qv,Buchmuller:2004nz} to satisfy the Davidson-Ibarra bound, placing it well outside the viable parameter space for a slowly-varying chain inflation potential. Consequently, generating the observed matter-antimatter asymmetry within this model must rely on  a low-scale  mechanism. 
For example, the asymmetry could be achieved via high-energy sphaleron transitions from out-of-equilibrium decays, which can operate at $0.1-1$ GeV \cite{Jaeckel:2022osh}, or through specific heavy moduli decays that remain viable down to $\sim 100$ MeV \cite{Buchmuller:2012tv}. 
Despite the inherent limitation of a low reheat temperature, chain inflation also exhibits an inherent advantage:
the non-adiabatic expansion and collision of bubble walls during the final transitions provide a
highly out-of-equilibrium environment, automatically fulfilling the third Sakharov condition. In Appendix~\ref{app:low_scale_baryogenesis} we sketch some generic requirements for baryogenesis after low-scale chain inflation and  propose some pathways to achieve it.  The ${\rm GeV}$ scales involved open an intriguing avenue for complementary signals in laboratory searches. We leave such an analysis for future work.

The predictive power of Eq.~\eqref{eq:ns_InflationScale_Relation}  is best seen once it is inverted to yield an analytical expression for the spectral index as a function of the inflationary energy scale. Solving Eq.~\eqref{eq:ns_InflationScale_Relation} for $n_s$ yields:
\begin{equation}\label{eq:nsvsV_cosine}
   n_s-1 =  \frac{2.4}{W_{-1}\left( -2.4 \left[ \frac{1.50 \times 10^{-9} \ \GeV}{V_*^{1/4}} \right]^4 \right)} \simeq 
   \frac{2.4}{\ln\left( 2.4 \left[ \frac{1.50 \times 10^{-9} \ \GeV}{V_*^{1/4}} \right]^4 \right)} \,,
\end{equation}
where $W_{-1}$ denotes the lower branch of the Lambert $W$ function. The approximation in the final step holds for arguments that are negative and close to zero. This  requires $V_*^{1/4} \gg 10^{-9} \, {\rm GeV}$, which is satisfied for all energy scales of interest. 

\begin{figure}[t]
    \centering
    \includegraphics[width=0.9\linewidth]{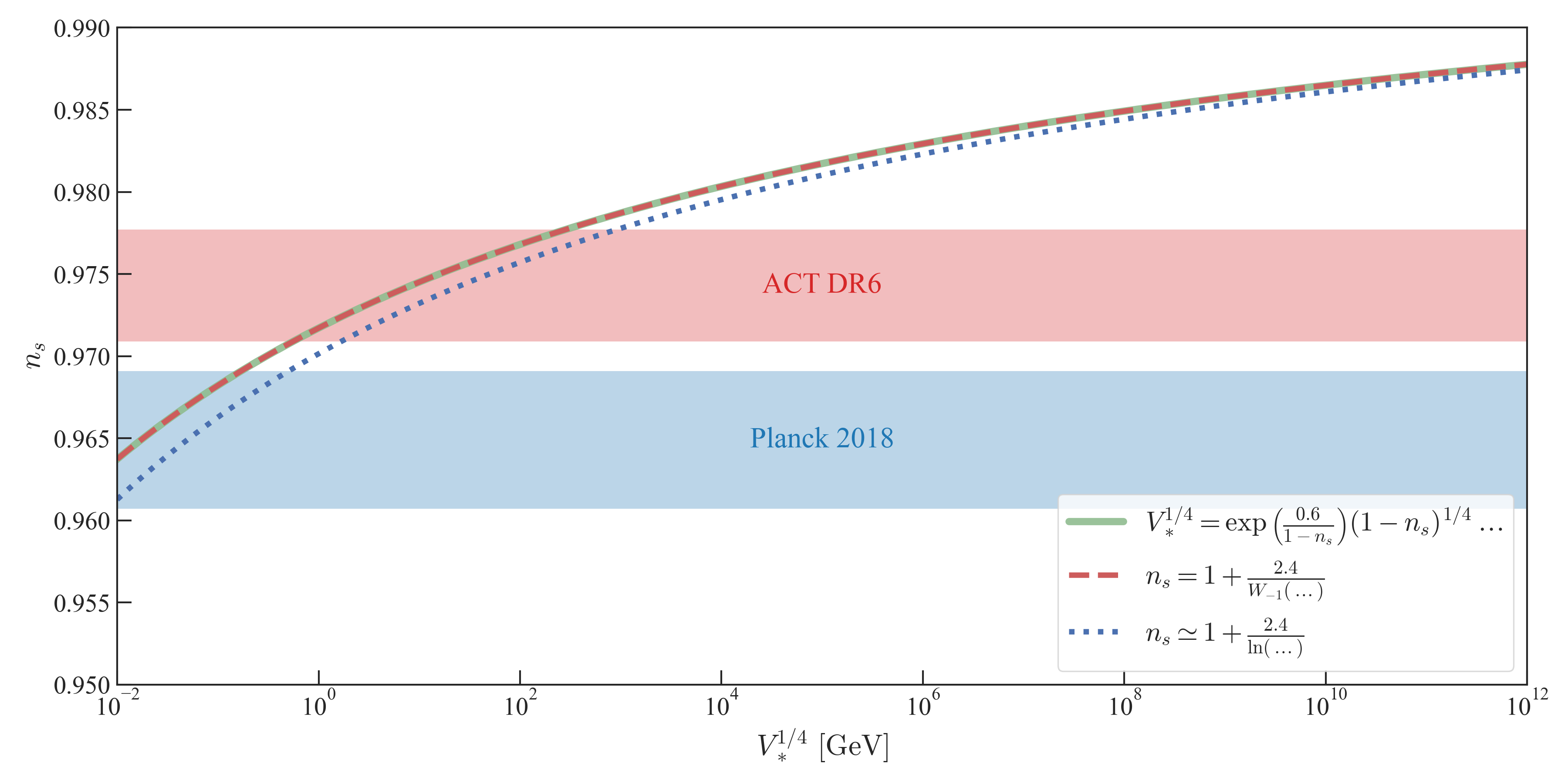}
    \caption{For the case of standard Chain Inflation ($\xi$=0), the analytical relation between the inflationary scale $V_*^{1/4}$ in the CMB window and the spectral index $n_s$, as derived in Eq.~\eqref{eq:ns_InflationScale_Relation} (solid green curve). The exact inversion using the Lambert $W$ function and its logarithmic approximation (Eq.~\eqref{eq:nsvsV_cosine}) are shown in red dashed and blue dotted curves, respectively. The relative error of the logarithmic approximation is $\lesssim 0.25\%$ across the entirety of the physically allowed energy regime. The blue and red bands show the $1\sigma$ regions for the \textit{Planck} and ACT constraints on $n_s$. 
    }
    \label{fig:LambertApproximation}
\end{figure}

As shown in Figure~\ref{fig:LambertApproximation}, the simple logarithmic approximation provides a strikingly simple and extraordinarily accurate expression for $n_s(V_*)$. Because $n_s$ is a monotonically increasing function of $V_*$, pushing the theory to high energy scales quickly violates CMB constraints. 

In this subsection we showed that the pure tilted cosine chain inflation model  is restricted to low-scale inflation with overall potential height 
$V_*^{1/4} \lesssim  3 \ \GeV\,$ at the $2\sigma$ level.  To obtain this bound, we used a combination of constraints in Eqs.~\eqref{eq:ns_PureTiltedCosine},~\eqref{eq:LowScale_Efolds},~\eqref{eq:N_efolds_62} and~\eqref{eq:Tclowscale}  as well as the relations $N_* = V_*/\Delta V$ to arrive at Eq.~\eqref{eq:ns_InflationScale_Relation}, which relates the potential height to the observed spectral index $n_s$.

We therefore conclude that single-field chain inflation governed by a pure tilted cosine potential is fundamentally incompatible with high-scale inflation. To break this rigidity and elevate the theory to energy scales capable of producing observable stochastic gravitational waves, a dynamical mechanism must be introduced to break the constancy of the tunneling rate, thereby providing non-zero $S_{i\geq0}$. In  Section~\ref{sec:NMC} we demonstrate that a non-minimal coupling to gravity naturally provides exactly such a mechanism.

\subsection{Running of the Spectral Index for the Pure Tilted Cosine Potential}
\label{sec:cosinerunning}

Future precision cosmological surveys, such as the Simons Observatory and CMB-S4, will place stringent constraints on the scale dependence of the primordial power spectrum. This scale dependence is parameterized by the running of the spectral index, $\alpha_s \equiv d n_s / d \ln k$. 
By using the constant tunneling rate of the pure tilted cosine  ($S_1=S_2=0$) in Eq.~\eqref{eq:Running_simplified}, we find the associated running of the spectral index expressed purely in terms of $n_s$
\begin{equation}\label{eq:Running_PureCosine}
    \alpha_s = -\frac{9}{5} (1-n_s)^2 \,.
\end{equation}
This represents a strict, falsifiable prediction for standard chain inflation with a tilted cosine potential. For an observed spectral index of $n_s \simeq 0.965$, the pure tilted cosine  predicts a running of $\alpha_s \approx -0.0022$. This value is entirely consistent with current {\it Planck}  constraints ($\alpha_s = -0.0045 \pm 0.0067$ \cite{Planck:2018jri}), but will serve as a definitive observational test for the theory given the projected $\sigma(\alpha_s) \sim 0.002$ sensitivity at 95\% confidence level from merging next-generation CMB surveys with large scale structure data~\cite{Bahr-Kalus:2022prj}. 
This subsection concludes our discussion of the pure tilted cosine in the absence of coupling to gravity.

\section{Non-Minimally Coupled Chain Inflation}\label{sec:NMC}

Whereas Section~\ref{sec:StandardChainInflation} studied the case of standard chain inflation with $\xi=0$, throughout the current Section~\ref{sec:NMC} we turn to non-minimally coupled chain inflation. 
Specifically, given the low-scale constraint of chain inflation governed by a pure tilted cosine potential, we move towards a modification that is well motivated, has few additional parameters (ideally only one), so as to avoid fine tuning, and  retains (most of) the analytical elegance  of the tilted cosine model. 
Here we introduce a one-parameter modification of the standard inflaton action through a non-minimal coupling with gravity, and detail its effects on the theory's predictions, as well as its theoretical implications on the model's parameters. 

From the perspective of quantum field theory in curved spacetime, a non-minimal coupling between the inflaton and the Ricci scalar of the form $\xi R\phi^2$ is not an ad-hoc theoretical addition, but rather a more generic feature generated by quantum corrections \cite{Callan:1970ze}. In the broader inflationary literature, non-minimal couplings have been highly successful in rescuing otherwise excluded phenomenological models, such as in Higgs inflation, where the coupling conformally flattens the steep quartic potential of the Standard Model Higgs to align perfectly with CMB constraints \cite{Bezrukov:2007ep} and
have been an active area of research in inflation and preheating (see e.g.~Refs.~\cite{Fakir:1990eg,Hertzberg:2010dc, Kaiser:2012ak, Kaiser:2013sna, DeCross:2015uza, DeCross:2016cbs, 
DeCross:2016fdz, Nguyen:2019kbm, vandeVis:2020qcp}). In our case we will not resort to the very large ${\cal O}(10^3-10^4)$ values of the non-minimal coupling required e.g.~for Higgs inflation. We will rather focus on $\xi={\cal O}(10)$, which, as we will show, has important consequences for the dynamics and predictions of chain inflation. 

\subsection{Action and Field Redefinition}
We consider the action for a scalar field $\phi$ with a non-minimal coupling (NMC) to the gravitational sector of the form $\xi R\phi^2 $ and 
the simplest
chain inflation potential, the one describing a pure tilted cosine, $V_{\text{chain}}(\phi)$ as in Eq.~\eqref{eq:tilted_cosine}. In the Jordan frame, the action is given by 
\begin{equation}
    S^{(J)} = \int d^4x \sqrt{-g} \left[ \frac{1}{2} \left(1 + \xi \phi^2\right) R - \frac{1}{2} g^{\mu\nu} \partial_\mu \phi \partial_\nu \phi - V_{\text{chain}}(\phi) \right],
    \label{eq:jordan_action}
\end{equation}
where $g_{\mu\nu}$ is the Jordan frame metric, $g$ is its determinant and we again set the reduced Planck mass to unity $M_{\rm Pl}=1$. To remove the non-minimal coupling term, we perform a conformal transformation to the Einstein frame metric $\tilde{g}_{\mu\nu}$, defined by
\begin{equation}
\label{eq:Omegadef}
    \tilde{g}_{\mu\nu} = \Omega^2(\phi) g_{\mu\nu}, \quad \text{with} \quad \Omega^2(\phi) = 1 + {\xi \phi^2} \, .
\end{equation}
Under this transformation, the Ricci scalar transforms as
\begin{equation}
    R = \Omega^2 \left[ \tilde{R} + 3\tilde{\square}\ln\Omega^2 - \frac{3}{2} \tilde{g}^{\mu\nu} \partial_\mu (\ln\Omega^2) \partial_\nu (\ln\Omega^2) \right].
\end{equation}
Substituting these relations into Eq.~\eqref{eq:jordan_action} and integrating by parts yields the Einstein frame action
\begin{equation}
    S^{(E)} = \int d^4x \sqrt{-\tilde{g}} \left[ \frac{1}{2} \tilde{R} - \frac{1}{2} F(\phi) \tilde{g}^{\mu\nu} \partial_\mu \phi \partial_\nu \phi - \tilde{V}(\phi) \right],
\end{equation}
where the  Einstein-frame potential is 
\begin{equation} \tilde{V}(\phi) = \frac{V_{\text{chain}}(\phi)}{ \Omega^4(\phi) }\, .\end{equation}  The non-canonical kinetic term coefficient $F(\phi)$ is given by
\begin{equation}
    F(\phi) = \frac{1}{\Omega^2} + \frac{3}{2} \left( \frac{d \ln \Omega^2}{d\phi} \right)^2 = \frac{1 + \xi \phi^2 + 6 \xi^2 \phi^2}{(1 + \xi \phi^2)^2}.
\end{equation}
To write this action in terms of a canonical scalar field $\chi$, we enforce the relation\footnote{From this point onward in the main text we use $\phi$ to denote  the canonically normalized field in the Jordan frame and its associated potential's parameters, while $\chi$ is reserved for the canonically normalized field in the Einstein frame and its analogous relevant quantities.}
\begin{equation}\label{eq:CanonicalNormalization}
    \frac{d\chi}{d\phi} = \sqrt{F(\phi)} = \frac{\sqrt{1 + (\xi+6\xi^2) \phi^2 }}{1 + \xi \phi^2}.
\end{equation}
This relation can be solved analytically as 
\begin{equation}\label{eq:Chi_Phi_Transformation}
    \chi(\phi)=\sqrt{\frac{1+6\xi}{\xi}}\sinh^{-1}\left (\sqrt{\xi+6\xi^2}\phi\right )-\sqrt{6}\tanh^{-1}\left(\frac{\sqrt{6}\xi\phi}{\sqrt{1+(\xi+6\xi^2)\phi^2}}\right) \, ,
\end{equation}
which reduces to the expected relation $\chi\approx\phi$ for $\phi\ll1$ to lowest order (the corrections are studied in the following section).

The final action in terms of the canonical field $\chi$ takes the standard Einstein-Hilbert form
\begin{equation}
    S^{(E)} = \int d^4x \sqrt{-\tilde{g}} \left[ \frac{1}{2} \tilde{R} - \frac{1}{2} \tilde{g}^{\mu\nu} \partial_\mu \chi \partial_\nu \chi - \tilde{V}(\phi(\chi)) \right].
\end{equation}

Recasting the action in the Einstein frame is crucial for the application of the semi-classical theory results for the decay of the false vacuum, as originally formulated in Refs.~\cite{Kobzarev:1974cp, Coleman:1977py,Callan:1977pt}. 
By describing the tunneling particle in terms of the canonically normalized field $\chi$ in the Einstein frame Lagrangian, we may treat the field as if it was minimally coupled to gravity and therefore apply the standard results from the literature on fast tunneling through a chain of false vacua.

\subsection{Non-minimal Coupling Effects in the Sub-Planckian Field Regime}

We can now evaluate the behavior of the model in the regime where the field displacement is sub-Planckian; as we will show this approximation is indeed generic.
We then evaluate the Euclidean bounce action that determines the tunneling rate in this regime, and find its dependence on the field value as it tunnels down the chain of potential minima.

Considering that typical chain inflation models  obey $f<10^{10}\ \GeV\sim10^{-8}M_{\rm Pl}$ \cite{Winkler:2020ape}, we see that the typical displacement in field space during observable inflation is of the order $\Delta\phi_{\rm total}\approx N_* \Delta\phi\approx10^6f\lesssim0.01M_{\rm Pl}$, i.e. the overall displacement of the field should be sub-Planckian. Later in this work, we confirm this estimate explicitly for all of the available parameter space in the theory; see Figure~\ref{fig:FieldTraversal_vs_InflationaryScale}. We can thus, without much loss of generality, work in the $\phi/M_{\rm Pl}\ll1$ regime. 
There are then two clearly distinct behaviors for the evolution of the Euclidean action with the field depending on the coupling strength $\xi$. We start by noting that, in the sub-Planckian regime, one can relate the variation of the field $\phi$ with respect to $\chi$ as
\begin{equation}
\label{eq:dphidchismall}
    \frac{d\phi}{d\chi}\approx1-\chi^2\left(3\xi^2-\frac{\xi}{2}\right) \,,
\end{equation}
where we have expanded Eq.~\eqref{eq:CanonicalNormalization} for small field values. 
This will have an impact on the distance between adjacent minima, as determined by $f_\phi$ in the Jordan frame and $f_\chi$ in the Einstein frame 
since 
\begin{equation}
    f_\chi\equiv\frac{\Delta\chi}{2\pi}=\frac{\Delta\phi}{2\pi}\frac{\Delta\chi}{\Delta\phi}
   \simeq  f_\phi   \frac{d\chi}{ d\phi}
    = \frac{f_\phi}{1-\left(3\xi^2-\frac{1}{2}\xi\right)\chi^2} \, ,
\end{equation}
where in the last step we used Eq.~\eqref{eq:dphidchismall}.
The above equation explicitly shows that the field displacement for each tunneling event in the Einstein frame (i.e.~in terms of the field $\chi$) is not constant, since $f_\chi$ is a function of $\phi$, even though $f_\phi$ is constant. Simply put, in the Jordan frame the potential is strictly periodic in $\phi$ (with intervals of $2\pi f$), whereas this periodicity is lost in the Einstein frame. This aperiodicity  allows for high-scale chain inflation in the case of  non-minimal coupling, even for an exact tilted cosine Jordan-frame potential.

The field redefinition, together with the overall rescaling of the potential by $\Omega^4(\phi)$, also dictate a change in the tilt parameter 
\begin{equation}
    \mu_\chi^3=-\frac{dV^{(E)}}{d\chi}\approx\frac{d}{d\chi}\left[\frac{\mu^3_\phi\phi}{(1+\xi\chi^2)^2}\right]\approx\frac{\mu^3_\phi}{(1+\xi\chi^2)^2}\left[\frac{d\phi}{d\chi}-\frac{4\xi\chi^2}{1+\xi\chi^2}\right]\approx\frac{1-\left(3\xi^2+\frac{7}{2}\xi\right)\chi^2}{(1+\xi\chi^2)^2} \mu^3_\phi \,,
\end{equation}
where we have denoted the Einstein and Jordan frame potentials, related via a rescaling by a factor of $\Omega^4$, with superscript $E$ and $J$ respectively. 
Following the relation between the Einstein and Jordan frame potentials, in order to represent the Einstein potential locally as a tilted cosine, we can define the  barrier parameter $\Lambda^4$  in the Einstein frame as
\begin{equation}     
    \Lambda_\chi^4\equiv \frac{\Lambda_\phi^4}{(1+\xi\chi^2)^2} \,.
\end{equation}
We can now treat the Einstein-frame potential locally as a tilted cosine potential, where  $f_\chi$ and $\Lambda_\chi$  are local parameters, which only change  significantly over many phase transitions.

All of this causes a change in the tunneling parameter $x$ due to the tilted cosine parameters being modified
\begin{equation}\label{eq:x_NMC}
    x_\chi=\frac{\mu^3_\chi f_\chi}{\Lambda^4_\chi}\approx\frac{\mu^3_\phi f_\phi}{\Lambda^4_\phi}(1-4\xi\chi^2)=x_\phi(1-4\xi\chi^2) \, ,
\end{equation}
such that in the sub-Planckian field regime the tunneling parameter decreases quadratically with the field value\footnote{
This quadratic decrease is particularly important. At
$\chi=0$, one has $x_\chi=x_\phi$, so choosing $x\equiv x_\phi=0.96$ places the
model at the approximate upper boundary of the allowed regime discussed after Eq.~\eqref{eq:calSofx}. As the field moves away from the origin, however,
$x_\chi$ decreases, so the potential barriers become comparatively more
pronounced and the model moves further away from the runaway regime.
Thus, if the model is marginally consistent at $\chi=0$, it is
necessarily safer at nonzero $|\chi|$.
}, with this growth becoming steeper linearly with $\xi$. As we have seen that $S_E=\frac{f^4}{\Lambda^4}\mathcal{S}(x)$ for the case where the potential  can be well approximated by a tilted cosine,
the total effect on the Euclidean bounce action is thus a combination of the modification of the $\Lambda^4$ and $f$-dependent pre-factor together with $x_\chi$
\begin{equation}\label{eq:NMC_ChangeInAction}
\begin{aligned}
    S_{E,\chi}=\frac{f_\chi^4}{\Lambda^4_\chi}\mathcal{S}(x_\chi)&\approx\frac{(1+\xi\chi^2)^2}{\left[1+\chi^2\left(\frac{\xi}{2}-3\xi^2\right)\right]^4}\frac{f_\phi^4}{\Lambda^4_\phi}\mathcal{S}\big\lvert_{x=x_\phi(1-4\xi\chi^2)}\approx S_{E,\phi}\left[1+\chi^2 B(\xi,x_\phi)\right] \, ,
\end{aligned}
\end{equation}
where we have defined
\begin{equation}
\label{eq:Bdef}
    B(\xi,x_\phi)\simeq12\xi^2-4\xi x_\phi\frac{\mathcal{S}'(x_\phi)}{\mathcal{S}
(x_\phi)} \,.
\end{equation}
Note that we have expanded $S_{E,\chi}(x_\chi)$ around $x_\chi=x_\phi$ (i.e. around $\chi = 0$ in Eq.~\eqref{eq:x_NMC})
in order to recombine all Jordan frame terms into the constant bounce action $S_{E,\phi}$. 
We have numerically confirmed the accuracy of our derived expression for $B(\xi,x_\phi)$. Across seven orders of magnitude of the coupling parameter ($\xi\sim10^{-3}-10^4$), the analytical approximation is in excellent agreement with the numerical data, falling well within the $1\sigma$ uncertainty bounds of the numerical fits for both the weak and strong coupling regimes. 

In Eq. \eqref{eq:NMC_ChangeInAction} one sees that $S_{E,\chi}$ is minimized as $\chi=0$.
The final expression on the right-hand side of Eq.~\eqref{eq:NMC_ChangeInAction} greatly simplifies the analysis of the effect of the NMC on the tunneling action, modifying what would otherwise be a constant $S_E$ over all minima for the pure tilted cosine potential. Considering the behavior of $\mathcal{S}(x)$ over the physically relevant range $0<x<1$, it is clear that $\mathcal{S}'(x)<0$, such that both the linear and quadratic $\xi$ terms in the equation above lead to $B>0$, i.e. a quadratic increase of the action with the field $\chi$, thus slowing down the tunneling chain. Since the lifetime of each vacuum follows from the transition rate $\Gamma(S_E)$, this effectively means that $\xi$ controls the steepness of the evolution of $\Gamma(\chi)$ and thus will directly affect the relevant CMB observables of the theory for a fixed set of chain potential parameters. 

\subsection{Bounce Action Expansion
}\label{subsec:NMC_BounceActionExpansion}

With both the action and the lifetime's modifications by the NMC established to be well approximated by our theoretical predictions, we are in a position to establish the physical implications of these results. 

Using the last part of Eq.~\eqref{eq:NMC_ChangeInAction}, we can find the value of the action at a distance $\chi$ from the pivot scale crossing field value $\chi_*$ as 
\begin{equation}
    \label{eq:SEquadratic}
    S_E(\chi_*+\chi)=\SZero\left[1+B(\chi_*+\chi)^2\right]=S_{E,*}+2B\SZero\chi_*\chi+B\SZero\chi^2  \, ,
\end{equation}
 where $\SZero \equiv S_E(\chi=0)$ and $S_{E,*}  \equiv S_E(\chi_*)=  S_{E,0} (1 + B \chi_*^2)$.
Note that $S_{E,*}\neq \SZero$ for $\chi_*\neq0$, although their values should be of the same order of magnitude due to the constraint on the slow variation of the action from CMB data. Using the notation $\Delta\chi_{\rm total}=\chi_c-\chi_*$ as before (the subscript $c$ corresponds to the field value at the end of inflation, as in Section~\ref{sec:SE}),  we can identify the necessary coefficients for the expansion of the action around the pivot scale field value $\chi_*$ according to their definition in Eq.~\eqref{eq:S_expansion} as 
\begin{equation}\label{eq:SCoeff_NMC}
    S_1=2B\SZero\chi_*\Delta \chi_{\rm total} \quad \quad \text{and} \quad \quad S_2=B\SZero\Delta \chi_{\rm total}^2 \, ,
\end{equation}
with all higher-order coefficients being zero. Note that reinstating the Planck mass into these equations amounts to dividing both of the above results for $S_{1,2}$ by $M_{\rm Pl}^2$, such that their magnitude is dependent on the relative size of the scale of the field displacement to the Planck scale over observationally relevant regions of the chain. The coefficients shown above are uniquely defined in terms of the potential parameters, since $\SZero$ is determined from the Jordan frame parameters $\{\mu^3_\phi,\Lambda^4_\phi,f_\phi\}$, $B$ is calculated from the coupling parameter $\xi$ and $x_\phi$, $\chi_c$ is calculated as described in the previous section and $\chi_*$ is determined from $\chi_c$ through the requirements of the observational constraints.

In the numerical implementation of the model and its observables, it will prove useful to consider the variable  
\begin{equation}
u=\chi/\chi_c \, ,
\end{equation}
 such that $u_*=\chi_*/\chi_c$.
 The parameter $u$ is always defined as the inflaton field value during chain inflation normalized  by the  value at the end of inflation (subscript $c$ refers to the end of inflation). In the absence of non-minimal couplings, we recover $u=\phi/\phi_c$.
 The first two coefficients of the Taylor expansion of the Euclidean action, defined in Section~\ref{sec:SE}, become
\begin{equation}\label{eq:SCoeff_NMC_Redefinition}
    S_2=B\SZero\chi_c^2(1-u_*)^2 \quad \quad \text{and} \quad \quad S_1=\frac{2u_*}{1-u_*}S_2 \, .
\end{equation}
Note that $u_*<1$ by definition, since we are considering a potential with a negative tilt, such that $\chi_*<\chi_c$. For the  case where the pivot scale is fixed at the origin of the potential ($u_*=0$), the non-minimal coupling term $\xi R\phi^2 $ term vanishes at the pivot scale, as does $S_1$,
 All of the theory's modifications will then be determined by the coefficient $S_2$, which is the second derivative of the Euclidean action with respect to the field. The importance of $S_2$ will be discussed in more detail in the following sections.  

Let us stress the importance of the value of $u_*$ in the non-minimally coupled model and contrast it with the minimally coupled case of Section~\ref{sec:StandardChainInflation}. In the standard minimally coupled case of a tilted cosine potential, the field value at the end of chain inflation can be arbitrarily chosen, since a shift in $\phi$ can be compensated by a different choice of the constant $C$ in Eq.~\eqref{eq:tilted_cosine}. In the case of a non-minimally coupled tilted cosine, this discrete shift symmetry is entirely lost through the gravitational coupling $\xi R \phi^2$. The latter defines an absolute origin for the field value $\phi$. This underlines the importance of the parameter $u_*$. For $u_*=0$  the non-minimal correction to the gravitational action (in the Jordan frame) vanishes at the pivot scale. For the case $u_*\ll -1$, the gravitational correction is maximal at the pivot scale and vanishes at the end of chain inflation, which occurs for $u=1$.

From the dependence of the spectral scalar index $n_s$ on $S_1$ given in Eq.~\eqref{eq:ns_Effective} and the imposition of a blue-tilted spectrum ($n_s<1$), one finds the upper bound $S_1<2$. This imposes an unavoidable constraint on the theory's parameters, which we write as $B\SZero\chi_*\Delta \chi_{\rm total}<1$. Of course, in practice the exact observed value of $n_s$ can (and must) be used to constrain the theory's parameter space to a closed region, instead of simply placing an upper bound on $S_1$.

\subsection{Modified Constraints in the Non-minimally Coupled Theory}

Having established how the non-minimal coupling induces a field-dependent modification of the Euclidean bounce action, we now translate this effect into constraints on the model parameters. The purpose of this subsection is to focus on the regions of parameter space that satisfy the basic requirements of chain inflation, rather than to present the full phenomenological consequences of the model. We impose constraints on the amplitude and spectral index of density perturbations from CMB observations, as well as impose
the requirement of sufficient expansion between the CMB pivot scale and the end of inflation. These conditions determine how the inflationary scale $V_*^{1/4}$, the coupling $\xi$, the tunneling parameter ($x$), and the relative pivot location $u_*=\chi_*/\chi_c$ are tied to one another. The detailed analytic estimates underlying the special $u_*=0$ branch, the perturbative-unitarity upper bound, and the large negative-$u_*$ universality regime are collected in Appendix \ref{app:AnalyticResults}; here we summarize their physical implications and use them to interpret the numerical parameter-space results.

In terms of the previously defined parameters, the primordial power spectrum amplitude constraint $A_s = 0.06 (\Gamma_*^{1/4}/H_*)^{-5/3}$, with $A_s=2.1\times10^{-9}$, can be rewritten using Eqs.~\eqref{eq:NMC_ChangeInAction} and~\eqref{eq:SCoeff_NMC_Redefinition} as 
 \begin{equation}\label{eq:NMC_AmplitudeConstraint}
      \underbrace{\Gamma_*/H_*^4}_{7.89\times10^{17}}=\frac{\Gamma_0\exp(-B\SZero \chi^2_c u_*^2)}{V_*^2/9}=\frac{9\Gamma_0}{V_*^2}\exp\left(-S_2\frac{u_*^2}{(1-u_*)^2}\right) \, ,
 \end{equation}
 where $\Gamma_0 $ is equal to the constant decay rate of a pure tilted  cosine without non-minimal coupling, i.e. $\Gamma(\chi=0)$.
Furthermore, in the above expression, we took  $H_*$ to only depend on the potential; we dropped the negligible effect of radiation at the pivot scale, since $V_*\gg\rho_r$. We should note that in all our numerical calculations, the radiation component was included for completeness, from which we confirmed that it has little to no effect on the results. For this model, the determination of $n_s$ shown in Eq.~\eqref{eq:ns_Effective} becomes 
\begin{equation}\label{eq:ns_Constraint_NMC}
 1-n_s=\frac{3.48\times10^4}{N_*}\left(1-\frac{S_2u_*}{1-u_*}\right) \,,
\end{equation}
where we have rewritten $S_1$ in terms of $S_2$ through Eq.~\eqref{eq:SCoeff_NMC_Redefinition} and used the relation $N_* = V_*/\Delta V$.  Although $\Delta V_\chi=2\pi\mu^3_\chi f_\chi$ is field-dependent in the Einstein frame due to the NMC, we have quantified its deviation from the constant Jordan frame $\Delta V_\phi$ for all cases of interest to this paper and found that this is at most a $\sim5\%$ difference\footnote{This can be readily seen by noting that the largest field value during observable inflation is roughly $\Delta\chi_{\rm total}=\sqrt{S_2/(B\SZero)}$ and that $\Delta V_\chi\approx\Delta V_\phi(1-6\xi\chi^2)$, such that $|1-\Delta V_\chi/\Delta V_\phi|\approx\frac{6\xi S_2}{12\xi^2\SZero}=\frac{S_2}{2\xi\SZero}$ for $\xi\gtrsim\mathcal{O}(1)$. As we will show later in this paper,   compatibility with CMB observations imposes $\xi\gtrsim30$, $\SZero\gtrsim3$ and $S_2\lesssim10$, meaning that $|1-\Delta V_\chi/\Delta V_\phi|\lesssim5\%$.}, thus justifying the continued usage of $N_*=V_*/\Delta V$ even when adding non-minimal coupling.

The number of $e$-folds between the pivot scale and the termination of chain inflation can be calculated similarly to Eq.~\eqref{eq:LowScale_Efolds}
 \begin{equation}\label{eq:HighScale_Efolds}
     \begin{aligned}
     \mathcal{N}_*&=
     \int_{\chi_*}^{\chi_c} d\phi\ \frac{H(\chi)}{1.4\Gamma_*^{1/4}\Delta\chi} 
     =
     \frac{N_*H_*}{1.4\Gamma_*^{1/4}(1-u_*)}\int_{u_*}^{1} du\ \sqrt{\frac{1-u}{1-u_*}}\exp\left[S_2\frac{u^2-u_*^2}{4(1-u_*)^2}\right]\,.
\end{aligned}
\end{equation}
In the last step of the above equation we used $H^2 = V/3M_{\rm Pl}^2$ instead of $V+\rho_{r}$. We also defined $N_* = V_*/\Delta V_\chi$ even though $\Delta V_\chi$ is not exactly constant due to the NMC. Both approximations are valid at the few percent level, as we discussed previously.  We have kept the variable in the integral as $u\equiv\chi/\chi_c$ to ensure that the equation is written fully in terms of the chosen parameters. The integral can be simplified even further by reintroducing the fractional field displacement 
\begin{equation}
    \tilde\chi=\frac{u-u_*}{1-u_*}=\frac{\chi-\chi_*}{\Delta\chi_{\rm total}} \Rightarrow \tilde\chi\in[0,1]\,.
\end{equation}
This redefinition immediately cancels out the $u_*$ dependence on both the integral bounds and the pre-factor, since $du=(1-u_*)dy$. Note that the fractional field displacement $\tilde\chi$ matches the analogous definition of $\tilde\phi$ introduced above Eq.~\eqref{eq:S_expansion}. The exponent in the NMC dilation term is also simplified 
\begin{equation}
  S_2\frac{u^2-u_*^2}{4(1-u_*)^2}=\frac{S_2}{4}\tilde\chi\left(2\frac{u_*}{1-u_*}+\tilde\chi\right) \, ,
\end{equation}
while the term in the square root becomes particularly more elegant
\begin{equation}
    \frac{1-u}{1-u_*}=\frac{(1-\tilde\chi)(1-u_*)}{1-u_*}=1-\tilde\chi \,.
\end{equation}
 Putting all of these changes together we obtain the much simpler constraint
\begin{equation}\label{eq:HighScale_Efolds_Redefined}
     \begin{aligned}
   {\cal N}_*=  \frac{N_*H_*}{1.4\Gamma_*^{1/4}}\int_{0}^{1} d\tilde\chi\ \sqrt{1-\tilde\chi}\exp\left[\frac{S_2}{4}\tilde\chi\left(2\frac{u_*}{1-u_*}+\tilde\chi\right)\right]\approx62+\ln\left(\frac{H_*}{T_c}\right) \, ,
\end{aligned}
\end{equation}
where for the last equality we used the standard relation for the number of $e$-folds given in Eq.~\eqref{eq:N_efolds_62}.
In the formulation of Eq.~\eqref{eq:HighScale_Efolds_Redefined} we 
 isolated the dependence on $u_*$ into one single term, which is present in the exponent of the NMC lifetime dilation term.

With this complete set of modified constraints established, the viable parameter space of the theory can be fully determined.

\section{CMB Observables }\label{sec:Results}

We now turn to the observable consequences of the non-minimally coupled chain inflation scenario. We will show that the modified tunneling rate opens a high-scale region of parameter space. We use this section to confront the model with cosmological data and to identify its characteristic signatures. We therefore focus on CMB observables, specifically on the scalar spectral index $n_s$ and the full scale-dependence of the  scalar power spectrum $\Delta^2_{\cal R}(k)$. In Section~\ref{subsec:GWSpectrum} we show the stochastic gravitational wave background sourced by bubble collisions, whose peak frequency and amplitude provide a complementary probe of the inflationary scale and the dynamics of the final transitions. In this way, the scalar and gravitational wave sectors together test both the tunneling history during inflation and the mechanism by which the chain ends.

\subsection{Scalar Spectral Index and  CMB constraints}\label{subsec:NMC_SpectralIndex}

Here we compute predictions for the spectral index of the CMB in chain inflation with non-minimally coupled gravity and compare to CMB data.
{In order to compute the function $n_s(V_*)$ we first choose specific values for $x$, $\xi$ and $u_*$,  since  each triple of values $x, \xi, u_*$ corresponds to  a single curve $n_s(V_*)$, as we will show in Figures~\ref{fig:ns_vs_V*_Full} and~\ref{fig:ns_vs_V*_ZoomGrid}. We then numerically solve the system of Eqs.~\eqref{eq:NMC_AmplitudeConstraint} and \eqref{eq:HighScale_Efolds_Redefined}. Specifically, for any chosen set of fixed parameters $\{x, \xi, u_*\}$ and for each $V_*$ the constraints~\eqref{eq:NMC_AmplitudeConstraint} and~\eqref{eq:HighScale_Efolds_Redefined}  uniquely fix the total number of transitions $N_*$ and the baseline Euclidean action $\SZero$, from which $n_s$ may be determined through Eq.~\eqref{eq:ns_Constraint_NMC}. By scanning over $V_*$ values, we map the entire curve $n_s(V_*)|_{x, \xi, u_*}$.

The above Eqs.~\eqref{eq:NMC_AmplitudeConstraint}, \eqref{eq:ns_Constraint_NMC} and~\eqref{eq:HighScale_Efolds_Redefined}  were derived using only a few well-controlled assumptions. We have used the fact that the field is sub-Planckian, which allows us to terminate the Taylor expansion of the Euclidean action at the term $S_2$ and expand the Euclidean action following Eq.~\eqref{eq:SEquadratic}. Finally, we approximated the number of transitions $N_*$ using the constant $\Delta\chi$ approximation, which we quantified to be accurate at the few percent level. We should note that in our numerical calculations the radiation energy density is included in the integral for the number of $e$-folds, which is missing from Eq.~\eqref{eq:HighScale_Efolds_Redefined}, and that the parameter $B$ that first appeared in Eq.~\eqref{eq:NMC_ChangeInAction}  is computed through Eq.~\eqref{eq:Bdef}.

Figures~\ref{fig:ns_vs_V*_Full} and \ref{fig:ns_vs_V*_ZoomGrid} show the scalar spectral index $n_s$ as a function of the inflationary scale $V_*^{1/4}$ for non-minimally coupled models. Different symbols correspond to different values of the ratio of the field value at the pivot scale and the end of inflation; $u_*=\chi_*/\chi_c=-10,-1,0,0.06$ (circle, square, triangle and rhombus respectively). As discussed in Section~\ref{subsec:NMC_BounceActionExpansion}, the gravitational coupling $\xi R \phi^2$ defines an absolute origin for the field value, so that $u_*$ has a physical significance it was lacking in the $\xi=0$ case. For $u_*=0$  the non-minimal correction to the gravitational action (in the Jordan frame) vanishes at the pivot scale. For the case $u_*\ll -1$, the gravitational correction is maximal at the pivot scale and vanishes at the end of chain inflation, which occurs for $u=1$. In the figures, the non-minimal coupling $\xi$ is set to $\xi=60$ and the tunneling parameter is $x=0.96$ for all points in Figure~\ref{fig:ns_vs_V*_Full}.
For  inflationary scales $V_*^{1/4}\lesssim 10^{10}\, {\rm GeV}$, all the points fall on top of the prediction of the simple tilted cosine model of Eq.~\eqref{eq:ns_InflationScale_Relation}.  However, for $V_*^{1/4}\gtrsim 10^{10}\, {\rm GeV}$, the non-minimal coupling changes the results significantly and allows for agreement between predictions and data for the case of the pure tilted cosine potential.

\begin{figure}
\centering
\includegraphics[width=\linewidth]{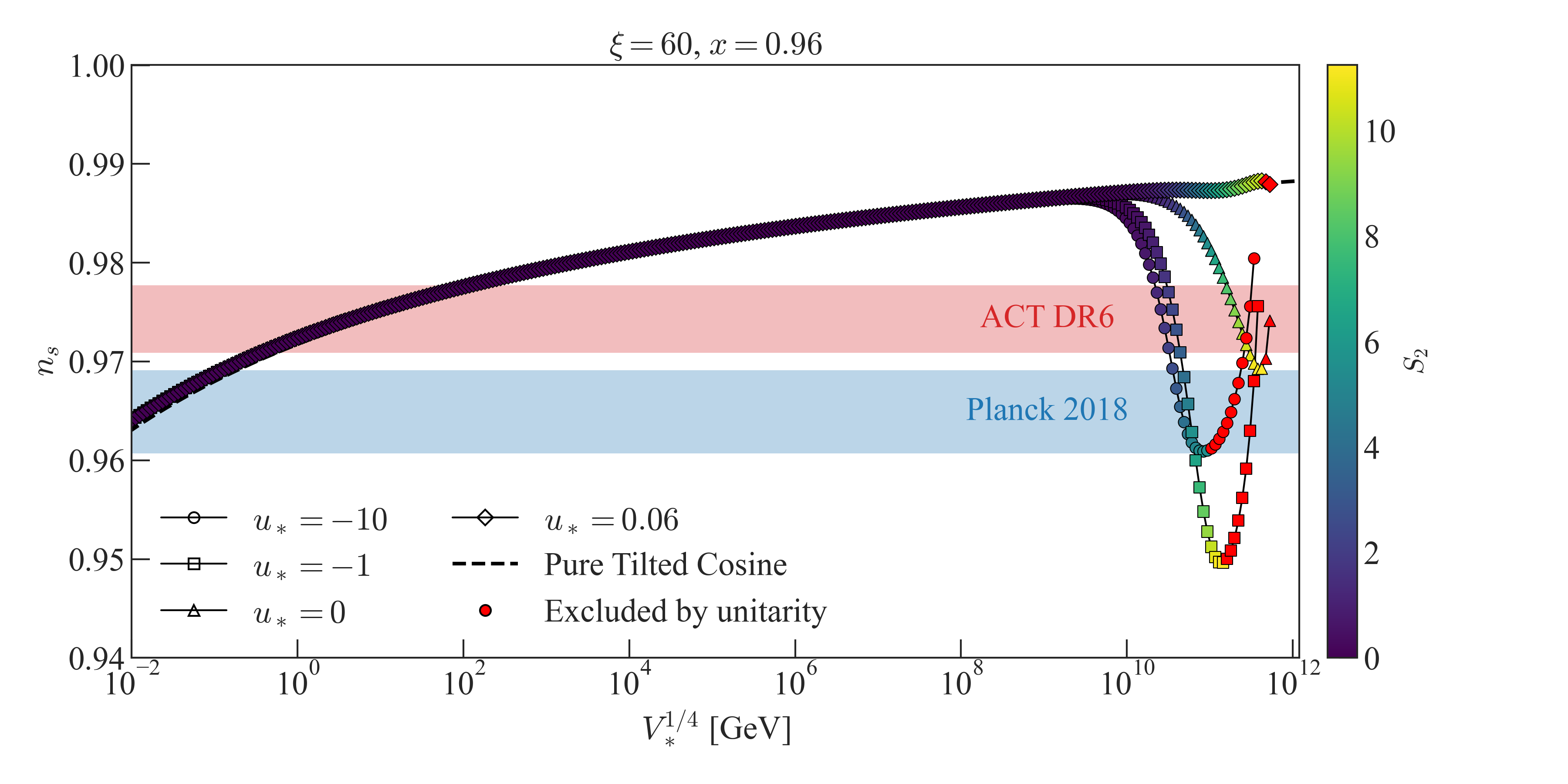}
\caption{  The scalar spectral index $n_s$ as a function of the inflationary scale $V_*^{1/4}$ for non-minimally coupled models. Different symbols correspond to different values of the ratio of the field value at the pivot scale and the end of inflation; $u_*=\chi_*/\chi_c=-10,-1,0,0.06$ (circle, square, triangle and rhombus respectively).
(As a reminder, the gravitational coupling $\xi R \phi^2$ defines an absolute origin for the field value, so that $u_*$ has  physical significance that it lacks in the $\xi=0$ case).
For exemplary purposes, the non-minimal coupling $\xi$ is set to $\xi=60$ and the tunneling parameter defined in Eq. \eqref{eq:x_definition} is $x=0.96$ for all points, where $x$ is the ratio of the energy difference between successive minima to the barrier height, and $x$ close to unity corresponds to fast tunneling. 
For  inflationary scales $V_*^{1/4}\lesssim 10^{10}\, {\rm GeV}$, all the points fall on top of the prediction of the simple tilted cosine model of Eq.~\eqref{eq:ns_InflationScale_Relation} (the dashed line of the simple tilted cosine is therefore covered by a broader solid line since so many models converge upon the same result). This overlapping effect results from the fact that small inflationary scales correspond to small overall field displacement, making the effects of the non-minimal coupling very suppressed and practically irrelevant.
For larger scales the NMC effects give rise to significant deviations from the pure tilted cosine result.
The color coding on the right indicates the value of $S_2$ for the different points, illustrating the intensity of the effects of the NMC on the observables of the theory.
$S_2$ is the second-order Taylor-expanded term of the Euclidean action and it controls the spectral index $n_s$ per Eqs.~\eqref{eq:ns_Constraint_NMC},~\eqref{eq:uStar0_V*} and~\eqref{eq:uStar_MinusInfty_V*}. As discussed in the main text, the minimum value of $n_s(V_*)$ is reached when $S_2$ is maximized. This effect is evident by the color-coding in this figure. It is worth noting that in the pure tilted cosine $S_2=0$ and thus $S_2$ encodes the effects of the NMC.
Red points are excluded by unitarity bounds. The horizontal colored bands show the 1$\sigma$ regions of the \textit{Planck} 2018 (blue) and ACT DR6 (red) constraints on $n_s$.  
}
\label{fig:ns_vs_V*_Full}
\end{figure}

The figures illustrate the unique solution for small values of $V_*^{1/4}$ (around $V_*^{1/4}\lesssim10^{8}$ GeV for the chosen values of $\xi\sim\mathcal{O}(10)$), where the value of $n_s(V_*)$ is independent of the non-minimal coupling $\xi$ and follows the tilted cosine result. This symmetry can be  understood as follows. For small values of the inflationary scale, the inflaton excursion $\Delta\phi_{\rm tot} \equiv|\phi_*-\phi_c|$ is also reduced, as is the scale $f$.
In our case, the inflaton either starts at the origin (for $u_*=0$) or ends at/close to the origin (for $u_*<0$). Thus the magnitude of the field excursion is also the maximum value of the field amplitude itself.
As the typical inflaton field value becomes smaller, it suppresses the effect of the non-minimal coupling. Referring back to Eq.~\eqref{eq:Omegadef} and restoring the Planck mass for clarity, the effect of the non-minimal coupling is seen through
\begin{equation}
\label{eq:NMClimit}
    \Omega^2 \sim 1+\xi\left (\frac{\Delta\phi_{\rm tot}}{ M_{\rm Pl}}\right )^2\simeq 1 \, , \quad {\rm for}~\sqrt\xi \Delta\phi_{\rm tot} \ll M_{\rm Pl} \, ,
\end{equation}
meaning that for small enough field values, much smaller than $M_{\rm Pl}/\sqrt{\xi}$, the effects of the non-minimal coupling on the inflationary dynamics are severely suppressed. As a reminder, $\Omega(\phi)$ is the function defining the conformal transformation to the Einstein frame metric, as defined in Section~\ref{sec:NMC}. 
}

\begin{figure}
    \centering
    \includegraphics[width=0.49\linewidth]{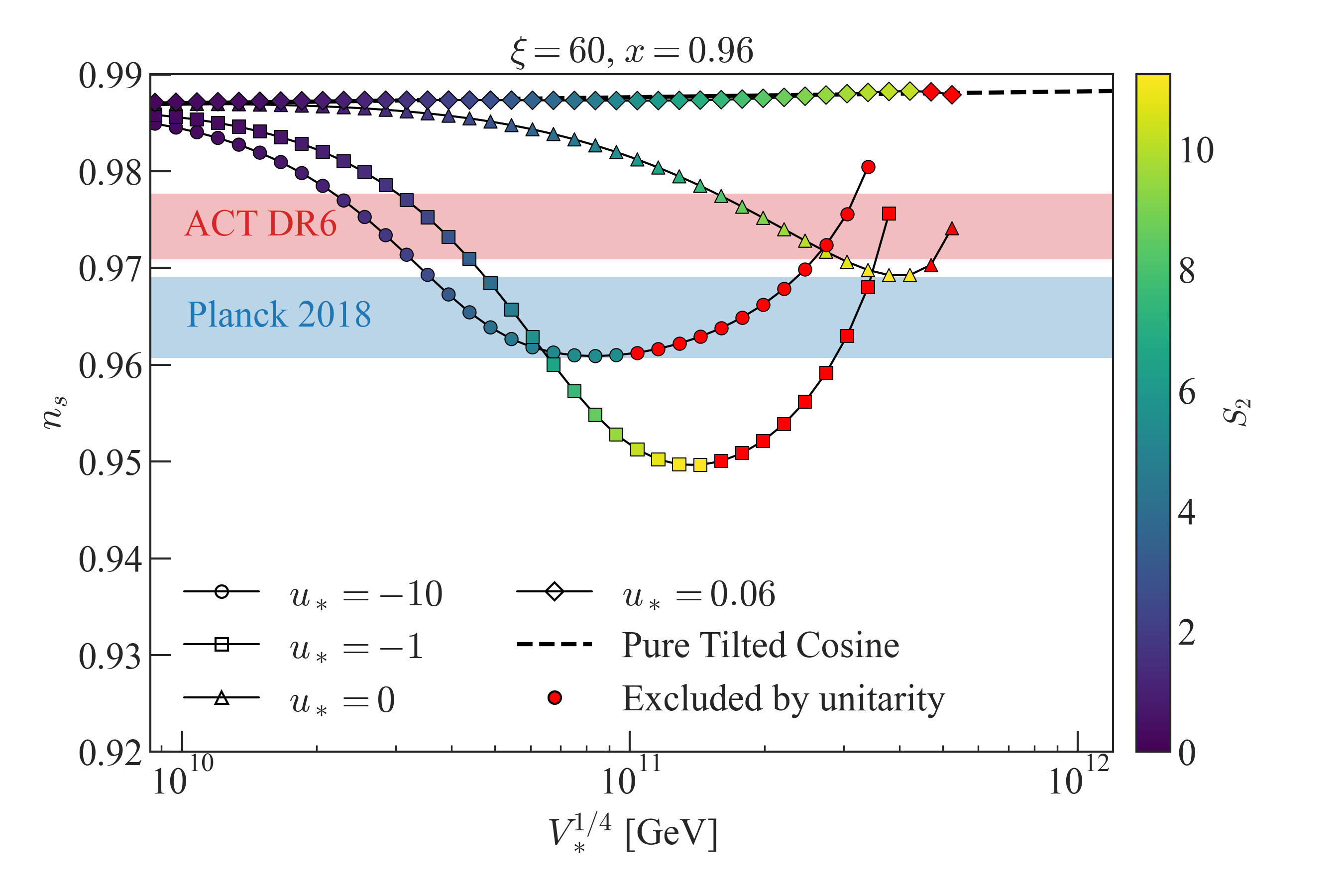}
    \includegraphics[width=0.49\linewidth]{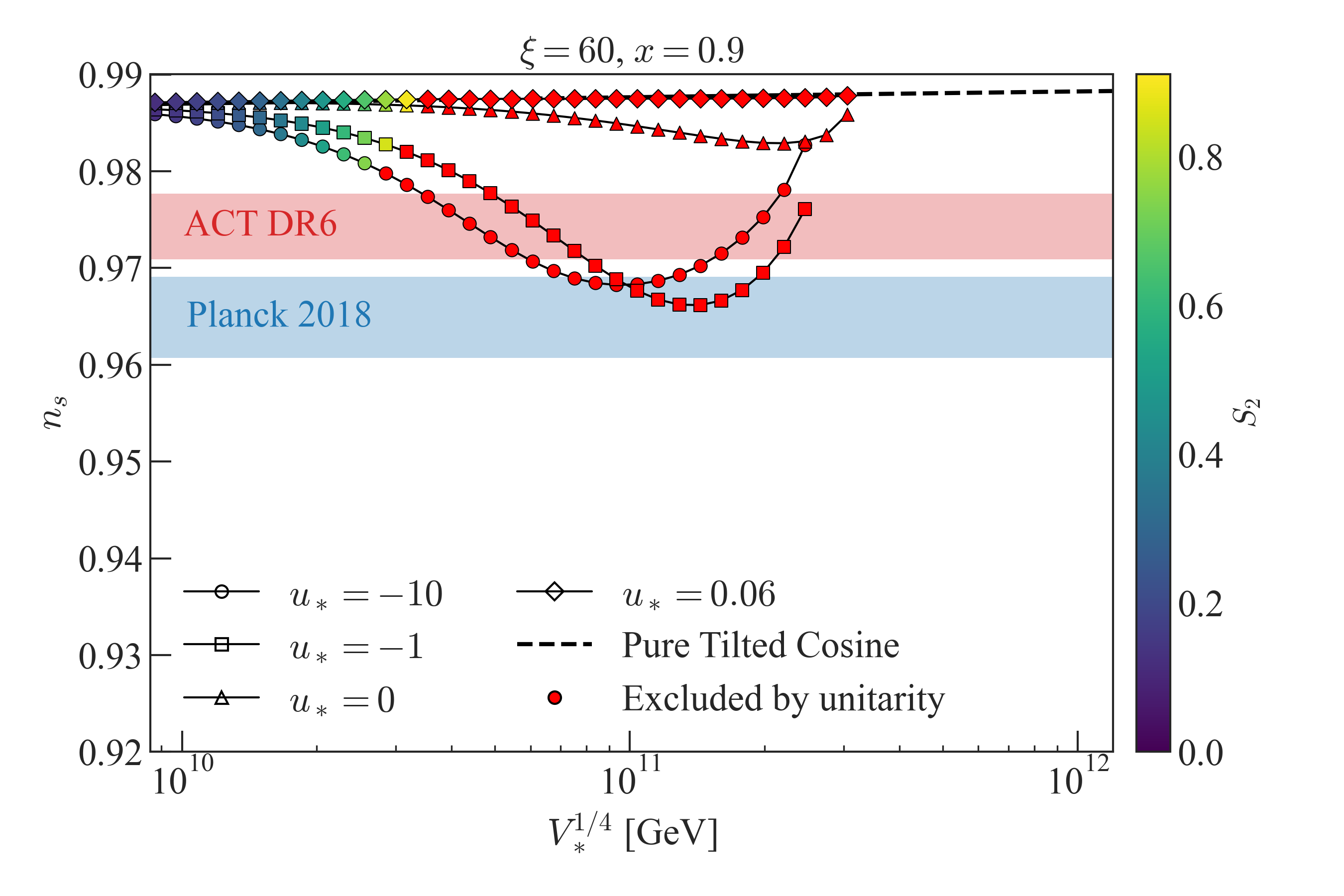}
    \includegraphics[width=0.49\linewidth]{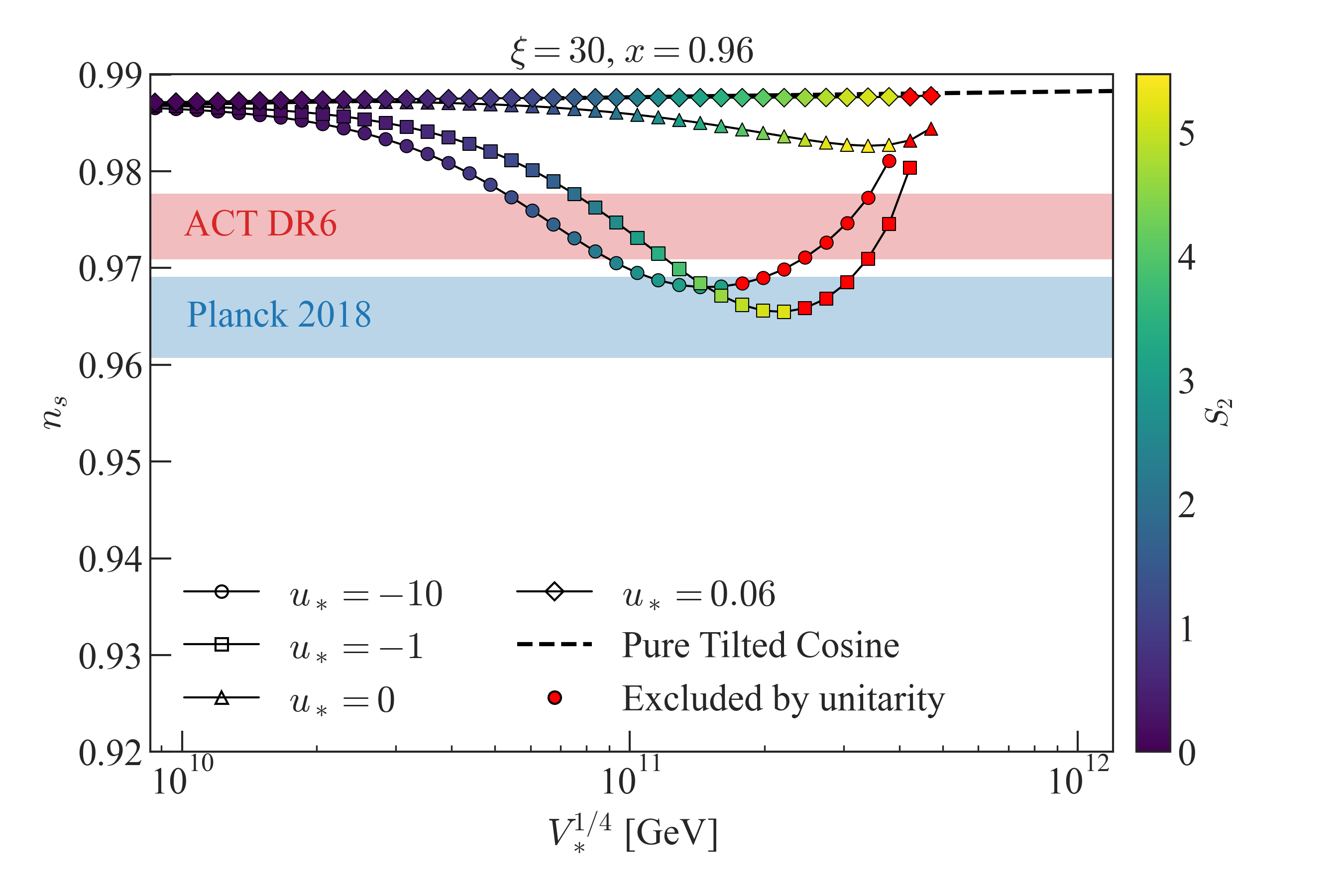}
    \includegraphics[width=0.49\linewidth]{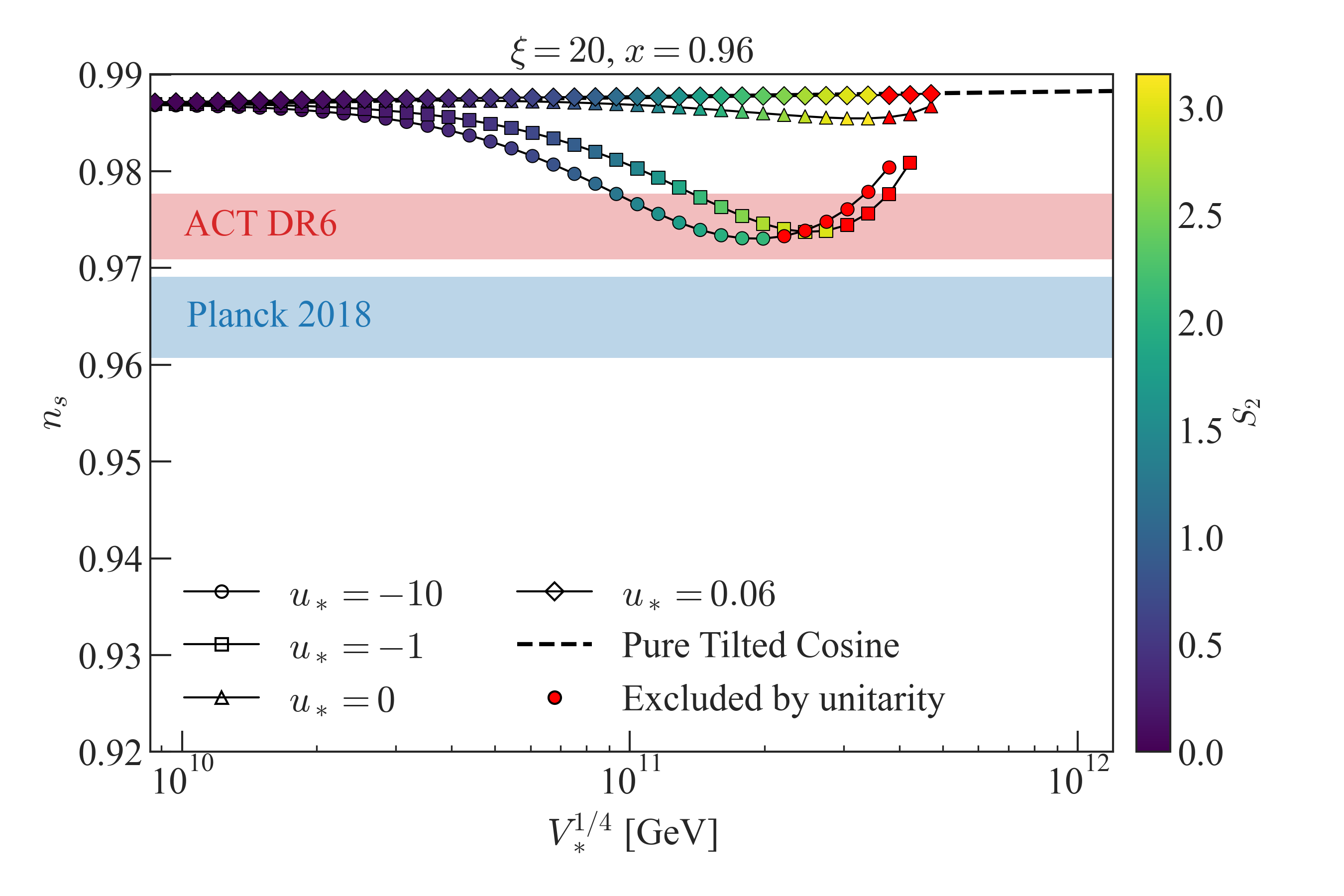}
    \caption{The scalar spectral index $n_s$ as a function of the inflationary scale $V_*^{1/4}$, showcasing the effect of the two dimensionless parameters that were fixed in Figure~\ref{fig:ns_vs_V*_Full}: the nonminimal coupling $\xi$ and the tunneling parameter $x$ defined in Eq. \eqref{eq:x_definition}. The figures are zoomed into the high-energy regime $10^{10}\,{\rm GeV}\lesssim V_*^{1/4}\lesssim 10^{12} \, {\rm GeV}$ for visual clarity.  The values of $u_*$  
     and color-coding follow Figure~\ref{fig:ns_vs_V*_Full}. Larger values of the nonminimal coupling $\xi$ lead to stronger deviations from the minimally coupled tilted cosine potential (dashed line) and thus allow for lower values of $n_s$. Slower tunneling (lower $x$) decreases the total observable length of the chain and thus reduces the deviations from the pure tilted cosine, leading to larger $n_s$.
    Finally, reducing $x$ leads to a larger part of parameter space being ruled out due to violating perturbative unitarity, as is explained in Appendix \ref{subsec:Unitarity_ustar0} below Eq.~\eqref{eq:SZero_unitarity_bound_main}. {Note (as seen in the upper right panel) that the range of $V_*^{1/4}$ corresponding to viable high-scale chain inflation, conforming both to  CMB data and unitarity, vanishes for $x=0.9$. We verified that high-scale chain inflation requires $x \ge 0.9$  for all values of $\xi\le 100$. }
    }
    \label{fig:ns_vs_V*_ZoomGrid}
\end{figure}

\begin{figure}
    \centering
    \includegraphics[width=0.49\linewidth]{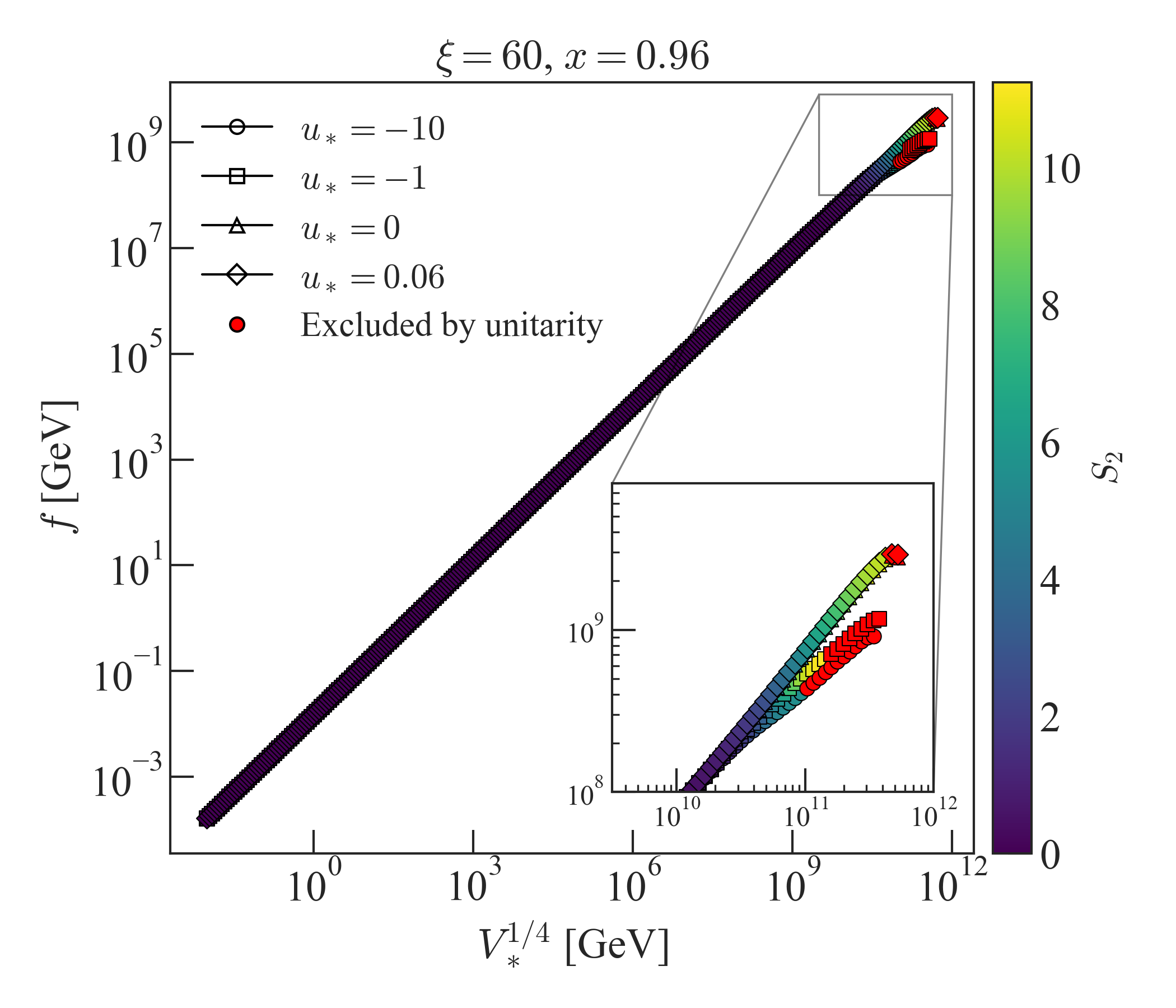}
    \includegraphics[width=0.49\linewidth]{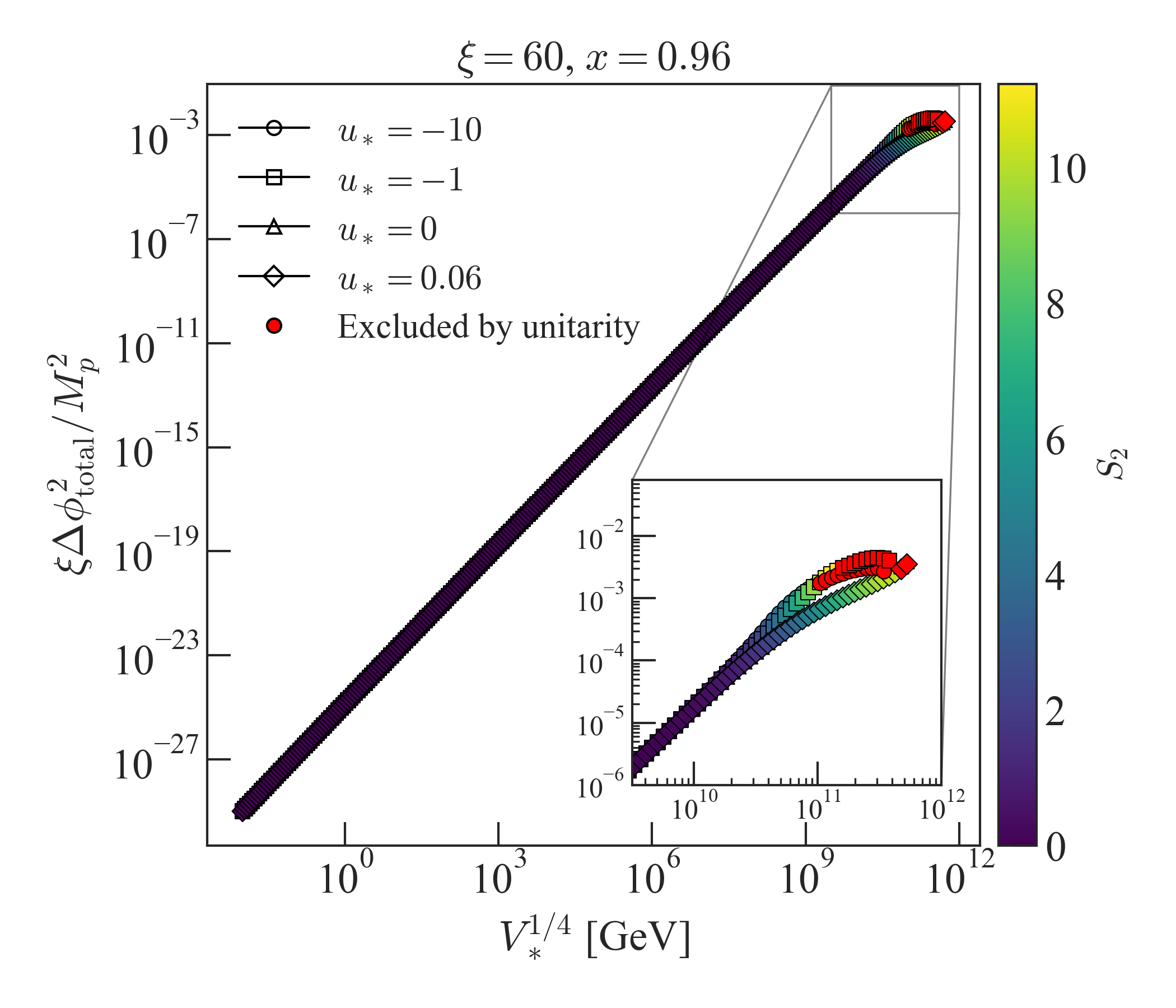}
    \caption{{\it Left:} The dependence of $f$ on the inflationary scale $V_*^{1/4}$. We observe an almost perfectly linear correlation, only slightly broken by NMC effects at high scales, with the separation of minima in the potential becoming $\Delta \phi\sim 2\pi f$ increasingly sub-Planckian for low inflationary potential scales. {\it Right:}
    The maximum correction to the gravitational action (see Eq.~\eqref{eq:Omegadef})
    as a function of the inflationary energy scale $V_*^{1/4}$. 
  We see that the model is always safely within the sub-Planckian regime, with NMC effects only emerging at high inflationary scales. It is also interesting to note, that  for small values  of $\xi\phi^2 \sim 10^{-3}M_{\rm Pl}^2$ can have significant effect on the chain inflation dynamics and observables. Higher values of $\xi\phi^2$ lead to unitarity violation, as seen by the red dots in the right panel. }
    \label{fig:FieldTraversal_vs_InflationaryScale}
\end{figure}

 We show the size of $f$ and $\Delta \phi_{\rm tot}$ in Figure~\ref{fig:FieldTraversal_vs_InflationaryScale}. There we observe that $f$ and $V_*^{1/4}$ are approximately linearly dependent, with this approximation only slightly breaking down at higher scales, for which the NMC effects becomes important. In fact, we can easily estimate this by noting that, since $V_*\approx N_*\Delta V\approx3\times10^4\Delta V/(1-n_s)$ and $\Delta V=2\pi f\mu^3=2\pi x\Lambda^4={2\pi xf^4\mathcal{S}(z)}/{\SZero}$, we can write
\begin{equation}
f=V_*^{1/4}\left(\frac{(1-n_s)\SZero}{3\times10^4\times2\pi x \mathcal{S}(x)}\right)^{1/4}\approx0.005\SZero^{1/4}V_*^{1/4}\approx0.015V_*^{1/4}\,,
\end{equation}
where we used the fact that $\SZero\sim100$ for low scales and fixed $x=0.96$ along with $n_s=0.965$. This agrees with the factor of ${\cal O}(0.01)$ observed between $f$ and $V_*^{1/4}$ in Figure~\ref{fig:FieldTraversal_vs_InflationaryScale}. Given the proportionality $\Delta\phi_{\rm total}=N_*\Delta\phi\propto f$, it is not surprising that we observe a similar linear dependence between the inflationary scale and the total field excursion.  By calculating the expansion parameter used in our application of the sub-Planckian regime ($\xi\phi^2/M_{pl}^2$ when restoring the Planck mass for clarity), we verify that its maximum value ($\xi\Delta\phi_{\rm total}^2/M_{pl}^2$) peaks at a maximum of $\sim10^{-2}$ at the highest possible inflationary scale (see the right panel of Figure~\ref{fig:FieldTraversal_vs_InflationaryScale}), while being exceedingly small $ (\lesssim10^{-25})$ for the low scales for which the tilted cosine was observationally viable. Figure~\ref{fig:FieldTraversal_vs_InflationaryScale} justifies the sub-Planckian approximation of Eq.~\eqref{eq:dphidchismall} across all relevant energy scales, and further shows how the tilted cosine is naturally recovered at low scales even in the presence of a non-minimal coupling to gravity.
One final note is that for the values of $\phi$ and $f$ shown in Figure~\ref{fig:FieldTraversal_vs_InflationaryScale}, $\chi$ is also sub-Planckian and in fact $\Delta\chi_{\rm tot}$ is of the same order of magnitude as $\Delta\phi_{\rm tot}$.

The behavior of the branches in Figures~\ref{fig:ns_vs_V*_Full} and \ref{fig:ns_vs_V*_ZoomGrid} can be understood analytically in the simple $u_*=0$ limit. In this case $S_1=0$, so the leading effect of the non-minimal coupling enters through the quadratic coefficient $S_2$ in the bounce action expansion. As shown in Appendix~\ref{subsec:Analytics_ustar0}, $S_2$ initially grows with the inflationary scale, causing the solutions to depart from the minimally coupled tilted cosine prediction. This growth is not indefinite: the scalar-amplitude constraint implies a turnover, with the largest departure occurring near $S_{E,0}\simeq 4$. This explains why the high-scale branches first move toward observationally allowed values of $n_s$ and then eventually turn back toward the minimally coupled curve. Equivalently, the minimum of $n_s(V_*)$ occurs where $S_2$ is maximized, as reflected by the color-coding in Figures 3 and 4. The opposite regime, $u_*\to-\infty$, provides a useful analytic reference point in which the predictions become insensitive to the precise pivot location, as discussed in Appendix \ref{subsec:Negative_uStar_and_Universality}.

The red points in Figures~\ref{fig:ns_vs_V*_Full} and \ref{fig:ns_vs_V*_ZoomGrid} correspond to solutions for which the semiclassical tunneling calculation is no longer under perturbative control. The relevant constraint can be phrased as a lower bound on the baseline Euclidean action $S_{E,0}$, or equivalently as an upper bound on the local curvature scale of the tilted cosine potential. The derivation is given in Appendix~\ref{subsec:Unitarity_ustar0}, while Appendix~\ref{sec:ActionBoundsAppendix} summarizes the viable range of Euclidean actions. This bound becomes significantly stronger for smaller $x$, which is why the phenomenologically interesting high-scale branch is naturally pushed toward the fast-tunneling regime with $x$ close to its upper allowed value. {In practice, for any value below $x\le0.9$, there is no high-scale solution for non-minimally coupled chain inflation with $\xi\lesssim100$ which falls into the CMB-allowed regime (whether by {\it Planck} or by ACT), while still obeying unitarity. This is shown in the upper right panel of Figure~\ref{fig:ns_vs_V*_ZoomGrid} for $\xi=60$ and $x=0.9$.}

In summary, the scalar spectral index provides the first direct test of the parameter space opened by the non-minimal coupling. At low inflationary scales the field excursion is small, the conformal factor remains close to unity, and the predictions reduce to those of the minimally coupled tilted cosine model. At higher scales, however, the field-dependent enhancement of the Euclidean bounce action modifies the tunneling history and allows the model to enter the \textit{Planck}- and ACT-compatible region. The viable high-scale branch is controlled by the interplay between the non-minimal coupling, the tunneling parameter $x$, and the pivot location $u_*$, while perturbative control of the semiclassical tunneling calculation removes part of the parameter space, especially for smaller $x$, as shown in Figures~\ref{fig:ns_vs_V*_Full} and \ref{fig:ns_vs_V*_ZoomGrid}. Thus the CMB spectral tilt already identifies a  viable high-scale regime, which we will further examine in more detail in the next section through the full scale dependence of the primordial curvature spectrum.

\subsection{
Scale Dependence of the Primordial Curvature Power Spectrum}

\begin{figure}
    \centering
    \includegraphics[width=0.8\linewidth]{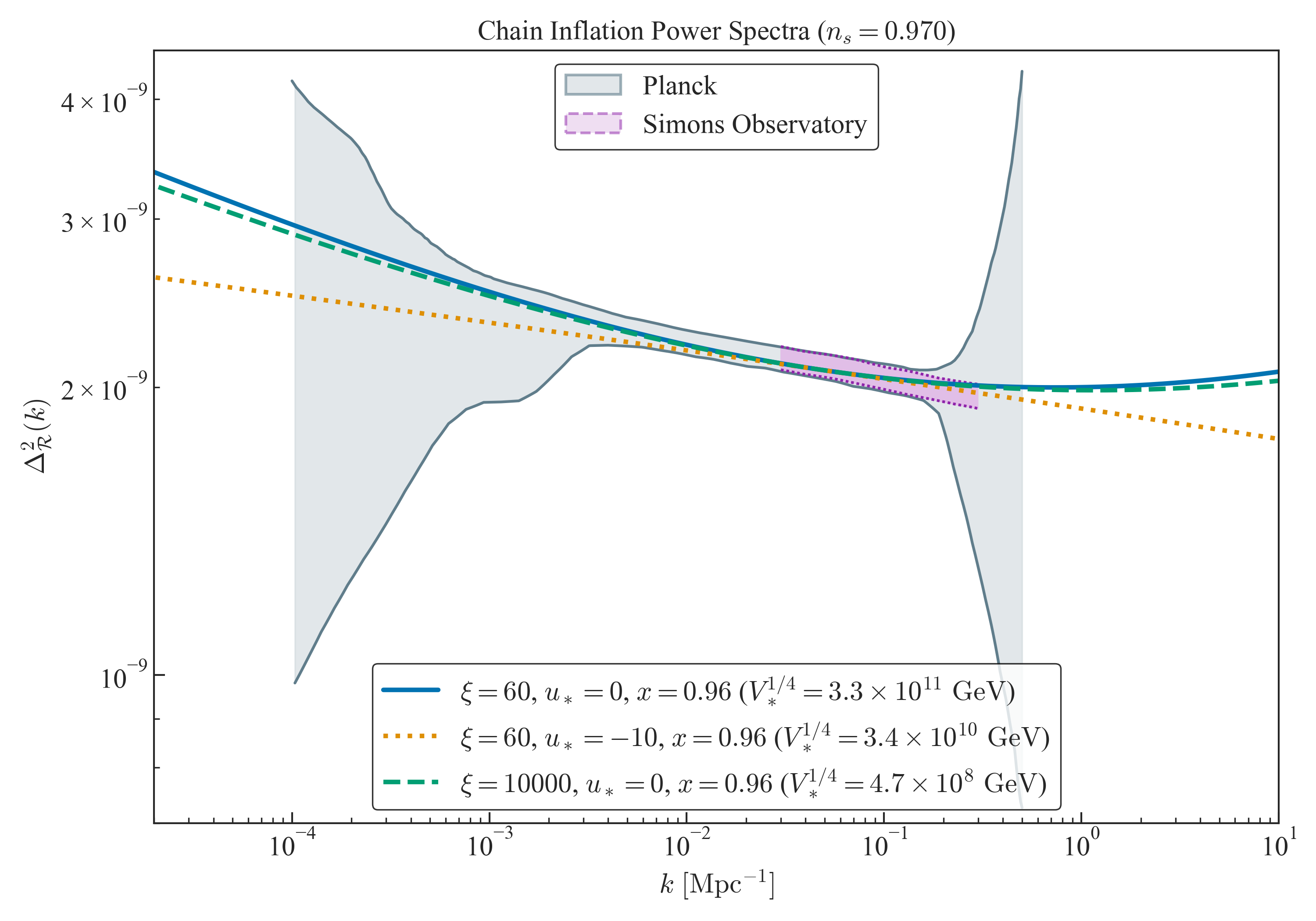}
    \caption{Predicted scalar power spectrum from NMC high-scale chain inflation. Curves within the gray region are compatible at $2\sigma$ with \textit{Planck} power spectrum reconstruction. Curves inside (outside) the pink band are expected to produce a signal (no signal) of a running spectral index at the Simons Observatory.
    }
    \label{fig:PowerSpectrum}
\end{figure}

{ Having constrained the scalar spectral index at the pivot scale, we now translate the
predictions of the non-minimally coupled model into the scale dependence of the primordial
curvature power spectrum over the entire range of CMB-observable modes. This is not meant to
introduce additional independent observables beyond $n_s$ and $\alpha_s$: as discussed
around Eq.~\eqref{eq:d_plogns}, higher-order derivatives of the spectrum are
parametrically suppressed for reasonable values of the  coefficients  of the Euclidean bounce action expansion (see Eq.~\eqref{eq:S_expansion}). Rather,
the purpose of this subsection is to display the predicted shape of $\Delta_{\mathcal R}^2(k)$
directly, compare it with the reconstructed \textit{Planck} power spectrum, and illustrate how future
measurements of scale dependence, such as those expected from the Simons Observatory~\cite{SimonsObservatory:2018koc},
can distinguish between different non-minimally coupled tunneling histories.

This representation is particularly useful because the two main NMC regimes identified
above lead to qualitatively different behavior away from the pivot scale. In the
$u_* \simeq 0$ branch (corresponding to the non-minimal coupling effect on gravity $\propto  \xi\phi^2R$ vanishing at the pivot scale), the non-minimal coupling can generate a sizeable positive running,
leading to a visible departure from a pure power-law spectrum across CMB scales. In the
large negative-$u_*$ regime (where the non-minimal coupling effect vanishes around the end of inflation), by contrast, the running is strongly suppressed and the spectrum
remains close to a power law even when the inflationary scale is high. Figure~\ref{fig:PowerSpectrum}
therefore serves both as a consistency check of the analytic $n_s,\alpha_s$ description and
as a direct visualization of the model predictions against current and projected CMB
sensitivity\footnote{{The forecast in Ref.~\cite{SimonsObservatory:2018koc} is presented in terms of $e^{-2\tau}P(k)$, such that to isolate the expected sensitivity on $P(k)$ we must factor out the uncertainty associated with the optical depth $\tau$, which we assume to be similar to \textit{Planck}. 
}} .}

To calculate $\Delta_{\mathcal{R}}^2(k)$ we may use the analytical approximation from Eq.~\eqref{eq:ScalarPowerSpectrumConstraint}. However, since at this point in the analysis both $\Gamma$ and $H$ are only determined in terms of the field value $\chi$, we must convert their functional dependence to the scale $k$ by calculating the field value at which each mode crosses the horizon.
By using the horizon-crossing condition $k(\chi)=a(\chi)H(\chi)$ we can write
\begin{equation}\label{eq:HorizonCrossing}
 k(\chi)=e^{\mathcal{N}(\chi)}\frac{H(\chi)}{H_*}k_*\, ,
\end{equation}
where $\mathcal{N}(\chi)$ is the number of $e$-folds between $\chi_*$ and $\chi$.

In Section~\ref{sec:cosinerunning} we derived the running of the scalar spectral index for a pure tilted cosine, computing $\alpha_s\simeq -0.002$ for $n_s=0.965$. 
When non-minimal couplings are considered, the above result changes. However, the running takes a particularly simple form, as $S_1$ can be written in terms of $S_2$ and $u_*$ through Eq.~\eqref{eq:SCoeff_NMC_Redefinition}. There are two primary cases for consideration, distinguished by the parameter $u_*=\chi_*/\chi_c$: the $u_*=0$ model, for which we have considerable analytical control, as seen in Section~\ref{subsec:Analytics_ustar0}, and the $|u_*|\gg1$ class of models, which includes the majority of the remaining viable parameter space under an elegant universality regime.

The former leads to a particularly simple expression for the running since $S_1=0$, and thus
\begin{equation}\label{eq:Running_ustar0}
    \alpha_s\Big\lvert_{u_*=0} = \frac{3}{5} (1 - n_s)^2 \left(  2S_2  - 3 \right) \, ,
\end{equation}
which is completely determined in terms of $n_s$ and $S_2$. Given $S_2>0$ and more precisely $S_2\sim 10$ for $n_s$ to fall within the observable allowed region for high inflationary scales (see Figure~\ref{fig:ns_vs_V*_Full}), we can estimate $\alpha_s\sim0.01$ for such models. This is significantly different from the low-scale value of $\alpha_s\sim-0.002$, not only due to the difference in size (by nearly an order of magnitude), but also due to having the opposite sign. This is precisely what we observe in Figure~\ref{fig:PowerSpectrum}, where we show the primordial scalar power spectrum predictions for both $\xi=60$ and $\xi=10^5$  in the $u_*=0$ regime.\footnote{The value $\xi=10^5$ was chosen to demonstrate that for $\xi\gg 1$, the exact value of $\xi$ is largely irrelevant. It is a common feature of non-minimally coupled models that for large $\xi$ the observables follow a universal behavior~\cite{Kallosh:2013tua,Kaiser:2013sna}.} Their behavior around the scales constrained by \textit{Planck} measurements is remarkably similar, as expected from the fact that the value of $S_2$ required to obtain the same $n_s$ is mostly unchanged even when considering incredibly different orders of magnitude for the coupling $\xi$ and thus can be considered a universal template for $\xi\gg 1$.
 With the Simons Observatory expected to measure the amplitude of the primordial scalar power spectrum across small scales ($k \sim 0.2$ Mpc$^{-1}$) with a precision of 0.4\% \cite{SimonsObservatory:2018koc}, the imprint of this running becomes resolvable. Relative to the standard pivot scale $k_* = 0.05$ Mpc$^{-1}$, an expected positive running of $\alpha_s \sim 0.01$ induces an enhancement of roughly 1\% in the primordial power at $k = 0.2$ Mpc$^{-1}$. Comparing this theoretical deviation directly to the forecasted 0.4\% observational uncertainty yields a single-bin signal-to-noise ratio of $2.4\sigma$ at this scale.

However, for the universality class $u_*\rightarrow-\infty$ (which practically is reached for $u_*\lesssim -1)$ the value of the running of the spectral index is highly suppressed, becoming even weaker than its corresponding magnitude for the standard minimally coupled model. In this case, $S_1=-2S_2$ and the running becomes
\begin{equation}\label{eq:Running_ustarInfty}
    \alpha_s\Big\lvert_{u_*\rightarrow-\infty} = -\frac{3}{5} (1 - n_s)^2\frac{3+ S_2^2}{(1 + S_2)^2}  \, ,
\end{equation}
which negates the amplification of the running seen in the $u_*=0$ regime due to the impact of the NMC through $S_1\neq0$ in the determination of $n_s$, which now plays a significant role in the dynamics of the model. Since in this universality class we require $S_2\sim3$ to recover the correct $n_s$, as seen in Figures~\ref{fig:ns_vs_V*_Full} and~\ref{fig:ns_vs_V*_ZoomGrid}, we estimate $\alpha_s\sim-0.0005$, which is smaller than the minimally coupled model's running by roughly a factor of 4. This is confirmed in Figure~\ref{fig:PowerSpectrum}, where we see that the $u_*=-10$ power spectrum is practically a straight line in the logarithmic scale of the plot, exhibiting effectively no detectable running. 
It also defines a clear method to falsify models and select between them, as the detection (or lack thereof) of a non-zero positive running of the spectral index in future experiments with higher sensitivity can clearly rule out significant regions of the parameter space of the model, mostly allowing us to distinguish between $u_*\simeq 0$ and $|u_*|\gg 1$.

\begin{figure}
    \centering
    \includegraphics[width=\linewidth]{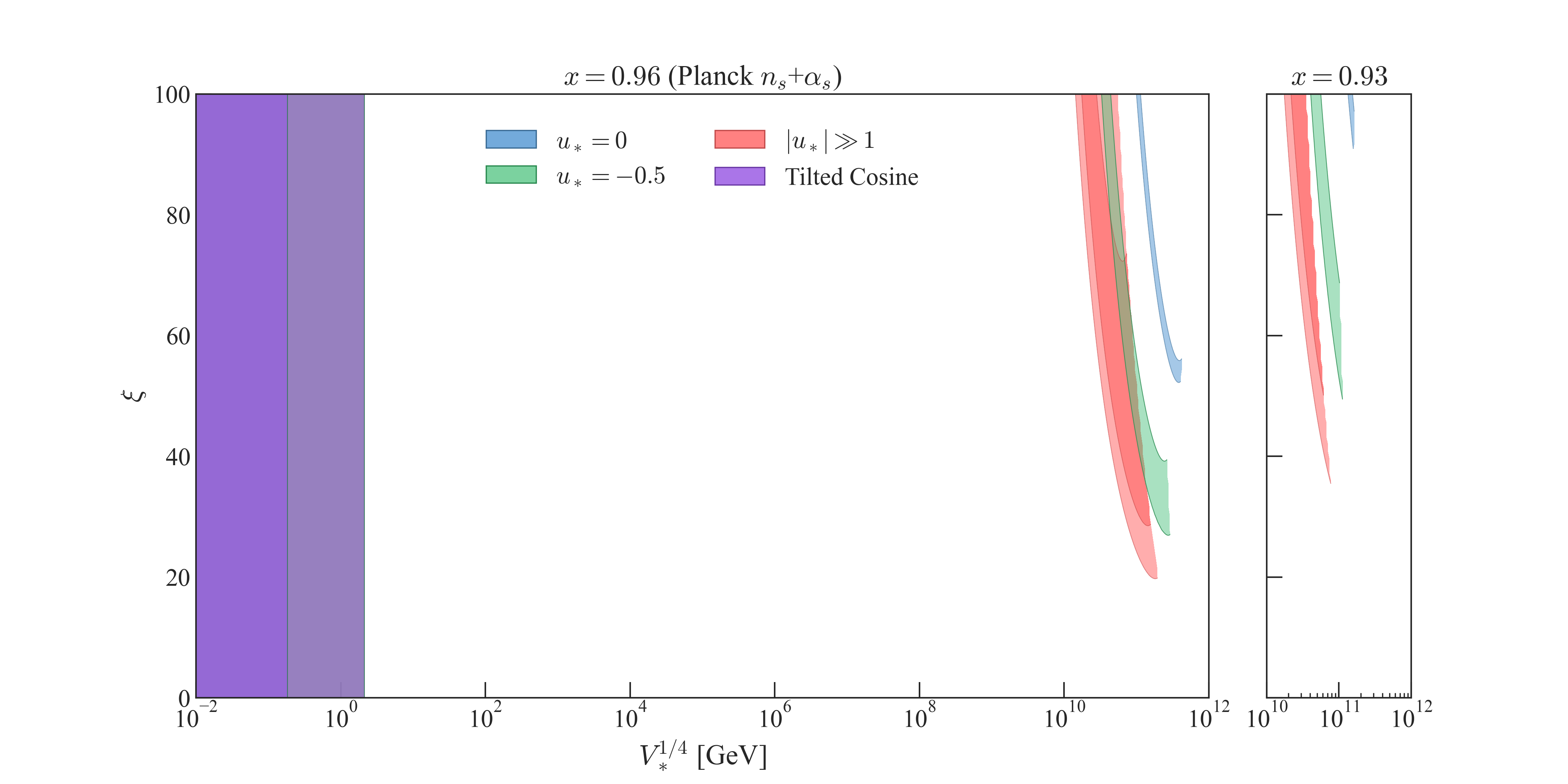}
    \caption{Parameter space plot showing the pairs of points $\{V_*^{1/4},\xi\}$, which lead to observables that fall within the $1\sigma$ and $2\sigma$ bounds of {\it Planck} for both $n_s$ and $\alpha_s$. One can see two viable regimes: a) purple vertical bands at low scales $V_*^{1/4}<3$ GeV, which are identical for NMC and minimally coupled tilted cosine models; b) high-scale $V_*^{1/4} \geq 10^{10}$ GeV models which are viable only in the case of NMC ($\xi \neq 0).$
    We distinguish the different classes of NMC models by the  horizon crossing field value $u_*=\chi_*/\chi_c$, shown in various colors as described in the legend (as a reminder, due to the NMC, there is physical significance to the value of this quantity).
    The dark (light) regions correspond to the $1\sigma$ ($2\sigma$)  bands.   The $u_*=0$ and $u_*=-0.5$ models are severely limited by their large running of the spectral index, and therefore only show compatibility at the $2\sigma$ level, therefore there is no $1\sigma$ contour present for these models. The tunneling parameter $x$ is the ratio of energy difference between successive vacua to barrier height, and $x$ close to unity corresponds to fast tunneling. We show results for $x=0.96$ (left) and $x=0.93$ (right), showing how slower tunneling models (lower $x$) have a strict reduction in their viable parameter space. Indeed we have confirmed that the allowed high-scale parameter space vanishes for $x\leq0.9$.  Regarding the low-scale viable models with $V_*^{1/4}<3$ GeV:
    Note that the tilted cosine bands in purple at low $V_*$ are the same for $x=0.96,0.93$);
    these low-scale bands remain open for $x\gtrsim 0.7$}.
    
\label{fig:ParameterSpacePlot_Planck}
\end{figure}

Figure~\ref{fig:ParameterSpacePlot_Planck} shows the parameter space in the $V_*-\xi$ plane, when requiring that $n_s$ and $\alpha_s$  be within the $2\sigma$ bounds of the latest {\it Planck} results. We see that the low-scale region $V_*^{1/4}\lesssim 3\, {\rm GeV}$ is independent of $\xi$, as it corresponds to the pure tilted cosine model, where the field excursion is so small that the effects of the NMC are negligible as discussed around Eq.~\eqref{eq:NMClimit}

On the contrary, the high-scale regime shows sharp dependence on $\xi$, $u_*$ as well as the tunneling parameter $x$. 
As discussed below Eq.~\eqref{eq:SZero_unitarity_bound_main}, the parameter space significantly opens up for $x\simeq 0.96$.
For large values of $\xi$ the allowed bands become almost vertical. This is expected, since the predictions of non-minimally coupled models of inflation typically become $\xi$-independent for $\xi\gg1$.
We see in Figure~\ref{fig:ParameterSpacePlot_Planck} that for smaller values of $x$, the allowed regions in the high energy part of the parameter space shrink significantly. This is visible for $x=0.93$. Indeed we confirmed that the allowed high-scale parameter space vanishes for $x\leq0.9$ for any value of $\xi\le 100$.

{Let us take a broader look at the perturbative-unitarity bound. As we start lowering $x$, the unitarity bound first ``removes" the high-scale NMC branch, since that
branch requires relatively small baseline actions, $S_{E,0}\sim {\cal O}(1-10)$. The
low-scale tilted cosine solution is much less sensitive to this constraint, because it is
realized at much larger bounce action, $S_{E,0}\sim 10^2$. Numerically, the
$V_*^{1/4}\sim 3\,{\rm GeV}$ solution remains perturbative for 
$x\gtrsim 0.7$. The detailed formulas are derived in Appendix~\ref{subsec:Unitarity_ustar0}.

{The physical reason why perturbative unitarity removes the high-scale NMC branch
first is that this branch relies on relatively small baseline bounce actions, which are associated with high inflationary scales. The
non-minimal coupling raises the inflationary scale by making the Euclidean action vary along the chain, but this effect is most effective when the baseline action
$S_{E,0}$ (defined in Eq.~\eqref{eq:baseline_action}) is not too large, since that corresponds to larger total field excursions and thus larger values of $\xi R\phi^2$. Perturbative unitarity, however, imposes the lower bound
on $S_{E,0}$ given in Eq.~\eqref{eq:SZero_unitarity_bound_main}. As $x$ is decreased, the
rescaled bounce action $\mathcal{S}(x)$ grows rapidly, and the lower bound
$S_{E,0}\gtrsim \sqrt{1-x^2}\mathcal{S}(x)/(8\pi)$ becomes increasingly restrictive. Comparing
this bound with the value of $S_{E,0}$ required by the scalar-amplitude constraint,
Eq.~\eqref{eq:VStar_SZero_unitarity_app}, shows that for sufficiently small $x$ the high-scale branch
is removed: the values of $S_{E,0}$ needed to realize large $V_*^{1/4}$ fall below the
unitarity-allowed range. By contrast, the low-scale tilted cosine solution is realized at a
much larger bounce action, $S_{E,0}\sim 100$, and therefore remains safely above the
unitarity bound throughout the fast-tunneling regime considered here.}

To summarize the dependence of our models on the tunneling parameter $x$:
the high-scale regime $V_*^{1/4} \gtrsim 10^{10}$ GeV is viable for $x>0.9$, while the low-scale regime $V_*^{1/4}\lesssim3$ GeV requires $x > 0.7$.

Further, as seen in Figures~\ref{fig:PowerSpectrum} and~\ref{fig:ParameterSpacePlot_Planck}, we have shown that the scale dependence of
$\Delta_{\mathcal R}^2(k)$ is sensitive to the location of the CMB pivot scale relative to
the origin selected by the non-minimal coupling. In the $u_*\simeq 0$ branch, the NMC
correction generates a sizeable positive running, producing a visible departure from a
pure power-law spectrum across CMB scales. This makes this branch testable by future
measurements of the running, such as those expected from the Simons Observatory. By
contrast, for large negative $u_*$, the running is strongly suppressed and the spectrum
remains close to a power law even at high inflationary scales. Thus a detection, or
non-detection, of a positive running of the scalar spectral index can partially distinguish between different tunneling
histories within the NMC model.

\section{Gravitational Wave Signatures}\label{subsec:GWSpectrum}

We now turn to the gravitational wave signatures of high-scale chain inflation with NMC. In contrast to the scalar observables discussed in the previous section, which probe the stochastic timing of tunneling events during the CMB window, the gravitational wave signal is sourced directly by the first-order phase transitions themselves. This makes it a complementary probe of the tunneling dynamics, the inflationary energy scale, and the mechanism by which the chain terminates. We first show that the usual vacuum tensor spectrum is far too small to be observable in the viable parameter space of the model. We then focus on the stochastic gravitational wave background generated by bubble collisions, whose peak frequency and amplitude can fall within the reach of future interferometers for the high-scale branch opened by the non-minimal coupling.

{\it Tensor Modes:}
The upper bound on the scale of chain inflation $V_*^{1/4}\lesssim10^{12}\ \GeV$, shown in Figures~\ref{fig:ns_vs_V*_Full} and~\ref{fig:ns_vs_V*_ZoomGrid} (see also Eq.~\eqref{eq:running_general} and Refs.~\cite{Freese:2021noj, Winkler:2020ape}),  even within the high-scale parameter space made possible by the NMC introduced in this work, has a direct consequence on the tensor-to-scalar ratio $r$, which is expected to be highly suppressed. 
Given the amplitude of the scalar power spectrum $A_s$, the tensor to scalar ratio is 
$r \equiv {\mathcal P_T}/{A_s}$
while the tensor power spectrum generated by vacuum fluctuations is
\begin{equation}
\mathcal P_T = \frac{2H^2}{\pi^2 M_{\rm Pl}^2} \, .
\end{equation}
Inverting this leads to
$H^2 = \frac{\pi^2}{2} A_s r M_{\rm Pl}^2$ where the Hubble scale 
during inflation is $H^2\simeq V_*/3 M_{\rm Pl}^2$. This relates the potential during inflation to the tensor to scalar ratio as
\begin{equation}
V_*^{1/4}
\simeq
10^{16}
\left(\frac{r}{0.01}\right)^{1/4}
{\rm GeV}.
\end{equation}
where we used For $A_s \simeq 2.1\times 10^{-9}$ and
$M_{\rm Pl}=2.435\times 10^{18}\,{\rm GeV}$. For $V_*^{1/4}\lesssim 10^{12}$ GeV, we get $r\lesssim 10^{-18}$, far below any upcoming detection attempt.

{\it Bubble Collisions:}
Fortunately, the first-order phase transitions (FOPTs) that occur during chain inflation give rise to bubble collisions that provide a powerful gravitational wave signal.  Depending on the parameters, this GW signal can peak precisely within the frequency range of pulsar timing arrays or ground/space-based interferometers \cite{Freese:2023szd}.

Given that chain inflation is composed of a large number of phase transitions, each contributing their own gravitational wave signal, the total output of the full chain, which we quantify by its fractional energy density $\Omega_{GW}$, can be described by the sum over all of these individual contributions
\begin{equation}\label{eq:sumspectrum}
    \Omega_{\text{GW}}(\nu) = \sum_i  \Omega_{\text{GW},i}(\nu) \,,
\end{equation}
where the subscript ``$i$" denotes the gravitational wave spectrum of the $i$th transition in the chain. Although this may seem like a complex superposition, the rapid inflationary expansion following the production of each GW signal leads to a severe dilution, such that the dominant contribution will be that of the final few transitions in the chain, with particular emphasis on the final one. This is particularly true in the case of GWs produced 
during a period of ``grateful exit" from chain inflation, as discussed in Appendix~\ref{sec:GracefulExitAppendix}.
During  the graceful exit period,   radiation domination may have already been established, even though the field may still tunnel through a few remaining barriers. These tunneling events produce significantly boosted GW signals.  In such cases, we parametrize the strength of the last (dominant) phase transition through the parameter $\alpha$, which we defined in Eq.~\eqref{eq:alpha} to be the ratio of the vacuum energy density to the energy density of radiation 
right before the transition.\footnote{If the final transition is described by inflationary dynamics, we can determine its duration  as a function of the vacuum lifetime, as written below in Eq.~\eqref{eq:betai} and described in Eq. 42 of Ref.~\cite{Freese:2023szd}} As we remain agnostic about the stopping mechanism apart from its nature as a chain-to-stop model, we can condense all information about this process in $\alpha$, which we treat  as a free parameter. 

The nature of each transition in the chain has a direct impact on the parameter $\beta$, which is  the inverse of the  time duration of each phase transition. In Appendix \ref{app:SlowlyVaryingGamma} we show that in our case the transition rate varies slowly, $|\dot\Gamma/H\Gamma\lesssim1|$. As shown in Ref.~\cite{Freese:2023szd}, 
one may take 
\begin{equation}
\label{eq:betai}
    \beta_i\simeq\begin{cases}
        2.8\Gamma_i^{1/4} \quad\quad \text{for transitions during chain inflation} \\
        \beta(\alpha) \quad\quad\quad \text{for transitions during graceful exit}
    \end{cases} \, 
\end{equation}
  The top expression is valid within the approximation of phase transitions with durations that are a fraction of the Hubble time ($\beta\gg H$), such that we neglect the expansion of the Universe during the transition itself and therefore use $\beta_i=2/\Delta t_i$, where $\Delta t_i$ is the lifetime of the $i$th minimum in the chain. Note that due to the large value of $\Gamma_*^{1/4}/H_*\sim10^4$, we expect this to hold for the duration of the bulk of chain inflation. 

When $\beta\gg H$  is not satisfied, the value of $\beta$ is strongly coupled to the background expansion, which in the radiation-dominated post-inflationary phase can be determined as a function $\beta(\alpha)$, with the explicit functional dependence $\beta(\alpha)$ being shown in Figure 1 of Ref.~\cite{Freese:2022qrl}. This is the case during the process of graceful exit, during which the transition rate is extremely suppressed to allow the inflaton to come to a halt and terminate the chain. Note that for $\alpha\lesssim1$, the value of $\beta(\alpha)$ is approximately independent of $\alpha$ and takes values $\beta(\alpha)/H\sim6-8$. If one allows for a second period of inflation after radiation originally overtakes the inflaton's energy, then values of $\alpha>1$ can be achieved, with $\beta$ decreasing with increasing $\alpha$, although one is limited by $\alpha\lesssim20$ to ensure proper percolation of the nucleated bubbles \cite{Freese:2022qrl}. Since the dominant GW signal will come from the final phase transition, as discussed previously, it is enough for us to focus on the bottom branch of Eq.~\eqref{eq:betai}, and therefore use $\beta(\alpha)$ when determining the inverse duration of the transition.

\subsection{Gravitational Wave Properties from Bubble Collisions}
The primary source of gravitational radiation in FOPTs are the collisions of the vacuum bubbles that form and expand as the inflaton moves through the chain. Assuming that all energy is released by the bubble collisions, we may use results previously provided in simulations by others for the properties of these GW.

{\it Peak Frequency of Gravitational Waves: }
The inverse duration of the FOPT $\beta$ determines the redshifted peak frequency of the GW signal, as extracted through simulations \cite{Huber:2008hg,Cutting:2020nla} as
\begin{equation}
\nu_{\text{peak},i}^0 = a_i \nu_{\text{peak},i} \simeq 0.2 a_i \beta_i \, ,
\label{eq:GWfrequency}
\end{equation}
 where $\nu_{\text{peak},i}^0$ denotes the peak frequency at present, while $\nu_{\text{peak},i}$ is its value at the moment of emission of the GWs, which is related to $\nu_{\text{peak},i}^0$ by the scale factor at the $i$th transition $a_i$, itself determined through $\mathcal{N}(\chi_i)$. Note that for the dominating source of a transition during graceful exit, since $\beta_i\propto H_i\propto V_*^{1/2}$ and $a_c\propto T_0/T_c\propto V^{-1/4}_*$, the peak frequency depends on the inflationary scale as
 \begin{equation}
     \nu_{\text{peak},i}^0\propto V_*^{1/4} \, .
 \end{equation}
 This dependence produces the main difference in the GW spectrum of low-scale chain inflation in the case of a pure tilted cosine potential with $\xi=0$ ($V_*^{1/4}\lesssim10 \ \GeV$ leading to $ \nu_{\text{peak},i}^0\sim \text{nHz}$) and one of a higher-energy model such as allowed in a NMC tilted cosine ($V_*^{1/4}\sim10^{11} \ \GeV$ leading to $ \nu_{\text{peak},i}^0\sim \text{kHz}$).
 { The effect of $V_*$ on the peak frequency of the GW spectrum of chain inflation was originally noted in Ref.~\cite{Freese:2023szd}, where the parameter space of chain inflation was mapped and it was determined that high-frequency (kHz) GWs can be obtained if the inflationary scale of a chain inflation model is $V_*^{1/4}\gtrsim10^{10}$ GeV (see Table II and Figure 7 of the same paper). High-frequency GWs are explicitly realized in the NMC tilted cosine  model we discuss in this work.}

\subsubsection{Gravitational Wave Spectrum and Amplitude  from Bubble Collisions}

We now  move to translate the bubble collision picture into a GW spectrum that can be compared against detector sensitivity curves. The relevant quantities are the transition timescale, the energy released in the final transitions, and the radiation bath present at the end of the chain, which together determine the peak frequency and amplitude of the signal.

The GW spectrum follows a broken power-law form as a function of  frequency~\cite{Kosowsky:1992rz,Kosowsky:1992vn,Caprini:2018mtu}
\begin{equation}
\Omega_{\text{GW},i}^{\text{bc}}(\nu) h^2 = \tilde{\Omega} h^2 \left( \frac{H_i}{\beta_i} \right)^2 \frac{\Delta V^2}{\rho_{\text{tot}}^i \rho_{\text{crit}}^0} a_i^4 \times \frac{(b+c)(\nu/\nu_{\text{peak},i}^0)^b}{c+b(\nu/\nu_{\text{peak},i}^0)^{b+c}} \,,
\label{eq:OmegaGWbc}
\end{equation}
where $h=H_0/(70 \ \text{km/s/Mpc})$ is the normalized Hubble constant.
The factor  $a_i^4$ accounts for the redshifting of the gravity wave amplitude from production until now ($a_i$ stands for the scale factor at the i-th
phase transition), and $\Delta V$ describes the energy released in one transition in the chain. The factor  $H_i/\beta_i$ describes the inverse duration of the phase transition relative to the Hubble rate, with ${\rho_{\text{crit}}^0}=3H_0^2$ and ${\rho_{\text{tot}}^i}=3H_i^2$ denoting the total energy density today and at the time of the $i$th phase transition respectively. Finally, $\{\tilde\Omega,b,c\}$ are model-dependent parameters, which we will discuss in more detail in what follows.

The remaining parameters entering this expression depend on the bubble regime under consideration. The most common estimate is the  ``envelope approximation", which assumes that all the energy is contained within a thin shell at the bubble wall, which is dissipated upon collision. However, this approximation applies to transitions within the thin-wall regime of tunneling, which is not expected to hold for the fast tunneling necessary for chain inflation. Conversely, one may employ the ``thick-wall approximation", for which there is a substantial propagation of the shear stress after the bubble collision. In order to be as general as possible in what concerns the particular mechanisms underlying bubble collisions in the mode, we present both regimes in our results.

For the envelope (thin-wall) approximation we use Eq.~\eqref{eq:OmegaGWbc} with $\{\tilde\Omega,b,c\}=\{0.077,2.8,1\}$ \cite{Huber:2008hg}, while for the thick-wall approximation we substitute  $\{0.027,0.7,2.2\}$ \cite{Cutting:2020nla}. The envelope and thick-wall approximations will strongly differ in the low and high-frequency tails of the spectrum, with the thick-wall having a much stronger presence in the low-frequency domain, while the envelope approximation leads to the opposite behavior. The larger value of $\tilde\Omega$ for the envelope approximation means that this spectrum will have a  slightly greater peak amplitude, although this value only differs by a factor of $3$ between the two cases (envelope vs thick-wall) and thus does not significantly affect  our analysis.

Given the slower nature of the final phase transitions compared to the typical rate during chain inflation and the dilution of all GW signals from earlier phase transitions, the GW spectrum from chain inflation is dominated by the gravitational radiation generated in the last few transitions. This is particularly true in the case of post-inflationary phase transitions, where the inverse duration is governed by the strength of the transition\footnote{{As a reminder, $\alpha$ is the ratio between the vacuum and radiation energy density. It is sometimes referred to in the literature as the ``strength of the transition",  because larger $\alpha$ means lower $\beta$ and therefore a stronger signal since $\Omega\propto1/\beta^2$.} }$\alpha$, which we treat as a free parameter in order to remain agnostic about the detailed mechanism through which the chain is terminated.

An additional dependence of the GW amplitude comes from the redshift factor
$a_i^4$ in Eq.~\eqref{eq:OmegaGWbc}. The dominant GW contribution comes from the final transition, at field value $\chi_{\rm last}$
and scale factor $a_{\rm last}$. These should be distinguished
from $\chi_c$ and $a_c$, the field value and scale factor at vacuum-radiation equality, at the end of 
 chain inflation. If the last tunneling event happens exactly at $\chi_c$, then $a_{\rm last}=a_c$;
 this case applies to chain-to-stop exit models with no additional phase transitions as well as to chain-to-roll exit models.
 Alternatively, as discussed previously in Section~\ref{sec:graceful_exit}, there may be an additional phase transition after the field reaches $\chi_c$
(following an extremely brief resurgence of vacuum domination); in that case the radiation redshifts between the end of inflation and the last (additional) phase transition.

Let $\rho_{r,c}$ denote the radiation density at equality, computed in Appendix~\ref{sec:RadiationAppendix}, and let
the remaining vacuum energy before the last transition be $\rho_{\rm v}\simeq\Delta V$. The
radiation density immediately before the last transition is then
\begin{equation}
\rho_r(a_{\rm last})=\rho_{r,c}\left(\frac{a_c}{a_{\rm last}}\right)^4 .
\end{equation}
Using the definition of the transition strength,
$\alpha=\rho_{\rm v}/\rho_r(a_{\rm last})$, we obtain
\begin{equation}
a_{\rm last}
=
a_c
\left(
\alpha\,\frac{\rho_{r,c}}{\Delta V}
\right)^{1/4}.
\end{equation}
Equivalently, since $\alpha_{\min}\equiv\Delta V/\rho_{r,c}$, this can be written as
\begin{equation}
a_{\rm last}=a_c\left(\frac{\alpha}{\alpha_{\min}}\right)^{1/4}.
\end{equation}
Thus $\alpha=\alpha_{\min}$ corresponds to a final transition occurring immediately at
vacuum-radiation equality, while larger values of $\alpha$ correspond to a later final
transition and hence a different redshift factor in Eq.~\eqref{eq:OmegaGWbc}.

The discussed redshift dependence is important because Eq.~\eqref{eq:OmegaGWbc} scales as $a_{\rm last}^4$.
Consequently, even at fixed microscopic transition parameters, changing the duration of the post-equality stage before the last tunneling event can appreciably change the observed GW amplitude.

\subsubsection{Gravitational Wave Spectrum and Amplitude from Plasma Sound Waves}
The analysis of the thick-wall and envelope approximations corresponds to the regime when GW production is dominated by bubble wall collisions.
The alternative regime is one in which the gravitational radiation is dominantly produced by sound waves in the primordial plasma. This plasma is assumed to have been produced by earlier bubble collisions, after which it can be thought of as being heated by the subsequent collisions. Given that only the bulk motion of the plasma can create anisotropic stress and the subsequent GWs, the strength of the GW radiation will be determined by how the vacuum energy is distributed between the heating of the plasma and the production of sound waves. We condense this information in the efficiency factor $\kappa_i$ for the $i$th transition. In the case where the bubble walls approach the speed of light, we may use the estimate \cite{Espinosa:2010hh}
\begin{equation}
\kappa_i \simeq \frac{\Delta V}{0.73 \rho_{\text{rad},i} + 0.083 \sqrt{\Delta V \rho_{\text{rad},i}} + \Delta V} \, .
\end{equation}

With the efficiency factor definition in place, we can express the GW spectrum due to sound-waves as \cite{Hindmarsh:2015qta,Caprini:2018mtu}
\begin{equation}
\Omega_{\text{GW},i}^{\text{sw}}(\nu) h^2 = 0.159 h^2 \left( \frac{H_i}{\beta_i} \right) \kappa_i^2 \frac{\Delta V^2}{\rho_{\text{crit}}^0 \rho_{\text{tot}}^i} a_i^4 \left( \frac{\nu}{\nu_{\text{peak},i}^0} \right)^3 \left( \frac{7}{4 + 3(\nu/\nu_{\text{peak},i}^0)^2} \right)^{7/2} \, .
\end{equation}
We see that the dependence on $\kappa_i^2$  will suppress the GW amplitude since $\kappa_i<1$. However, this is approximately compensated by the linear dependence on $H_i/\beta_i$, which in general will be $\ll1$, comparatively to the quadratic dependence present in the formula for the bubble-collision spectrum.

With the peak frequency and amplitude now expressed in terms of the chain inflation parameters and the strength of the final transition, we can directly assess whether the resulting stochastic background is observable. We therefore turn next to the projected sensitivities of future gravitational wave experiments.

\begin{figure}[t]
    \centering
    \includegraphics[width=0.8\linewidth]{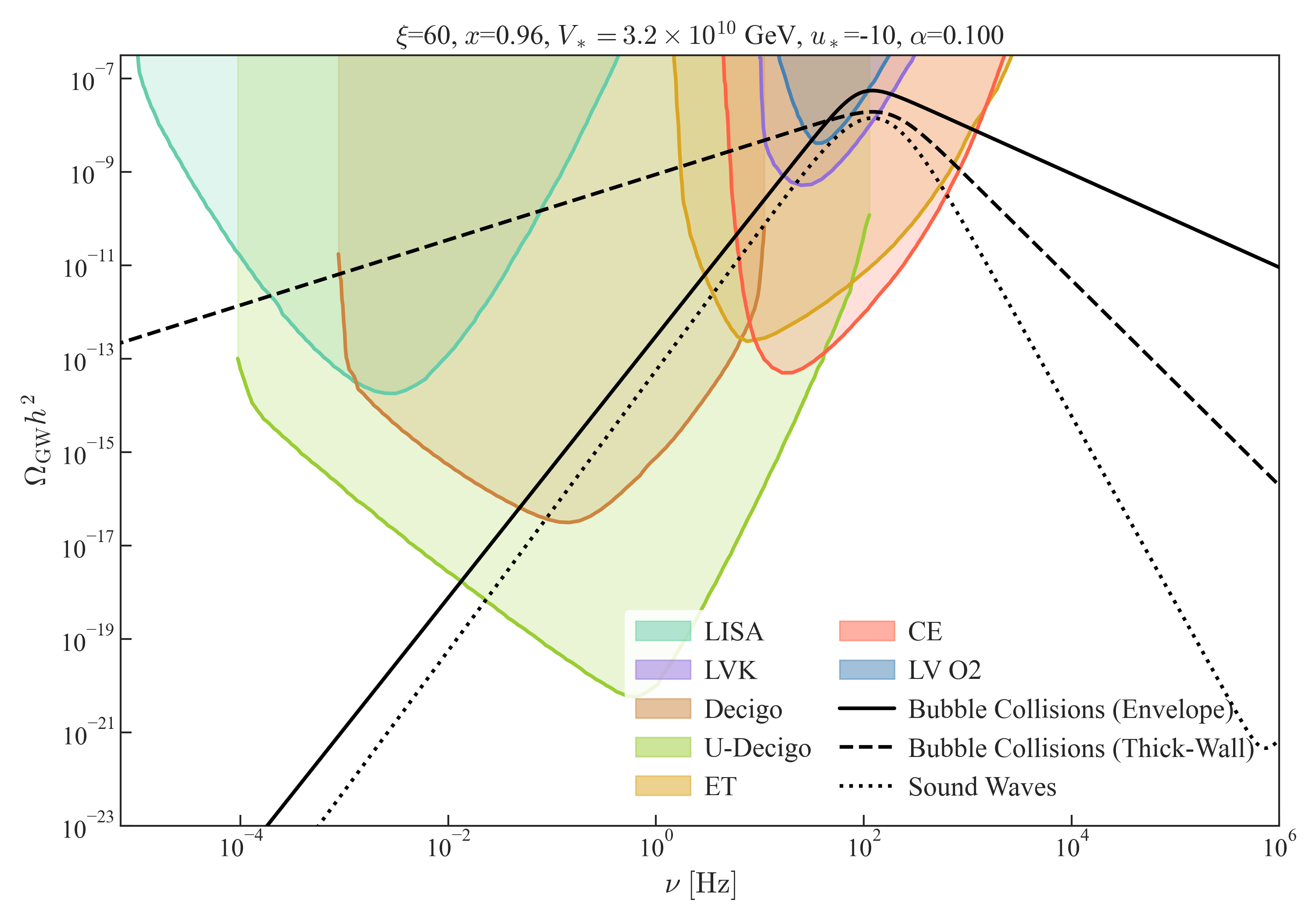} 
    \caption{Predicted stochastic gravitational wave background $\Omega_{GW} h^2$ as a function of frequency $\nu$ from NMC high-scale chain inflation for model parameters as indicated in the figure. The high-scale nature of the model allows for the production of gravitational waves within range of future ground and space-based interferometers. The color-shaded regions indicate the parameter space testable by future GW experiments. Black curves indicate predictions due to bubble collisions in the thin-wall limit (solid lines), bubble collisions in the thick-wall regime (dashed lines), and plasma effects (dotted lines). }
    \label{fig:GW_spectrum}
\end{figure}

\subsection{Detectability in Future Experiments}

In this section we compare the predicted GW background in our model to existing and upcoming GW data. 
The inflationary scale $V_*$, affecting the GW spectrum, is restricted to values that lead to CMB-compatible observables. 
Within the model under study, this limits us to two regions of the parameter space: low-scale and high-scale solutions. As we have shown, in the case of low-scale inflation, $V_* ^{1/4}\lesssim10^9$ GeV, NMC chain inflation reduces to the same results as for chain inflation with a minimally coupled tilted cosine potential.
The stochastic GWs from ordinary chain inflation have been extensively studied in Refs.~\cite{Freese:2022qrl,Freese:2023szd} for both low and high-scale inflation cases. The previous analyses~\cite{Freese:2022qrl,Freese:2023szd}  were performed in a model-independent way, treating the expansion coefficients of the Euclidean bounce action $S_i$ (see Eq.~\eqref{eq:S_expansion}), as free parameters.   Ref.~\cite{Freese:2022qrl} showed that chain inflation with $V_*^{1/4} \sim$ MeV-GeV produces a GW spectrum peaking around $\nu\sim\mathcal{O}(\text{nHz)}$, which can explain the signals reported by
the NANOGrav, Parkes, and European Pulsar Timing Array (EPTA) experiments \cite{EPTA:2021fqa,EPTA:2021crs,Goncharov:2021oub,Antoniadis:2022pcn};
in fact the bubble collision explanation offers a slightly better fit to data than do merging Supermassive Black Holes (although with low statistical significance.).

In the current work, we present a concrete realization of the high-scale chain inflation regime examined in Refs.~\cite{Freese:2022qrl,Freese:2023szd}, which can be achieved by adding a non-minimal coupling with $\xi\simeq {\cal O}(10)$ to a pure tilted cosine potential. We can thus use future GW experiments to constrain, not only  the expansion parameters $S_i$, but the actual microscopical parameters of the Lagrangian; the features of the potential (energy scale, barrier height, etc.) and the non-minimal coupling to gravity.

To determine the possibility of detection of  high-frequency GW signals in our model, we compare the theoretical predictions of the model with the sensitivities of future ground and space-based interferometry experiments.

\begin{itemize}
   \item  Regarding ground-based observatories, we consider the current advanced detector network, specifically the estimated aLIGO-aVirgo sensitivity of run O2 (LV O2) and the combined aLIGO-aVirgo-KAGRA design sensitivity (LVK) \cite{Aso_2013,Somiya_2012,Aasi_2015,Harry_2010,Acernese_2015}, which probes $\Omega_{GW}h^2\gtrsim 10^{-9}$ at a peak sensitivity of around $\nu\sim10$ Hz. We also consider proposed next-generation facilities awaiting full construction funding, namely the Cosmic Explorer (CE) \cite{Reitze:2019iox} and Einstein Telescope (ET) \cite{Punturo:2010zz}. Both next-generation experiments have a peak sensitivity around $\nu\sim10$ Hz and could possibly detect stochastic signals as weak as $\Omega_{GW}h^2\sim 10^{-13}$.
    \item Regarding space-based experiments, we include the approved Laser Interferometer Space Antenna (LISA) \cite{LISA:2017pwj}, which is most sensitive around the $\nu\sim$ mHz frequency band and probes signals as low as $\Omega_{GW}h^2\sim10^{-14}$. Moreover, we consider the proposed Deci-Hertz-Interferometer-gravitational wave Observatory (Decigo) \cite{Kawamura:2006up,Kuroyanagi:2014qza,Seto:2001qf}, as well as its conceptual successor, Ultimate-Decigo (U-Decigo). These proposed missions peak around the deci-Hertz frequency range, with Decigo projecting a sensitivity of $\Omega_{GW}h^2\sim10^{-15}$ and U-Decigo theoretically boosting this to $\Omega_{GW}h^2\sim10^{-20}$.
\end{itemize}

  In Figure~\ref{fig:GW_spectrum} we show an example of the GW spectrum in the high-energy chain inflation with the inclusion of a NMC. We chose $\xi=60$ and $u_*=-10$ (well within the large $|u_*|$ universality regime), such that an observationally compatible spectral index is achieved at a scale of $V_*^{1/4}\approx3\times10^{10}\ \GeV$. This particular set of parameters generates a GW spectrum with 
 a peak frequency at present of around $\nu\sim100$ Hz, placing it
within the projected sensitivities of future experiments. In particular, its peak is within the LVK region regardless of the GW production mechanism that we considered. The envelope approximation places the spectrum's peak within the LV O2 region, while for the thick-wall approximation with the same parameters the peak of the spectrum lies barely outside (to the right of) the LV O2 region, with the low-frequency tail of the spectrum crossing through the LV O2 region. On the low-frequency tail of the thick-wall spectrum, we also find that there is an intersection with the projected LISA sensitivity region, hinting at the possibility of a simultaneous detection in both the mHz and kHz frequency range.

{ The example shown in Figure~\ref{fig:GW_spectrum} should be contrasted with earlier phenomenological studies
of chain inflation, where the tunneling history was treated in a model-independent way
through an effective expansion of the bounce action, parametrized by coefficients such as
the $S_i$~\cite{Freese:2023szd}
. In such a description, one may continuously dial the evolution of the tunneling
rate and hence obtain a broad range of possible inflationary scales and GW peak frequencies.
Here, instead, we have specified a concrete microscopic mechanism: an exact tilted cosine potential
with a non-minimal coupling to gravity. This additional structure makes the
prediction more restrictive. Within the parameter range considered in this work, the viable
solutions organize into two disconnected regimes: the low-scale tilted cosine branch, where
the NMC effect is negligible and $V_*^{1/4}\lesssim 3\,{\rm GeV}$, and the high-scale
NMC branch, where the field-dependent amplification of the bounce action allows for
$10^{10}\,{\rm GeV} \lesssim V_*^{1/4}\lesssim 10^{12}\,{\rm GeV}$. We do not find a continuous band of viable
solutions connecting these two regimes, as can be also seen in Figure~\ref{fig:ns_vs_V*_Full}. These regions of parameter space can only be brought closer together by increasing the coupling $\xi$, which for $\xi\lesssim100$ limits us to the aforementioned high-scale limits.

This has a direct consequence for the GW signal. The low-scale branch reproduces the
standard tilted cosine expectation of a signal in the nHz range, relevant for pulsar-timing
arrays, while the high-scale NMC branch shifts the peak into the Hz--kHz range targeted
by interferometric detectors; the latter is visualized in Figure~\ref{fig:GW_spectrum}. The detailed position and amplitude
of the peak are then controlled by the microscopic parameters of the model and by the
properties of the graceful exit, as we will see in Section~\ref{sec:GWparameterdependence}. Thus, combining the CMB constraints on $n_s$ and
$\alpha_s$ with the shape and amplitude of the stochastic GW spectrum provides a way to
restrict, and potentially infer, the underlying parameters $\xi$, $u_*$, $x$, and $\alpha$. Finally, it is interesting to note  that the thick-wall solution for the GW signal (which is a better approximation for large values of $x$ than the thin wall solution, and thus better suited for our case)  leads to a spectrum that decays slowly enough at low frequencies, so that it can be in principle detected concurrently by some kHz experiment, such as LVK or CE/ET, as well as by LISA.}

In the high-scale regime, made possible for a pure tilted cosine by the non-minimal coupling,
the GW signal  appears in the frequency bands targeted by interferometric experiments. The precise detectability, however, depends on how the peak amplitude and frequency vary with the microscopic and phenomenological parameters of the model, which we analyze next.

\subsection{Dependence of Gravitational Wave Spectrum on Model Parameters}
\label{sec:GWparameterdependence}

Having established the overall properties of the GW spectrum of the high-scale parameter space of the NMC chain inflation model, we now describe   how the parameters in the theory affect the GW spectrum. Our model has four parameters: the non-minimal coupling $\xi$, the relative position of the pivot scale field value $u_*\equiv\chi_*/\chi_c$, the strength of the phase transition $\alpha$, and the energy scale of inflation at the pivot scale $V_*$.
In practice, once a fixed value for $n_s\sim0.970$ has been set by CMB observations, only three parameters remain; given $n_s$, we take $V_*$ to be fixed in terms of the other three parameters,
as in Figure~\ref{fig:ns_vs_V*_Full}.

Figure~\ref{fig:GWSpectrum_VariationWithParameters} plots the position of the peak of the thick-wall spectrum $\Omega_{\rm GW}h^2$ as a function of frequency $\nu$, and illustrates the variation of the peak with each of the parameters $\xi$, $u_*$, and $\alpha$. 
In the plot, we have fixed $n_s=0.97$.

The impact of the nonminimal coupling $\xi$ on the GW spectrum is relatively straightforward: larger $\xi$ leads to lower frequency GW waves. Since we are fixing the  value of $n_s$ for all models in this parameter analysis, raising $\xi$ will lead to lower inflationary scales $V_*$, since  NMC effects  will become important 
at a lower field-value, therefore at a lower value of $f$ and eventually at a lower energy scale $V_*$; see  Figure~\ref{fig:FieldTraversal_vs_InflationaryScale}.  In the case of $\xi$, we confirm that increasing it leads to a shifting of the GW  peak to lower frequencies, associated with the lower inflationary scales that are required for observational compatibility, as in Figure~\ref{fig:ns_vs_V*_ZoomGrid}. In the example shown by the middle horizontal band in Figure~\ref{fig:GWSpectrum_VariationWithParameters}, where we keep $\alpha=0.1$ and $u_*=0$ fixed, the peak shifts from $\nu\sim$ kHz for $\xi=60$ to $\nu\sim10$ Hz for $\xi=10^3$. This means that the model will have its peak GW emission   within the CE and ET sensitivity regions for $\xi=\mathcal{O}(10^2)$.

\begin{figure}
    \centering
    \includegraphics[width=\linewidth]{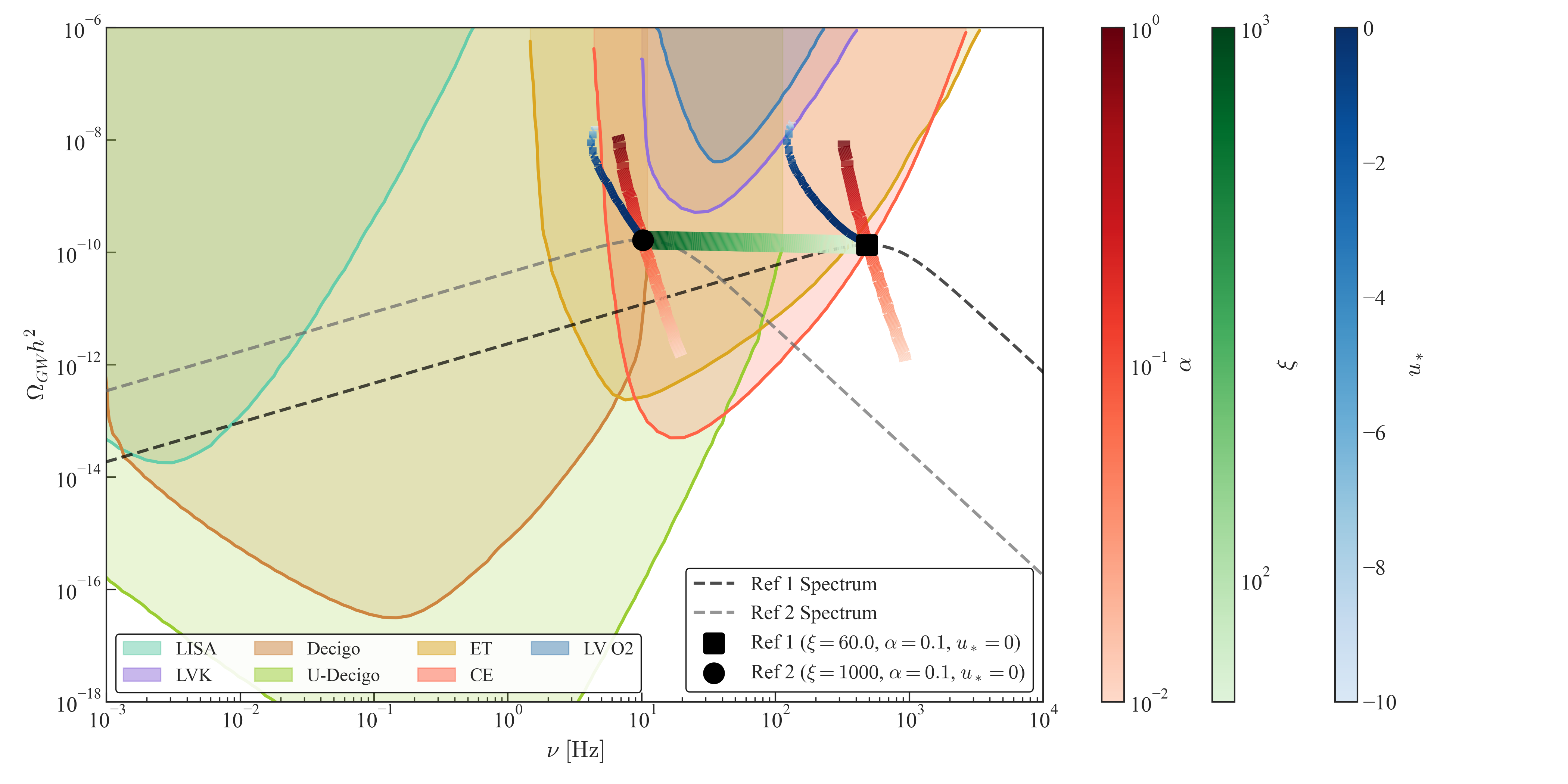}
    \caption{The variation of the peak of the thick-wall GW spectrum in a NMC chain inflation model with the parameters ${\alpha,\xi,u_*}$. Starting from the two black dots defined in the legend, we vary each of these three parameters along red ($\alpha)$,  green ($\xi$) and blue ($u_*$), curves as shown. The color-shaded regions indicate the parameter space testable by future GW experiments as labeled. For each set of parameters, we have fixed $n_s=0.970$ as a target to determine the required high-energy inflationary scale $V_*$. }
    \label{fig:GWSpectrum_VariationWithParameters}
\end{figure}

The primary effect of $\alpha$ is  to alter the peak amplitude of the
GW spectrum: larger $\alpha$ produces larger GW peak amplitude. The reason is twofold. First, the strength of the final transition is
correlated with the duration parameter $\beta$ during graceful exit. In this regime
$\beta/H$ should be evaluated as a function of $\alpha$, as in Ref.~\cite{Freese:2022qrl}, rather than
identified with the approximately constant value used for the intermediate transitions
during the chain. Increasing $\alpha$ decreases $\beta/H$, which enhances the factor
$(H/\beta)^2$ appearing in the bubble-collision spectrum of Eq.~\eqref{eq:OmegaGWbc}. Second, $\alpha$
determines when the last transition occurs relative to vacuum-radiation equality: larger $\alpha$ corresponds to a larger value of $a_{\rm last}$, and
therefore enhances the redshift factor $a_{\rm last}^4$ in Eq.~\eqref{eq:OmegaGWbc}. These effects both
increase the observed GW amplitude.

The parameter $\alpha$ also somewhat affects the peak frequency. From Eq.~\eqref{eq:GWfrequency}, one has
$\nu^0_{\rm peak}\propto a_{\rm last}\beta(\alpha)$. Thus increasing $\alpha$ has two
competing effects: the larger value of $a_{\rm last}$ tends to increase the observed
frequency, while the smaller value of $\beta(\alpha)$ tends to decrease it. In the
parameter range shown in Figure~\ref{fig:GWSpectrum_VariationWithParameters}, the latter effect dominates, so the peak shifts
slightly toward lower frequencies as $\alpha$ is increased. This is visible in the nearly
vertical red band of Figure~\ref{fig:GWSpectrum_VariationWithParameters}: for fixed $\xi=60$ and $u_*=0$, increasing
$\alpha$ from $0.01$ to $1$ raises the peak amplitude from
$\Omega_{\rm GW}h^2\sim 10^{-12}$ to $\Omega_{\rm GW}h^2\sim 10^{-8}$, while shifting
the peak frequency from $\nu\sim 1\,{\rm kHz}$ to $\nu\sim 0.3\,{\rm kHz}$. Stronger
final transitions can therefore move the signal into the CE and ET sensitivity regions
even for $\xi=\mathcal{O}(10)$.

{Finally, the relative pivot-scale parameter $u_*$ has two main effects on the GW
spectrum. Recall that $u_*$ controls the location of the CMB pivot scale relative to the
point where the non-minimal coupling vanishes in the Jordan frame, Eq.~\eqref{eq:jordan_action}: the latter occurs at the pivot scale for $u_*=0$, or closer to the end of inflation for $u_*\ll -1$. First,
as shown in Figure~\ref{fig:ns_vs_V*_ZoomGrid}, moving $u_*$ from $u_*=0$ toward large negative values lowers the
inflationary scale $V_*$ required to obtain observationally compatible values of $n_s$.
Since the observed peak frequency scales approximately as
$\nu_{\rm peak}^0\propto V_*^{1/4}$, this lowers the peak frequency of the GW spectrum.
This is visible in Figure~\ref{fig:GWSpectrum_VariationWithParameters}, where the $u_*$ curve shifts left as $u_*$ is decreased toward
the large-negative-$u_*$ regime.
Second, changing $u_*$ also affects the amplitude of the GW spectrum through the
radiation density at the end of inflation. As discussed in Appendix~\ref{sec:RadiationAppendix}, moving toward
large negative $u_*$ can increase the final radiation density $\rho_{r,c}$ by up to roughly
an order of magnitude relative to the $u_*=0$ case. The relevant scale factor for the
last bubble-collision burst is $a_{\rm last}$. Using the definition of
$\alpha$ in Eq.~\eqref{eq:alpha}, together with $\alpha_{\min}=\Delta V/\rho_{r,c}$ from
Eq.~\eqref{eq:alpha_min}, one finds
\begin{equation}
a_{\rm last}^4
=
a_c^4\,\alpha\,\frac{\rho_{r,c}}{\Delta V}
=
a_c^4\,\frac{\alpha}{\alpha_{\min}} \,  .
\end{equation}
Thus, for fixed $\alpha$, the simultaneous increase in $\rho_{r,c}$ and decrease in
$\Delta V$ associated with decreasing $u_*$ can enhance the redshift factor entering the
observed GW amplitude. This can boost the peak amplitude by as much as a factor of
$\mathcal{O}(100)$, as seen in Figure~\ref{fig:GWSpectrum_VariationWithParameters}. Such an enhancement can place the peak of the
GW spectrum within the LVK sensitivity region even for $\alpha\sim 0.1$, with the possibility of entering the LV O2 region for stronger final transitions with larger $\alpha$.}

In this section, we explored the gravitational wave implications of the high-scale chain inflation regime 
with non-minimal coupling of the tunneling field to gravity. 
We have shown that the model predicts a correlated set of scalar and gravitational wave signatures, making chain inflation with NMC testable through complementary cosmological observables.

\section{Conclusions and Discussion}\label{sec:Conclusions}

In this work, we studied the chain inflation model, in which the inflaton field tunnels through a series of minima of ever lower energy.  We began
by revisiting the original chain inflation from both an analytic and phenomenological perspective. 
Building on the effective bounce action expansion used to describe variations of the tunneling rate along the chain, we developed a systematic analytic treatment of the resulting cosmological observables. In particular, we derived closed-form expressions for the scalar spectral index, its running, and their dependence on the microscopic parameters controlling the tunneling dynamics. Applied to the minimally coupled tilted cosine potential, this framework yields, for the first time, an analytic derivation of the relation between the scalar tilt and the inflationary scale, $n_s(V_*^{1/4})$, in this model. We showed that the constant bounce action of the pure tilted cosine (in which properties of the potential are identical for every tunneling event albeit at lower energies) locks the scalar spectral index to the duration of inflation, forcing the model into a low-scale regime, $V_*^{1/4}\lesssim 3\,\GeV$, once current CMB constraints on $n_s$ are imposed.  We note that, for example in the context of the string landscape, the pure tilted cosine is merely an approximation to what may be a more complicated set of tunneling potentials (in which case the restriction to low scales may not hold up).

We then studied the effect of a non-minimal coupling between the tunneling field and gravity. Such a coupling is not an ad hoc deformation: in curved spacetime, the operator $\xi \phi^2R$ is allowed by symmetry and is generically required as a counterterm in interacting scalar field theories. This makes non-minimally coupled chain inflation a natural extension of the minimally coupled setup, especially at the high curvatures relevant for the early Universe. In the Einstein frame, the coupling flattens the effective potential and induces a field-dependent amplification of the Euclidean bounce action. As a result, the tunneling rate evolves across the chain, allowing the expansion to become concentrated in different portions of the field traversal and breaking the rigid connection between $n_s$ and the inflationary scale.

{This mechanism opens a viable high-scale regime of chain inflation, with
$V_*^{1/4}\sim 10^{11}\,{\rm GeV}$, which remains compatible with current measurements
of the scalar spectral index. The allowed parameter space is shaped by the interplay
between the non-minimal coupling $\xi$, the tunneling parameter $x$, and the relative
location of the pivot scale $u_*$. Perturbative control of the tunneling calculation imposes an important theoretical
constraint on this high-scale branch, because the high-scale NMC solutions require
relatively small baseline bounce actions. In particular, the unitarity
condition  becomes increasingly restrictive
as $x$ is lowered. For the range $\xi\lesssim 100$ studied here, this removes the high-scale NMC
solutions for $x\lesssim 0.9$, pushing the viable high-scale regime toward fast tunneling,
with $x$ close to its upper allowed value.
This should be distinguished from the low-scale tilted cosine branch, which effectively
coincides with the minimally coupled solution. At low scales the
field excursion is so small that the non-minimal correction is negligible, and the solutions
reduce to those of the minimally coupled tilted cosine model.  Numerically, the low-scale branch remains viable down to
approximately $x\gtrsim 0.7$. Thus the condition $x\gtrsim 0.9$ should be understood as
a unitarity-driven requirement for high-scale NMC chain inflation, not as a general lower
bound on the tilted cosine chain.}

An important distinction from familiar applications of non-minimal couplings to slow-roll
inflation is the way in which the coupling enters the observable dynamics. In models such
as Higgs inflation, the non-minimal coupling is usually used to flatten the classical potential
and thereby sustain slow roll, often requiring parametrically large values of $\xi$. In the
present case, the field evolution is dominated by tunneling rather than classical rolling. The
non-minimal coupling therefore matters primarily because it makes the Euclidean bounce
action field dependent, which in turn changes the tunneling rate exponentially. This provides
a much more sensitive lever arm: moderate values $\xi=\mathcal{O}(10)$ are sufficient to
produce observable effects. At low scales, the field excursion is too small for the conformal
factor to matter, while at high scales the NMC-induced amplification of $S_E$ accumulates
over the chain and opens the high-scale branch with distinctive CMB running and
interferometer-band gravitational waves.

As shown in Figures~\ref{fig:ns_vs_V*_Full} and~\ref{fig:ns_vs_V*_ZoomGrid}, for the example of $\xi=60$ and $x=0.96$, the  departure from the minimally coupled tilted cosine predictions becomes important for
$V_*^{1/4}\gtrsim10^{10}\,\GeV$, so that $n_s$ can dip down into the observationally viable range compatible with  CMB measurements.  More generally\footnote{In the analytically tractable $u_*=0$ branch, the departure from the minimally coupled tilted cosine predictions becomes important once the quadratic coefficient $S_2$ approaches order unity. Using the estimate $S_2\simeq(\xi V_*^{1/4}/10^{12}\,\GeV)^2$,  given in Eq.~\eqref{eq:S2_uStar0}, this occurs for 
$V_*^{1/4}\gtrsim10^{10}\,\GeV$, when $\xi\gtrsim20$. 
Indeed, the numerical solutions for any $u_*$ show the same qualitative onset of NMC effects at high scales, while the relative importance of $S_1$ and $S_2$ depends on the pivot location $u_*$.}, we find this same onset of NMC effect at such high scales when $\xi\gtrsim20$.  
Furthermore, the running of the scalar spectral index, $\alpha_s$ provides a further observational handle into distinguishing between $u_*\simeq 0$ and $u_*\ll -1$. As a reminder $u_*\simeq 0$ corresponds to cases where the non-minimal correction to the gravitational part of the action vanishes  at the pivot scale, whereas for $u_*\ll -1$ the same non-minimal correction vanishes close to the end of inflation. Interestingly, in the former case the running is typically $\alpha_s\sim 10^{-2}$, making it a prime target for future CMB surveys. In the latter case, the running is suppressed, with representative values $\alpha_s\sim -5\times 10^{-4}$, below current observational forecasts.

{Previous studies of high-scale chain inflation treated the tunneling history largely
in a phenomenological way, parameterizing the variation of the Euclidean action through effective
expansion coefficients~\cite{Freese:2023szd}. In that approach, the inflationary scale and the resulting
GW peak frequency can be scanned in a relatively model-independent way. In the present
work we have instead embedded this phenomenological possibility in a concrete microscopic
model: a tilted cosine chain supplemented by a non-minimal coupling to gravity.  This added microscopic structure makes the predictions
more restrictive: the viable solutions do not continuously populate all intermediate
inflationary scales, but instead organize into two disconnected regimes, the minimally
coupled low-scale tilted cosine branch with $V_*^{1/4}\lesssim 3\,{\rm GeV}$ and the
high-scale NMC branch with $V_*^{1/4}\sim 10^{10}-10^{11}\,{\rm GeV}$. Consequently, the
associated GW signal is also split between two characteristic frequency ranges, the nHz
regime for the low-scale branch and the Hz--kHz interferometer regime for the high-scale
NMC branch.}

The gravitational wave sector provides a complementary probe of the same modified tunneling history. The usual vacuum tensor signal is negligible throughout the viable parameter space. However, bubble collisions from the first-order phase transitions near the end of the chain can generate a stochastic gravitational wave background with potentially observable amplitude.
Because the non-minimal coupling opens a high-scale branch far above the minimally
coupled tilted cosine solution, the peak frequency is split between two characteristic
regimes: the nHz range for the low-scale branch, $V_*^{1/4}\sim{\rm MeV-GeV}$, and
the Hz--kHz range for the high-scale branch, $V_*^{1/4}\gtrsim 10^{10}\,{\rm GeV}$.
For low inflationary scales, the produced GW signal can account for the NANOGrav and
other PTA-motivated parameter ranges discussed in Ref.~\cite{Freese:2022qrl}. In the high-scale NMC
branch, the resulting GW signal can fall within the projected reach of future
interferometers such as the Einstein Telescope and Cosmic Explorer.
 The amplitude and peak frequency depend on the strength of the final transition, the radiation density at the end of inflation, and the location of the pivot scale, making the gravitational wave signal an independent probe of the same tunneling dynamics that controls the scalar spectrum.

Several avenues remain open for future work. The analytic methodology developed here provides a foundation for studying more general chain inflation potentials beyond the ideal tilted cosine, including chains in which the barrier height, spacing, or tilt vary along the tunneling path. Such deformations can also break the low-scale lock of the exact tilted cosine potential and may lead to distinct CMB observables, offering a complementary route to high-scale chain inflation~\cite{DisorderedChainInflation}.

Importantly, the current analysis  assumed that the energy difference between successive vacua was released entirely as radiation through bubble collisions. 
 However, simulations have revealed a rich post-collision phenomenology, including complex bubble wall dynamics and oscillons~\cite{Cutting:2018tjt,Pirvu:2023plk}.
 Furthermore, the by-products of bubble collisions (whether in the form of radiation or not), can significantly affect the tunneling rate and possibly allow classical motion over the tops of barriers. In this paper we treat the dynamics of the phase transition by using the standard zero-temperature vacuum tunneling rate. Appendix~\ref{app:backreaction} provides a brief discussion of back-reaction effects on tunneling dynamics and describes avenues for future research, that would help reduce the current uncertainties.
{
Furthermore, the generic ``tunneling catastrophe" bound $x\lesssim 0.96$ was obtained in Ref.~\cite{Cline:2011fi} from $1+1$ dimensional classical field simulations. Extending the analysis to $3+1$ dimensions would provide a more robust determination of the runaway threshold. 
}
On the gravitational wave side, a more detailed treatment of the late-time vacuum-radiation dynamics, finite-temperature effects on rapid tunneling, bubble-wall evolution, and the efficiency with which vacuum energy is converted into radiation would sharpen the prediction for the final phase transitions in the cascade. More broadly, the analytic framework developed in this work opens a systematic way to classify chain inflation models according to their tunneling history and according to their correlated CMB and gravitational wave signatures.

Overall, non-minimally coupled chain inflation provides a theoretically motivated and observationally testable high-scale realization of tunneling-driven inflation. The same curvature coupling that is generically expected in scalar field theories on curved backgrounds (such as our Friedmann-Robertson-Walker Universe) dynamically modifies the vacuum lifetimes, opens the high-scale parameter space, and produces correlated CMB and gravitational wave signatures. Future measurements of the scalar running, together with searches for stochastic gravitational waves in the Hz--kHz range, will therefore be able to test not only the viability of this scenario, but also the detailed tunneling history of the early Universe.

\section*{Acknowledgments}
K.F. holds the Jeff \& Gail Kodosky Endowed Chair at the University of Texas,
Austin. K.F. and E.I.S. are grateful for support from this Chair. K.F. and E.I.S. acknowledge
support from the U.S. Department of Energy, Office of Science, Office of High Energy Physics
program under Award Number DESC-0022021. K.F. also acknowledges support from the
Swedish Research Council (Contract No. 638-2013-8993).
The work of M.B.V. is supported by FCT (Fundação para a Ciência e Tecnologia, Portugal) through the grant 2024.00457.BD with DOI identifier https://doi.org/10.54499/2024.00457.BD. Centro de Fisica do Porto is partially funded by Fundação para a Ciência e Tecnologia (FCT) under the grant UID04650-FCUP.

\begin{appendices}

\section{Vacuum-radiation Dynamics}\label{sec:RadiationAppendix}

{In this appendix we justify the treatment of the radiation component used throughout the main text. During chain inflation, each tunneling event converts a fraction of vacuum energy into radiation, while the background expansion simultaneously redshifts the radiation bath. The main text uses the corresponding tracking solution to estimate both the end of inflation and the radiation density relevant for the gravitational wave signal. Here we check when this tracking approximation is valid and quantify how much the radiation density at vacuum-radiation equality differs from its value at the CMB pivot scale.}

The tracking solution \eqref{eq:trackingSolution} can be taken to be true if the condition
\begin{equation}
    \left|\frac{\dot\rho_r}{\rho_r}\right|\approx\left|\frac{\Delta\tilde\phi}{\Delta t}\frac{d\ln\rho_r}{d\tilde\phi}\right|=\left|\frac{1.4\Gamma^{1/4}}{N_*}\frac{d\ln\rho_r}{d\tilde\phi}\right|\ll4H
\end{equation}
is satisfied. To test the validity of this condition, we parametrize the Hubble rate as $H^2=(V_*(1-\tilde\phi)+\rho_{r,*})/3$, 
where we have taken $V(\phi_*)=V_*$ and $V(\phi_c)\approx0$, while also using the fractional field displacement $\tilde\phi$ and approximating the radiation density to be roughly of the order of its value at the pivot scale $\rho_{r,*}$. This potential is a good local approximation for generic chain inflation potentials, and it is a nearly exact representation for the tilted cosine model discussed later in this work. Using $\Delta V\approx V_*/N_*$, we also find
\begin{equation}
    \rho_{rad,*} = \frac{1.4 \left( \frac{V_*}{N_*} \right) \Gamma_*^{1/4}}{4 H_*} = \frac{V_*}{4} \left( \frac{1.4 \Gamma_*^{1/4}}{N_* H_*} \right)=\frac{3V_*}{5}\frac{1-n_s}{2-S_1}\approx\frac{0.018}{2-S_1}V_* \,,
\end{equation}
where we have used Eq.~\eqref{eq:ns_Effective} and took $n_s\approx0.970$. This allows us to write
\begin{equation}
    H(\tilde\phi)\approx H_*\sqrt{(1-\tilde\phi)+\rho_{r,*}/V_*} \,.
\end{equation}
For simplicity, and with the results of Section~\ref{sec:NMC} in mind, we restrict our attention to a transition rate evolving with only $S_{1,2}\neq0$, although we stress that including higher-order terms can be done through an entirely analogous method, adding only length to the expressions and additional theory parameters. We can thus write
\begin{equation}
    \frac{d\ln\Gamma}{d\tilde\phi}=-S_1-2S_2\tilde\phi \quad \quad \quad \quad \quad \frac{d\ln H}{d\tilde \phi}=-\frac{1}{2(1-\tilde\phi+\rho_{rad,*}/V_*)} \, ,
\end{equation}
such that 
\begin{equation}
    \frac{d \ln \rho_{r}}{d\tilde{\phi}} =\frac{1}{4}\frac{d\ln\Gamma}{d\tilde\phi}-\frac{d\ln H}{d\tilde\phi}= -\frac{1}{4}(S_1 + 2S_2\tilde{\phi}) + \frac{1}{2(1 - \tilde{\phi} + \rho_{rad,*}/V_*)}
\end{equation}
and therefore the tracking condition becomes
\begin{equation}\label{eq:TrackingCondition}
\begin{aligned}
    &\left| \frac{1.4 \Gamma^{1/4}(\tilde{\phi})}{4 H(\tilde{\phi}) N_*} \left[ -\frac{S_1 + 2S_2\tilde{\phi}}{4} + \frac{1}{2(1 - \tilde{\phi} + \frac{0.018}{2-S_1})} \right] \right|\\
    &=\frac{0.018}{2-S_1}\frac{1}{\sqrt{1 - \tilde{\phi} + \frac{0.018}{2-S_1}}} \exp\left(-\frac{S_1\tilde{\phi} + S_2\tilde{\phi}^2}{4}\right) \left| -\frac{S_1 + 2S_2\tilde{\phi}}{4} + \frac{1}{2(1 - \tilde{\phi} + \frac{0.018}{2-S_1})} \right| \ll 1 \,,
\end{aligned}
\end{equation} 
where in the final step we used $\frac{1.4 \Gamma^{1/4}_*}{4 H_* N_*}=\frac{3}{5}\frac{1-n_s}{2-S_1}$ from Eq.~\eqref{eq:ns_Effective} with $n_s\approx0.970$ to simplify the constant pre-factor.

If we wish to find the vacuum-radiation equilibrium point $\tilde\phi_{eq}$ within the tracking solution, we must solve
\begin{equation}
    \rho_{\rm v}(\tilde\phi_{eq})=\rho_{r}(\tilde\phi_{eq})\Rightarrow V_*(1-\tilde\phi)= \frac{1.4 \Delta V \sqrt{3}\Gamma(\tilde\phi_{eq})^{1/4}}{4 \sqrt{\rho_{vac}(\tilde\phi_{eq}) + \rho_{r}(\tilde\phi_{eq})}} \,.
\end{equation}
After simplification this is equivalent to the condition
\begin{equation}\label{eq:Vac_Rad_Equality}
    (1-\tilde\phi_{eq})^{3/2} = \frac{1}{\sqrt{2}} \frac{0.018}{2-S_1}\exp\left( -\frac{S_1\tilde\phi_{eq} + S_2\tilde\phi_{eq}^2}{4} \right)\, ,
\end{equation}
which may be solved numerically for $\tilde\phi_{eq}$ for a given $S_1$ and $S_2$. The radiation density at this point can then be calculated as
\begin{equation}\label{eq:Rad_AtEquality}
    \frac{\rho_{r}(\tilde\phi_{eq})}{\rho_{r,*}}=\frac{V_*(1-\tilde\phi_{eq})}{\rho_{r,*}}=\frac{2-S_1}{0.018}(1-\tilde\phi_{eq}) \,.
\end{equation} 

\subsection{Tilted Cosine}

{We first apply the tracking analysis to the pure tilted cosine model, for which the tunneling rate is constant along the chain, $S_1=S_2=0$.}
We find that the tracking condition \eqref{eq:TrackingCondition} is broken at around $\tilde\phi\gtrsim0.98$, i.e. towards the latter stages of the tunneling chain, up to which point one may consider the radiation density to obey the tracking solution. Note that in this case, since $\Gamma=\Gamma_*$ for all field values, the violation of the tracking condition follows uniquely from the decrease of the Hubble rate $H(\tilde\phi)$ as the inflaton descends along the chain potential.

For the tilted cosine, we find through Eq.~\eqref{eq:Vac_Rad_Equality} that vacuum-radiation equality occurs at $\tilde\phi_{eq}=0.965$, which we note occurs before the breaking point of the tracking condition at $\tilde\phi\sim0.98$, thus justifying the usage of that approximation. 

We find ${\rho_{r}(\tilde\phi_{eq})}/{\rho_{r,*}}\approx4$, such that the radiation density at the end of inflation should be calculated through its tracking value near the end of the chain and not taken to be fixed at its pivot scale value even for completely static chain parameters ($\dot\Gamma=0$), contrarily to what was originally suggested in Ref. \cite{Freese:2023szd}. This earlier termination of inflation in the tilted cosine model has little effect on the observable accumulation of $e$-folds, since this decoherence occurs at the end of the chain, where the $e$-fold accumulation is negligible in comparison to what the Universe inflates near the pivot scale. The consequences on the reheating temperature $T_c\propto\rho_r^{1/4}$ are not significant, since this is only altered by an $\mathcal{O}(1)$ factor of $4^{1/4}\approx1.4$.

\subsection{NMC Chain Inflation with \texorpdfstring{$u_*=0$}{u*=0}}
{We next consider the non-minimally coupled model in the $u_*=0$ branch. This case is useful because the non-minimal coupling slows the tunneling rate toward the end of the chain, precisely where vacuum-radiation equality is approached.}

In the case of the NMC model, the effects of these corrections  to the simple constant attractor solution of radiation are also mostly negligible at the level of CMB observables, a fact that holds for both $u_*=\chi_*/\chi_c=0$ and $u_*\rightarrow-\infty$ classes of models. In fact, for observationally compatible models with $u_*=0$ we find $S_2\sim5-10$ (and $S_1=0$ by definition), which ensures that this condition holds pretty much all the way until $\tilde\phi_{eq}\approx1$ (i.e. the end of the chain), since the transition rate is decreased towards the end of the chain. Therefore, even though the dilation of the vacuum lifetimes can lead to an increased amount of $e$-fold accumulation at the bottom of the potential, the same effect forces the moment of vacuum-radiation equality to approach the exact end of the chain, such that no significant amount of $e$-folds is ignored and our predictions are valid. In this case, we determine the ratio ${\rho_{r}(\tilde\phi_{eq})}/{\rho_{r,*}}\approx0.7$, which leads to a negligible impact on both the CMB and GW observables of the theory, as well as indicating that the pivot scale attractor value is an acceptable approximation.

\subsection{NMC Chain Inflation with \texorpdfstring{$u_*\rightarrow-\infty$}{u*->-infty}}

{Finally, we examine the opposite limiting regime, $u_*\to-\infty$. In this branch most of the $e$-fold accumulation occurs near the pivot scale, while the tunneling rate accelerates toward the end of the chain, making it important to verify that the radiation treatment remains self-consistent.}

Oppositely to what is found for $u_*=0$, the $u_*\rightarrow-\infty$ universality class of models leads to an accumulation of $e$-folds of expansion at the top of the chain, near the pivot scale, with an acceleration of the vacuum tunneling towards the end of the chain. Observational compatibility at the high-energy scales requires $S_2=-S_1/2\sim2-3$. This leads to a premature vacuum-radiation equality at $\tilde\phi_{eq}\approx0.977$, which is nevertheless a more suitable outcome than the value of $\tilde\phi_{eq}\approx0.965$ found for the tilted cosine. However, since vacuum lifetimes decrease towards the end of the chain, this leads to the $e$-fold accumulation being negligible at this point, such that an earlier termination of the chain has effectively no effect on the CMB observables of the theory. Importantly, we find ${\rho_{r}(\tilde\phi_{eq})}/{\rho_{r,*}}\approx10$, which means that fixing the attractor solution to its pivot scale value would underestimate the final radiation density by one order of magnitude.

\section{Gracefully Exiting Chain Inflation}\label{sec:GracefulExitAppendix}

{In the main text we remain largely agnostic about the microscopic mechanism that terminates the tunneling chain. In this appendix we summarize two simple classes of graceful-exit scenarios that capture the relevant possibilities for the late-time dynamics and for the gravitational wave signal: a chain-to-stop transition, in which the final tunneling events release the last vacuum energy, and a chain-to-roll transition, in which the tunneling phase gives way to a smooth descent toward the minimum.}

After the inflaton completes enough transitions through the chain to satisfy the necessary number of $e$-folds of expansion, we must gracefully exit chain inflation, as the Universe must become radiation-dominated to ensure that the standard hot Big Bang cosmology follows and gives rise to the components and structure of the Universe we live in. The reheating of the Universe is brought about by the radiation bath, which is automatically achieved by the bubble collisions during the phase transitions, while the emergence of radiation domination must follow from some stopping mechanism for the tunneling chain. There are two main classes for such exits, which we describe as ``chain-to-stop" and ``chain-to-roll".

\subsection{Chain-to-stop}

The first of these, ``chain-to-stop", consists of a scenario where the barriers of the potential increase, thus increasing the Euclidean bounce action $S_E$ and decreasing the tunneling rate $\Gamma$, up to a point where the lifetime of the vacuum becomes so long that there are effectively no more transitions within the observable age of the Universe and the inflaton essentially stops, leading to the radiation-dominated epoch, as shown in the left panel of Figure~\ref{fig:StoppingMechanisms}. Depending on the specific model, there may be up to a few slower additional phase transitions after the onset of radiation domination. However, the lifetime of these intermediate vacua must not be large enough to affect post-inflationary epochs such as BBN. An example of a chain-to-stop mechanism implemented through an auxiliary scalar field and inspired by the relaxion mechanism \cite{Graham:2015cka} has been discussed in Refs. \cite{Freese:2023szd,Freese:2021rjq,Freese:2022qrl}. In this example, the auxiliary field quickly raises the inflaton's potential barriers as it gets displaced from its original position in a Higgs-like symmetry breaking potential. In this scenario, the large majority of the chain can be treated as completely decoupled from the auxiliary field, with only its termination being dictated by the relaxion-inspired coupling mechanism. This preserves the structure of the results presented thus far and in what follows, as a great part of the observational signatures of chain inflation derive from the bulk of the chain, namely the point at which the pivot scale crosses the horizon.

\subsection{Chain-to-roll}
Alternatively, a ``chain-to-roll" scenario involves a decrease of the barriers in the potential up to a point where there are no more maxima to tunnel through. Once this happens, the inflaton rolls along the remaining path to the global minimum of the potential, after which it can oscillate around this point and decay into radiation, as shown in the right panel of Figure~\ref{fig:StoppingMechanisms}. In fact, some of the final phase transitions may be skipped, as once the barriers become small enough there may be a tunneling catastrophe, leading the field to tunnel directly to the bottom of the potential \cite{Cline:2011fi}. This has little effect on the observational aspects of the theory, as even without a tunneling catastrophe the field tunnels so quickly through this region that its boundary can be taken as the terminating point of the tunneling chain. This can be achieved, for example, in natural chain inflation \cite{Freese:2021noj}, where corrections to the axion potential in supergravity endow it with small quasiperiodic oscillations over the standard natural inflation potential, such that chain inflation can occur along the global potential's shape, terminating at the bottom of the baseline axion potential.

\begin{figure}
    \centering
    \includegraphics[width=\linewidth]{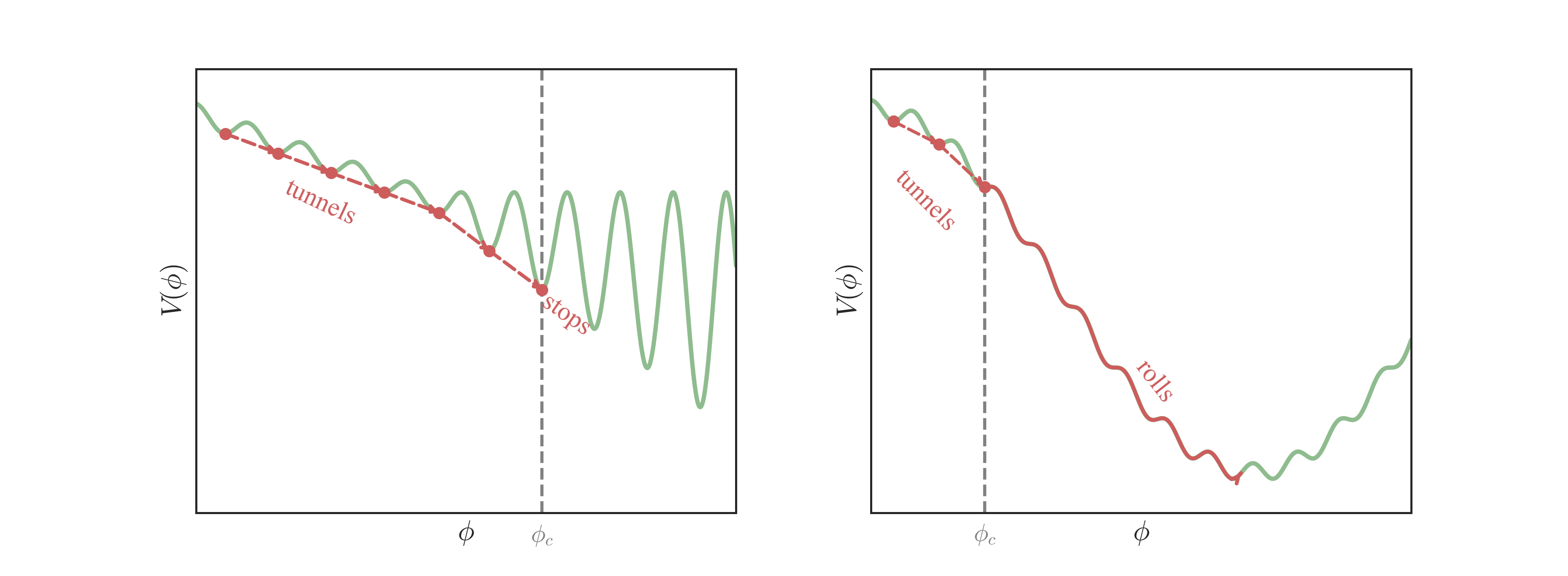}
    \caption{Schematic representation of graceful exit mechanisms in chain inflation. Left panel: The ``chain-to-stop" scenario, where the potential barriers increase, lowering the tunneling rate until transitions effectively cease and the inflaton stops. Right panel: The ``chain-to-roll" scenario, where the potential barriers decrease until no maxima remain, allowing the inflaton to roll to the global minimum of the potential.}
    \label{fig:StoppingMechanisms}
\end{figure}

\section{Simplifying Bounce Action with Potential Symmetries} \label{sec:ActionSymmetryAppendix}

{The analytic estimates in the main text  use the fact that the Euclidean bounce action for a locally tilted cosine potential can be written as a dimensionless function of the tunneling parameter $x$, multiplied by an overall factor $f^4/\Lambda^4$. In this appendix we derive this scaling explicitly by rescaling the field and Euclidean radial coordinate in the bounce equation.}

The bounce solution $\phi(\rho)$ is a solution to the Euclidean equation of motion\footnote{Note that here $\rho$ is not the energy density of the vacuum, but rather the radial coordinate in 4-dimensional Euclidean space.}
\begin{equation}
    \frac{d^2\phi}{d\rho^2} + \frac{3}{\rho}\frac{d\phi}{d\rho} = V'(\phi) \,.
    \label{eq:eom}
\end{equation}
We consider the local form for a general potential relevant for chain inflation
\begin{equation}
    V(\phi) = \mu^3\phi + \Lambda^4 \cos\left(\frac{\phi}{f}\right) \,,
\end{equation}
which yields the derivative
\begin{equation}
    V'(\phi) = \mu^3 - \frac{\Lambda^4}{f}\sin\left(\frac{\phi}{f}\right) = \frac{\Lambda^4}{f} \left[ x - \sin\left(\frac{\phi}{f}\right) \right] \,,
\end{equation}
where we have defined the dimensionless parameter $x \equiv f\mu^3/\Lambda^4$. 

To demonstrate the scaling behavior, we introduce a dimensionless field $\psi$ and a dimensionless radial coordinate $\tilde{\rho}$
\begin{equation}
    \psi \equiv \frac{\phi}{f} \,, \quad \tilde{\rho} \equiv \frac{\Lambda^2}{f}\rho \,.
\end{equation}
Substituting these variables into Eq.~\eqref{eq:eom}, the derivatives transform as $d/d\rho = (\Lambda^2/f) d/d\tilde{\rho}$. The equation of motion becomes
\begin{equation}
    \left(\frac{\Lambda^2}{f}\right)^2 \frac{d^2\psi}{d\tilde{\rho}^2} + \frac{3}{\tilde{\rho}} \left(\frac{\Lambda^2}{f}\right)^2 \frac{d\psi}{d\tilde{\rho}} = \frac{\Lambda^4}{f^2} \left[ x - \sin(\psi) \right] \,.
\end{equation}
The dimensional pre-factor $\Lambda^4/f^2$ factors out completely, leaving a universal dimensionless equation for $\psi$
\begin{equation}
    \frac{d^2\psi}{d\tilde{\rho}^2} + \frac{3}{\tilde{\rho}}\frac{d\psi}{d\tilde{\rho}} = x - \sin(\psi) \,.
\end{equation}
The $O(4)$-symmetric Euclidean action $S_E = 2\pi^2 \int\mathcal{L}_E\  \rho^3 d\rho$ also scales according to the change of variables. The integral measure scales as $\rho^3 d\rho = (f/\Lambda^2)^4 \tilde{\rho}^3 d\tilde{\rho}$, and the Euclidean Lagrangian density $\mathcal{L}_E \sim (\partial_\rho \phi)^2 + V(\phi)$ scales as $\Lambda^4$. Thus, the total action scales as
\begin{equation}
    S_E \propto \left(\frac{f}{\Lambda^2}\right)^4  \Lambda^4 \mathcal{S}(x) = \frac{f^4}{\Lambda^4} \mathcal{S}(x) \,,
\end{equation}
where $\mathcal{S}(x)$ is the dimensionless action resulting from the integration of the universal equation for $\psi$.

\section{Low-Scale Baryogenesis in Standard Chain Inflation}
\label{app:low_scale_baryogenesis}

In the standard, perfectly periodic realization of chain inflation with a tilted cosine potential, the kinematic constraints required to match the CMB spectral index force the inflationary energy scale into the  low-energy regime ($V_*^{1/4} \lesssim 3 \text{ GeV}$). Because the radiation temperature at the end of the cascade is related to the energy released per step by $T_c \simeq 4.15\,\Delta V^{1/4}$, the Universe reheats to temperatures strictly bounded by $T_c \lesssim 300 \text{ MeV}$, with typical realizations lying closer to $T_c \sim {\cal O}(1-10) \text{ MeV}$, depending on the exact value of $n_s$.

This low thermalization scale entirely precludes conventional high-temperature mechanisms such as standard Thermal Leptogenesis ($T \gtrsim 10^9 \text{ GeV}$) and Electroweak Baryogenesis ($T \sim 100 \text{ GeV}$). Because electroweak sphalerons are exponentially decoupled at temperatures $T \ll 100 \text{ GeV}$, a successful baryogenesis mechanism in this scenario typically features explicit baryon  number ($B$) violation at the GeV scale. Furthermore, because any asymmetry $Y_B$ generated earlier in the chain is exponentially diluted by the subsequent accelerated expansion ($Y_B^{\rm final} \sim Y_B^{\rm produced} e^{-3{\cal N}_{\rm after}}$), the generation of the baryon asymmetry must be intimately tied to the final few vacuum transitions.

Despite these severe thermal limitations, chain inflation provides a distinct advantage: the non-adiabatic expansion and collision of bubble walls during the final transitions provide a highly out-of-equilibrium environment, automatically fulfilling the third Sakharov condition.

\subsection{Model-Independent Yield and BBN Constraints}

A useful, model-independent estimate of the final baryon yield can be obtained by assuming a fraction $f_X$ of the vacuum energy released in the final transition is non-thermally deposited into a baryogenesis-relevant species $X$, with a characteristic mass $m_X$. The number density of these parent particles is $n_X \simeq f_X \rho_{\rm vac} / m_X$. If each $X$ decay or transfer event produces a net baryon number asymmetry $\epsilon_B$, the generated baryon number density is $n_B \simeq \epsilon_B f_X \rho_{\rm vac} / m_X$.

Assuming the released vacuum energy is subsequently converted into the final radiation bath ($\rho_{\rm vac} \simeq \rho_{\rm rad}$), the relation between the radiation density and the entropy density ($s$) at reheating evaluates to $\rho_{\rm rad}/s \approx \frac{3}{4} T_c$. The final baryon-to-entropy yield is therefore:
\begin{equation}
    Y_B \equiv \frac{n_B}{s} \simeq \frac{3}{4} \epsilon_B f_X \frac{T_c}{m_X} \, .
    \label{eq:low_scale_chain_baryogenesis_yield}
\end{equation}

Matching the observed value $Y_B \simeq 8.7 \times 10^{-11}$ requires a combined efficiency of:
\begin{equation}
    \epsilon_B f_X \simeq 1.1 \times 10^{-10} \left( \frac{m_X}{T_c} \right) .
\end{equation}
We take $f_X\sim 10^{-3}$
 as a plausible benchmark, since non-thermal particle production during strong first-order phase transitions can be efficient in concrete bubble-wall and bubble-collision scenarios, although the precise value is model dependent.
The required CP asymmetry is then only $\epsilon_B \sim 10^{-6}$, which is readily achievable via loop-suppressed interactions.

The primary phenomenological hurdle is instead temporal. The heavily compressed cosmological timeline between the end of inflation and Big Bang Nucleosynthesis (BBN) imposes stringent limits on the mediator's lifetime. Reheating at $T_c \sim 10 \text{ MeV}$ corresponds to a cosmic time of $t \sim 10^{-2} \text{ s}$. 
To avoid modifying the neutron-to-proton ratio around weak freeze-out and the onset of
BBN, we conservatively require $X$ to decay or transfer its asymmetry on a timescale
$\tau_X\lesssim0.1-1\,{\rm s}$.
 For a simple two-body decay with coupling $y$, the decay rate $\Gamma_X \sim y^2 m_X / 8\pi$ requires a coupling of:
\begin{equation}
    y \gtrsim  10^{-11} \left( \frac{100 \text{ MeV}}{m_X} \right)^{1/2}  \left ( \frac{1\, {\rm s}}{ \tau_X} \right)^{1/2} .
\end{equation}
This places a model-dependent lower bound on the effective portal coupling connecting X to the Standard Model.

Provided these general yield and timing constraints are met, we can describe four phenomenological pathways  to realize low-scale baryogenesis in chain inflation.

\subsection{Specific Realizations}

\vspace{0.2cm}
\underline{1.\textit{ Non-Thermal Particle Decay (Dark Baryogenesis / Mesogenesis):}} \\
In this scenario, the mediator $X$ undergoes CP-violating and explicitly $B$-violating decays into Standard Model quarks alongside a stable dark sector fermion $\chi$, which carries a compensating baryon number to guarantee absolute proton stability~\cite{Elor:2018twp}. 

While elegant, this exact mechanism places a highly predictive, restrictive constraint on the Dark Matter (DM) relic abundance. In simple co-genesis models, the yield of the dark fermion  mirrors the baryon yield ($Y_\chi \approx Y_B$). Its fractional density compared to the total DM density is therefore a simple ratio of masses:
\begin{equation}
    f_{DM} \equiv \frac{\Omega_\chi}{\Omega_{DM}} \approx \left( \frac{\Omega_B}{\Omega_{DM}} \right) \frac{m_\chi}{M_{\rm Pl}} \approx \frac{m_\chi}{5 M_{\rm Pl}} \, .
\end{equation}
Kinematical constraints for particle production and proton stability restrict the dark fermion mass to $M_{\rm Pl} < m_\chi < m_X \lesssim \Delta V^{1/4}$. For an optimistic step size of $\Delta V^{1/4} \sim 2 \text{ GeV}$, the dark fermion produced in this standard chain inflation scenario can account for at most  $\sim 40\%$ of the total Dark Matter abundance ($f_{DM} \lesssim 0.4$), mandating a secondary DM production mechanism.

\vspace{0.2cm}
\underline{2.\textit{ Spontaneous Baryogenesis via a Chemical Potential:}} \\
Alternatively, the baryon asymmetry can be generated dynamically by leveraging the rapid time evolution of the tunneling field, $\dot{\phi} \neq 0$ \cite{Cohen:1988kt}. By introducing a derivative coupling to the SM baryon current, $\mathcal{L} \supset \frac{1}{M_*} \partial_\mu \phi J_B^\mu$, the rolling field acts as an effective chemical potential $\mu_B \sim \dot{\phi}/M_*$. 
During the phase transition, the wall velocity $v_w$ and width $L_w$ dictate $\dot{\phi} \sim \Delta \phi / (L_w/v_w)$. Provided there is an explicit $B$-violating operator active inside the bubble wall, the plasma will preferentially populate baryon states over anti-baryons.

If the $B$-violating interactions are efficient during the transition, the plasma tracks
the equilibrium asymmetry induced by the chemical potential, giving
$Y_B \sim c\,\mu_B/T$, where $c$ is a dimensionless number
set by the baryon-carrying degrees of freedom relative to the total plasma degrees of
freedom. If instead these interactions are inefficient, the yield is further suppressed by
the ratio of the $B$-violating rate to the relevant time scale, schematically
$Y_B\sim c\,(\mu_B/T)(\Gamma_{\Delta B}/H)$, or
$Y_B\sim c\,(\mu_B/T)\Gamma_{\Delta B}\Delta t_{\rm wall}$ for wall-localized
production.

\vspace{0.2cm}
\underline{3.\textit{ ``Cold" Bubble Wall Filtering:}} \\
Standard electroweak baryogenesis relies on CP-violating interactions between plasma
particles and an advancing bubble wall~\cite{Hall:2019rld}. In chain inflation, the expanding bubble walls can
similarly act as CP-asymmetric filters. However, because sphalerons are inactive at
$T_c\lesssim300\,{\rm MeV}$, the wall itself, or interactions localized near it, must
mediate explicit \(B\)-violation. Moreover, for the typical low-scale realizations with
$T_c\lesssim \Lambda_{\rm QCD}$, the ambient thermal plasma is hadronic rather than
partonic. Thus the standard diffusion picture involving free thermal quarks scattering off
the wall does not directly apply. A viable mechanism would instead have to operate through
hadronic $B$-violating interactions, or through nonthermal partons produced directly by
wall collisions. The efficiency of this mechanism depends sensitively on the wall velocity,
wall width, CP-violating reflection asymmetries, and the microscopic structure of the
$B$-violating operator.

\vspace{0.2cm}
\underline{4.\textit{ Non-Equilibrium High-Energy Sphalerons:}} \\
A more exotic possibility is to exploit electroweak sphaleron transitions far below the
electroweak temperature through highly non-thermal scattering events. 
Although 
sphalerons are exponentially suppressed at $T_c \ll 100\,{\rm GeV}$, individual hard
scatterings can in principle overcome the sphaleron barrier if their center-of-mass energy
satisfies $E_{\rm cm}\gtrsim E_{\rm sph}\sim 7\,{\rm TeV}$~\cite{Jaeckel:2022osh}. This provides a
possible route to $B+L$ violation in a macroscopically cold plasma.

In the minimal low-scale tilted cosine chain, however, this mechanism is difficult to
realize. For a hard particle of energy $E$ scattering on a thermal particle of energy
$\sim T$, one has $E_{\rm cm}^2\sim 2ET$. At the typical reheating temperature
$T_c\sim 10\,{\rm MeV}$, reaching $E_{\rm cm}\sim 7\,{\rm TeV}$ would require
$E\sim 10^9\,{\rm GeV}$, far above the characteristic energy released in GeV-scale final
transitions. Therefore, non-equilibrium sphaleron baryogenesis is not generic in the
minimal low-scale chain inflation setup. It would require an additional ultra-heavy
non-thermal component, an earlier high-scale injection event not subsequently diluted, or
other nonminimal dynamics capable of producing sufficiently energetic Standard Model
particles near the end of the chain.
 
\subsection{Summary}
While generating the correct baryon asymmetry $Y_B \sim 10^{-10}$ is parametrically
possible in low-scale chain inflation, the kinematics and proximity to BBN impose severe
restrictions. Successful baryogenesis requires prompt dynamics, typically
$\tau \lesssim 0.1-1\,{\rm s}$, and any asymmetric dark matter candidate produced
in the simplest co-genesis realization is naturally restricted to be a sub-component of the
total dark matter abundance. Moreover, because the reheating temperature is far below the
electroweak scale, the mechanism must either involve explicit low-scale $B$ violation,
operate through hadronic or non-thermal degrees of freedom near the bubble walls, or invoke
additional  dynamics capable of producing ultra-energetic particles for
non-equilibrium electroweak sphalerons.

Thus, although low-scale baryogenesis is not excluded in standard tilted cosine chain
inflation, it is highly constrained and model dependent. At the same time, the GeV-scale
energies involved make this scenario an interesting target for explicit model building, with
possible complementary signatures in laboratory searches for low-mass mediators, displaced
decays, baryon-number violation, or sub-component asymmetric dark matter. As demonstrated
in the main text, introducing a non-minimal coupling to gravity provides a different route: it naturally elevates the inflationary scale,
$V_*^{1/4}\sim10^{11}-10^{12}\,{\rm GeV}$, thereby bypassing the severe
low-temperature limitations of the minimal scenario and allowing standard high-scale
mechanisms such as thermal leptogenesis.

\section{Analytic Results for Non-Minimally Coupled Chain Inflation}\label{app:AnalyticResults}

{In the main text we used numerical solutions of the scalar amplitude, spectral index, and $e$-fold constraints to identify the viable parameter space of non-minimally coupled chain inflation. The purpose of this appendix is to provide analytic control over the main features of those solutions. We focus on three related questions: why the $u_*=0$ branch develops a characteristic turnover in $n_s(V_*^{1/4})$, how perturbative unitarity bounds the maximum viable inflationary scale, and why the large negative-$u_*$ regime approaches a universal behavior. These estimates allow us to build intuition for the model and understand the qualitative structure of the numerical branches shown in the main text.}

\subsection{Analytics for \texorpdfstring{$S_2$}{S2} in the \texorpdfstring{$u_*=0$}{u*=0} Regime}\label{subsec:Analytics_ustar0}

{We begin with the particularly transparent case in which the CMB pivot scale crosses the horizon at the origin of field space, $u_*=\chi_*/\chi_c=0$.
Here, the end of inflation happens at $\chi_c>0$, leading to a larger accumulation of $e$-folds towards the latter stages of the chain.
For $u_*=0$, the linear coefficient in the bounce action expansion vanishes, $S_1=0$, so the leading modification of the tunneling history is controlled by the quadratic coefficient $S_2$. This makes the $u_*=0$ branch a useful analytic laboratory for understanding how the non-minimal coupling drives the model away from the minimally coupled tilted cosine prediction.}

We start by invoking Eq.~\eqref{eq:SCoeff_NMC} to show that $u_*=0$ leads to $S_1=0$. This proves that the leading effect of the non-minimal coupling on the tunneling dynamics is controlled entirely by the quadratic coefficient $S_2$. The spectral index constraint in Eq.~\eqref{eq:ns_Constraint_NMC} then reduces to the same form as in the minimally coupled tilted cosine model,
\begin{equation}
1-n_s=\frac{3.48\times10^4}{N_*}\,,
\end{equation}
while the number of $e$-folds is modified by the $S_2$-dependent dilation of the vacuum lifetimes along the chain.

The size of this correction can be estimated analytically. Starting from Eq.~\eqref{eq:SCoeff_NMC}, we write
\begin{equation}
S_2=B(\xi,x)\SZero(2\pi fN_*)^2 \, ,
\end{equation}
where $N_*$ is the number of transitions between $\chi_*$ and $\chi_c$. We can express $f$ in terms of the value of the Euclidean action in the minimally coupled regime
\begin{equation}
f^4=\Lambda^4\frac{\SZero}{\mathcal{S}(x)}\, .
\end{equation}
Since $\Delta V=2\pi\mu^3 f=2\pi x\Lambda^4$ and $N_*=V_*/\Delta V$, we also have
\begin{equation}
\Lambda^4=\frac{V_*}{2\pi xN_*}\, ,
\end{equation}
and therefore
\begin{equation}
f^4=\frac{V_*}{2\pi xN_*}\frac{\SZero}{\mathcal{S}(x)} \, .
\end{equation}
Substituting this into the expression for $S_2$ gives
\begin{equation}\label{eq:S2_VStar_SZero_app}
S_2 = \frac{4\pi^2 B(\xi, x)}{\sqrt{2\pi x \mathcal{S}(x)}}  N_*^{3/2} \SZero^{3/2} \frac{V_*^{1/2}}{M_{\rm Pl}^2} \,.
\end{equation}
Neglecting $\mathcal{O}(1)$ factors, we find the parametric dependence of $S_2$ on the non-minimal coupling and the inflationary scale
\begin{equation}\label{eq:S2_uStar0}
S_2\simeq B(N_*\SZero)^{3/2}\left(\frac{V_*^{1/4}}{M_{\rm Pl}}\right)^2
\simeq\left(\frac{\xi \, V_*^{1/4}}{10^{12}\ \GeV}\right)^2 \, ,
\end{equation}
where we have used $B\simeq12\xi^2$, appropriate for the relevant values $\xi\gtrsim10$, together with $\SZero\sim\mathcal{O}(10)$ for the energy scales where the NMC begins to dominate and $N_*\sim\mathcal{O}(10^6)$. This scaling shows that the NMC effects become important once $S_2$ approaches order unity, which occurs at parametrically lower inflationary scales for larger values of $\xi$. For $\xi\gtrsim20$, non-trivial deviations from the minimally coupled tilted cosine prediction are expected around $V_*^{1/4}\gtrsim10^{10}\ \GeV$, as seen in Figures~\ref{fig:ns_vs_V*_Full} and~\ref{fig:ns_vs_V*_ZoomGrid}.

However, the growth of the NMC effects with the inflationary scale is not unbounded, as the Euclidean action $\SZero$ itself has a dependence on $V_*$ that suppresses its value for large enough scales. We can see this clearly by analyzing the dependence of the minimally coupled transition rate $\Gamma_0$, given in Eq.~\eqref{eq:GeneralGamma}, on $V_*$ and $\SZero$
\begin{equation}
\begin{aligned}
    \ln\Gamma_*\big\lvert_{u_*=0}=\ln \Gamma_0 &= 2\ln(\Lambda^4) - \ln(f^4) + 2\ln(\SZero) - \SZero + \mathcal{C}(x)\\
        &= \ln(\Lambda^4) + \ln(\SZero) - \SZero + \ln \mathcal{S}(x) + \mathcal{C}(x) \\
        &\simeq \ln(V_*) + \ln(\SZero) - \SZero + \mathcal{C}(x,N_*) \, ,
\end{aligned}
\end{equation}
where we have absorbed all variables of roughly constant magnitude into $\mathcal{C}$. Note that $\ln (N_*)$ is of relatively constant value since $N_*\propto(1-n_s)^{-1}$ and $n_s$ varies on much smaller scales than $V_*$. Inserting this result into the CMB amplitude constraint, which imposes $\Gamma_*/H_*^4=\mathcal{C}$, and using the fact that $H^4_*\propto V_*^2$, we find that
\begin{equation}
    \ln(V_*) + \ln(\SZero) - \SZero - 2\ln\left( V_* \right) = \mathcal{C}(x,N_*,A_s) \, ,
\end{equation}
leading to
\begin{equation}
V_* \propto \SZero e^{-\SZero}\, .
\end{equation}
This justifies our earlier approximation of treating $\SZero$ as roughly independent of $V_*$ around the scales at which the NMC is small or negligible, as then $\SZero\gg1$ and the exponential suppression dominates, such that
\begin{equation}
    \SZero\simeq-\ln\left(\frac{V_*}{M_{\rm Pl}^4}\right) + \mathcal{C}(x,N_*,A_s) \, .
\end{equation}
This indicates that the variation of $\SZero^{3/2}V_*^{1/2}$ in $S_2$ is dominated by the $V_*$ term, confirming our assumption. Conversely, once we reach large enough energy scales, we must take the full dependence of $V_*$ on $\SZero$ into account, which is equivalent to writing 
\begin{equation}
    S_2\propto \SZero^{3/2}V_*^{1/2}\propto \SZero^2 e^{-\SZero/2} \,.
\end{equation}
Therefore $S_2$ won't grow uninterruptedly as we increase $V_*$, since it eventually reaches a maximum value determined by
\begin{equation}
\frac{dS_2}{d\SZero}=0 \Rightarrow \SZero=4 \, .
\end{equation}
This explains the turnover of the high-scale branches in Figures~\ref{fig:ns_vs_V*_Full} and~\ref{fig:ns_vs_V*_ZoomGrid}. As $V_*^{1/4}$ is increased, the NMC correction first grows and pushes $n_s$ away from the minimally coupled tilted cosine prediction towards lower values. Once $S_2$ reaches its maximum, the correction begins to decrease and the solutions turn back toward the minimally coupled curve. This behavior is also visible in the color coding of the figures, where the minimum of $n_s(V_*)$ coincides with the largest value of $S_2$.

The full dependence of $V_*$ on $\SZero$ can be determined as
\begin{equation}\label{eq:VStar_SZero_uStar0}
    V_*^{1/4}  =  \left[ \frac{(1-n_s) \mathcal{S}(x)(1-x^2)\SZero}{x} \right]^{1/4} \exp\left(-\frac{\SZero}{4} - \frac{3.95}{x^{2.9}}\right)\times6.94 \times 10^{13} \ \GeV \,,
\end{equation}
which, when combined with the above result for the value of $\SZero$ corresponding to the maximum $S_2$, gives us the precise value of $V_*$ where this maximum deviation from the tilted cosine occurs
\begin{equation}\label{eq:VStar_maxDeviation_uStar0}
    V_*^{1/4} \Big\lvert_{S_2=\text{max}} =  \left[ \frac{(1-n_s) \mathcal{S}(x)(1-x^2)}{x} \right]^{1/4} \exp\left(- \frac{3.95}{x^{2.9}}\right)\times 3.61 \times 10^{13} \ \GeV \, .
\end{equation}
Note that due to the weak dependence on $n_s$, this is effectively independent of $\xi$, which is confirmed in Figure~\ref{fig:ns_vs_V*_ZoomGrid}, where it can be seen that the curve corresponding to $u_*=0$ always has its lowest $n_s$ value in the NMC region for the same value of $V_*$.

Using the relation between $V_*$ and $\SZero$ in Eq.~\eqref{eq:VStar_SZero_uStar0}, we can write $S_2$ from Eq.~\eqref{eq:S2_VStar_SZero_app} as
\begin{equation}
\begin{aligned}\label{eq:uStar0_S2}
    S_2 =& 1.03\times10^8\left(\frac{V_*^{1/4}}{M_{\rm Pl}}\right)^2 \frac{B(\xi, x) }{\sqrt{x \mathcal{S}(x)} (1-n_s)^{3/2}}\\
    &\times\left[ -W_{-1} \left( -\frac{V_*}{M_{\rm Pl}^4} \frac{1.52\times10^{18} x}{\mathcal{S}(x) (1-x^2) (1-n_s)} \exp\left( \frac{15.8}{x^{2.9}}\right) \right) \right]^{3/2} \,.
\end{aligned}
\end{equation}
Combining this with the value of $V_*$ that maximizes the effect of the NMC shown in Eq.~\eqref{eq:VStar_maxDeviation_uStar0}, we obtain the maximum value of $S_2$ in the $u_*=0$ regime
\begin{equation}\label{eq:MaxS2_uStar0}
    S_{2,\text{max}} \approx 0.18 \frac{B(\xi, x)}{1-n_s} \frac{\sqrt{1-x^2}}{x} \exp\left( - \frac{7.9}{x^{2.9}} \right) \,.
\end{equation}
For a fixed $x$ and $\xi$, the maximum value of $S_2$ depends on the value of $n_s$, although in the viable range $n_s\in[0.96,0.98]$ this changes $S_{2,\text{max}}$ by at most a factor of 2.

For the $u_*=0$ case, the $e$-fold constraint in Eq.~\eqref{eq:HighScale_Efolds_Redefined} can also be evaluated analytically by neglecting the small contribution of radiation to the expansion rate ($\rho_r/V_*\ll1$). Combining it with the remaining constraints gives
\begin{equation}\label{eq:uStar0_V*}
    \begin{aligned}
        V_*^{1/4}=0.79 M_{\rm Pl}\left({1-n_s}\right)^{1/4} \times\exp\left(\frac{0.55}{1-n_s}\ {}_2F_2\left(\frac{1}{2}, 1; \frac{5}{4}, \frac{7}{4}; \frac{S_2}{4}\right) -62\right) \, ,
    \end{aligned}
\end{equation}
where $_2F_2$ denotes the generalized hyper-geometric function. This is a generalized version of Eq.~\eqref{eq:nsvsV_cosine}, taking into account the NMC $\xi$, albeit in the special $u_*=0$ case. Although the combination of this equation with Eq.~\eqref{eq:uStar0_S2} does not allow for an analytically solvable expression of $n_s(V_*,x,\xi)$, one can substitute $S_2(V_*,n_s,\xi,x)$ into the expression above and numerically find the contour in the $V_*-n_s$ plane that satisfies the resulting constraint. By doing this, we confirm the precise numerical results for $u_*=0$ shown in Figures~\ref{fig:ns_vs_V*_Full} and \ref{fig:ns_vs_V*_ZoomGrid}.

Throughout the numerical analysis of the constraints used in generating the results on Figures~\ref{fig:ns_vs_V*_Full} and~\ref{fig:ns_vs_V*_ZoomGrid}, we do not impose the condition of no fine-tuning, which is fundamental to the argument that limits the inflationary scale to $V_*^{1/4}\lesssim10^{12}$ GeV for generic chain inflation models, as originally discussed in Ref.~\cite{Winkler:2020ape} and shown visually in Figure 2 of Ref.~\cite{Freese:2021noj}. However, even though we do not enforce it, we found that this limitation imposes itself through the increasing numerical stiffness that follows from the necessity of carefully canceling out the increasing individual contributions of $\dot \Gamma$ and $\dot H$ when determining $n_s$, as seen from Eq.~\eqref{eq:ScalarSpectralIndex_Raw}. Therefore, our numerical solver quickly becomes unable to reach a stable solution and no result is found. This is why the lines in Figure~\ref{fig:ns_vs_V*_Full} terminate very close to the original bound of $V_*^{1/4}\lesssim10^{12}$ GeV. Nevertheless, as we will show in what follows, there is a more stringent physical constraint that bounds the maximum inflationary scale in the NMC model.

\subsection{Upper Bound on Inflationary Scale from Perturbative Unitarity}\label{subsec:Unitarity_ustar0}

{The high-scale solutions found in the main text rely on a controlled semiclassical description of tunneling between adjacent minima. This requires not only that the Euclidean action be large enough for the bounce calculation to be meaningful, but also that the local interactions in the tilted cosine potential remain perturbative. In this subsection we translate the perturbative unitarity condition into a lower bound on the baseline Euclidean action $S_{E,0}$ and, consequently, into an upper bound on the inflationary scale.}
{We focus on the particular case of models with $u_*=0$, as we can determine their associated perturbative unitarity bounds fully analytically, while for all other models this is done numerically. Nevertheless, the following serves as a methodological guide for the general determination of such limits on the theory.}

Perturbative unitarity requires that the semiclassical tunneling calculation remain under control. Since the decay rate is computed as $\Gamma=\mathcal{A}\ e^{-S_E}$, this approximation assumes that the Euclidean bounce is a good saddle point and that quantum corrections around it are sub-leading. If $S_E$ becomes too small, the exponential hierarchy is lost, higher-order corrections are no longer parametrically suppressed, and the tunneling rate cannot be reliably computed within the semiclassical expansion. We therefore impose a lower bound on $S_E$ as a perturbative unitarity condition.

The high-scale branch in the $u_*=0$ regime is restricted by imposing perturbative unitarity. Since the non-minimal coupling increases the Euclidean action away from the origin, $S_E(\chi)\geq \SZero$, it is sufficient to impose the unitarity bound on the baseline action $\SZero$. 

The origin of this lower bound is simple. Perturbative unitarity constrains the size of the dimensionless quartic interaction around a minimum of a tilted cosine potential. Expanding around a local minimum $\phi_{\rm min}$, this requirement can be written as
\begin{equation}
\left.\frac{d^4 V}{d\phi^4}\right|_{\phi=\phi_{\rm min}} < 8\pi \, .
\end{equation}
For a tilted cosine potential (in any frame, taking $\Lambda$ and $f$ as the local parameters) this leads to
\begin{equation}
\frac{\Lambda^4}{f^4}\sqrt{1-x^2}<8\pi \, .
\end{equation}
Rearranging this inequality yields
\begin{equation}
\frac{f^4}{\Lambda^4} > \frac{\sqrt{1-x^2}}{8\pi} \, .
\end{equation}
On the other hand, the baseline Euclidean action for the tilted cosine potential is
\begin{equation}
\label{eq:baseline_action}
\SZero=\frac{f^4}{\Lambda^4}\mathcal{S}(x) \, .
\end{equation}
Combining the previous two equations leads to
\begin{equation}\label{eq:SZero_unitarity_bound_main}
\SZero = \frac{f^4}{\Lambda^4}\mathcal{S}(x) \geq \frac{\sqrt{1-x^2}}{8\pi}\mathcal{S}(x) \, .
\end{equation}
Thus perturbative unitarity places an upper bound on the local curvature scale $\Lambda^4/f^4$ of the tilted cosine potential, which is equivalently a lower bound on the Euclidean bounce action. For the largest tunneling parameter considered here, $x\simeq0.96$, this gives approximately $\SZero\gtrsim3.5$, as discussed in Appendix~\ref{sec:ActionBoundsAppendix}. This value is close to the analytic estimate $\SZero=4$ at which the NMC correction $S_2$ is maximized, explaining why the unitarity bound cuts off the $u_*=0$ branch near its point of largest deviation from the minimally coupled tilted cosine result, as seen in Figure~\ref{fig:ns_vs_V*_Full}.

The bound of Eq.~\eqref{eq:SZero_unitarity_bound_main} becomes rapidly stronger as $x$ is lowered. For example, it gives $\SZero\gtrsim12$ for $x=0.9$ and $\SZero\gtrsim40$ for $x=0.8$. Since the maximal NMC departure occurs near $\SZero=4$, smaller values of $x$ prevent the theory from reaching the region where the non-minimal coupling has its largest effect before perturbative unitarity is lost. This is why the viable high-scale parameter space is naturally pushed toward fast tunneling, with $x$ close to its upper allowed value.

We now translate this lower bound into an upper bound on $V_*$. In the $u_*=0$ regime, the NMC correction vanishes at the pivot scale, so $\Gamma_*=\Gamma_0$. Starting from Eq.~\eqref{eq:GeneralGamma}, imposing the scalar-amplitude constraint in Eq.~\eqref{eq:GammaH_Ratio}, and using $H_*^4\simeq V_*^2/(9M_{\rm Pl}^4)$, one obtains
\begin{equation}\label{eq:VStar_SZero_unitarity_app}
V_* = \frac{\mathcal{S}(x)(1-x^2)}{8.76\times10^{16}\times8\pi^3 xN_*} \exp\left(13.15-\frac{15.8}{x^{2.9}}\right) \SZero e^{-\SZero} M_{\rm Pl}^4 \, .
\end{equation}
Here the numerical coefficient comes from the observed scalar amplitude and the fitted prefactor in the tunneling rate. Since $u_*=0$ implies $S_1=0$, the spectral-index relation gives
\begin{equation}
\label{eq:Nstarforunitarity}
N_*=\frac{3.48\times10^4}{1-n_s}\, .
\end{equation}
Substituting this into Eq.~\eqref{eq:VStar_SZero_unitarity_app} gives $V_*$ as a function of $\SZero$, $x$, and $n_s$.

The largest allowed inflationary scale is obtained by taking the smallest value of $\SZero$ consistent with perturbative unitarity. Using Eq.~\eqref{eq:SZero_unitarity_bound_main}, we find
\begin{equation}\label{eq:VStarMax_unitarity_main}
V^{1/4}_{*,\max}(x,n_s) = \left[ \frac{(1-n_s)(1-x^2)^{3/2}\mathcal{S}^2(x)}{x} e^{\left(-\frac{15.8}{x^{2.9}}-\frac{\sqrt{1-x^2}\mathcal{S}(x)}{8\pi}\right)} \right]^{1/4} \times 3.13\times10^{13}\ \GeV \, .
\end{equation}
This expression is independent of the non-minimal coupling $\xi$. The reason is that, for $u_*=0$ the amplitude constraint is evaluated at the origin where the NMC correction to the bounce action vanishes. The coupling $\xi$ controls how strongly the solution deviates from the minimally coupled curve through $S_2$, but the endpoint of the high-scale branch is fixed by the minimum allowed value of $\SZero$.

For a representative value $n_s\simeq0.970$ and the range $0.8\lesssim x\lesssim0.96$, Eq.~\eqref{eq:VStarMax_unitarity_main} gives
\begin{equation}
7\times10^6\ \GeV \lesssim V_{*,\max}^{1/4} \lesssim 4.5\times10^{11}\ \GeV \, .
\end{equation}
This explains the red points in Figures~\ref{fig:ns_vs_V*_Full} and~\ref{fig:ns_vs_V*_ZoomGrid}: the $u_*=0$ solutions are excluded precisely when the value of $\SZero$ required by the CMB amplitude constraint falls below the perturbative unitarity bound. The constraint is weaker near $x\simeq0.96$, while slower tunneling, corresponding to smaller $x$, lowers the maximum allowed inflationary scale and removes much of the phenomenologically interesting high-scale branch.

\subsection{Raising the Minimum Coupling with \texorpdfstring{$u_*<0$}{u*<0} and Universality}\label{subsec:Negative_uStar_and_Universality}

{We now consider the complementary regime in which the pivot scale lies far from the origin, $u_*\to -\infty$, while the end of inflation occurs near the origin in field space. This limit provides a useful analytic reference point because the predictions become insensitive to the precise value of $u_*$ and depend primarily on the asymptotic form of the bounce action expansion. It therefore captures the universal behavior of the large negative-$u_*$ branches discussed in the main text.}

\begin{figure}
    \centering
    \includegraphics[width=0.49\linewidth]{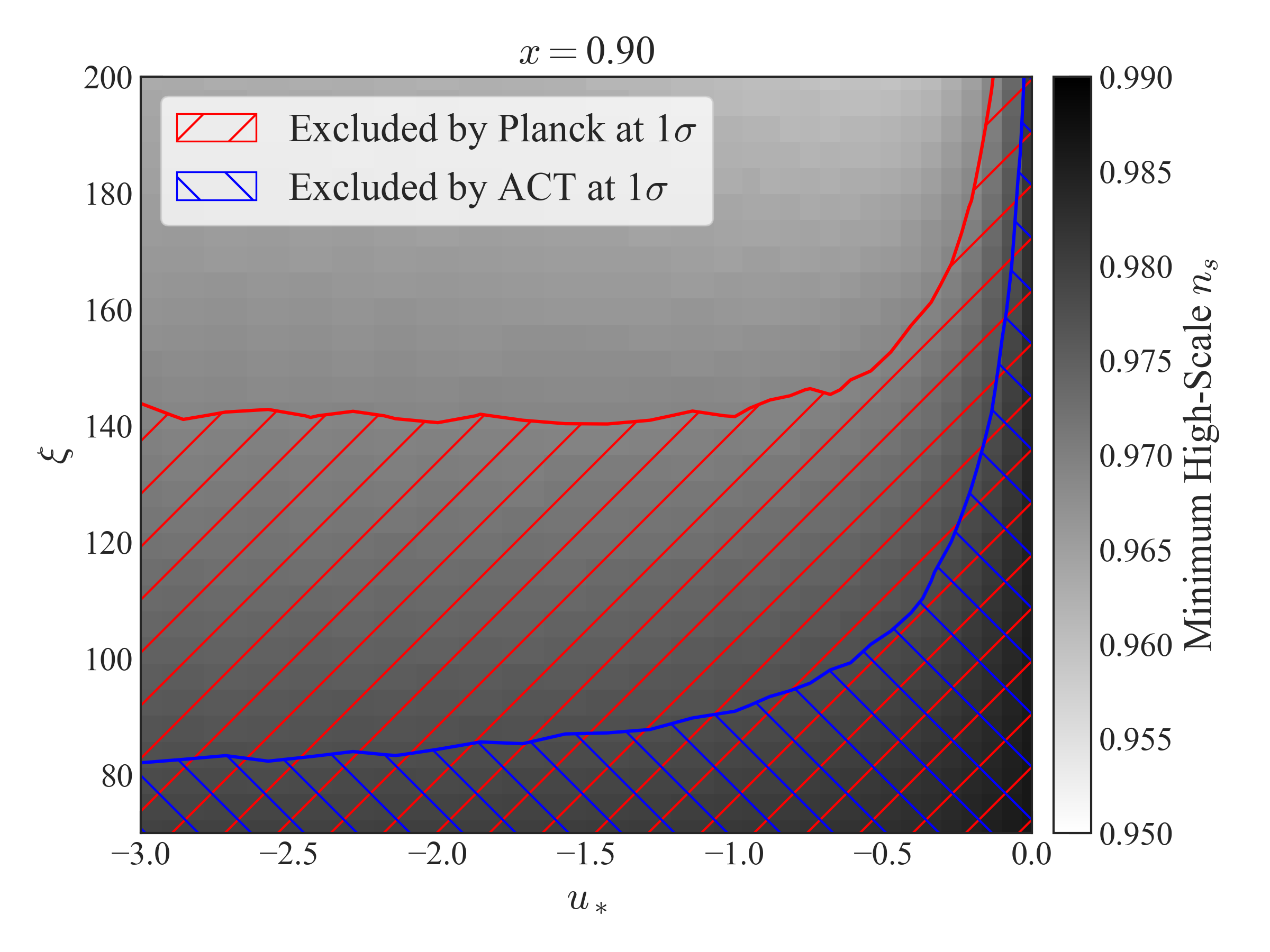}
    \includegraphics[width=0.49\linewidth]{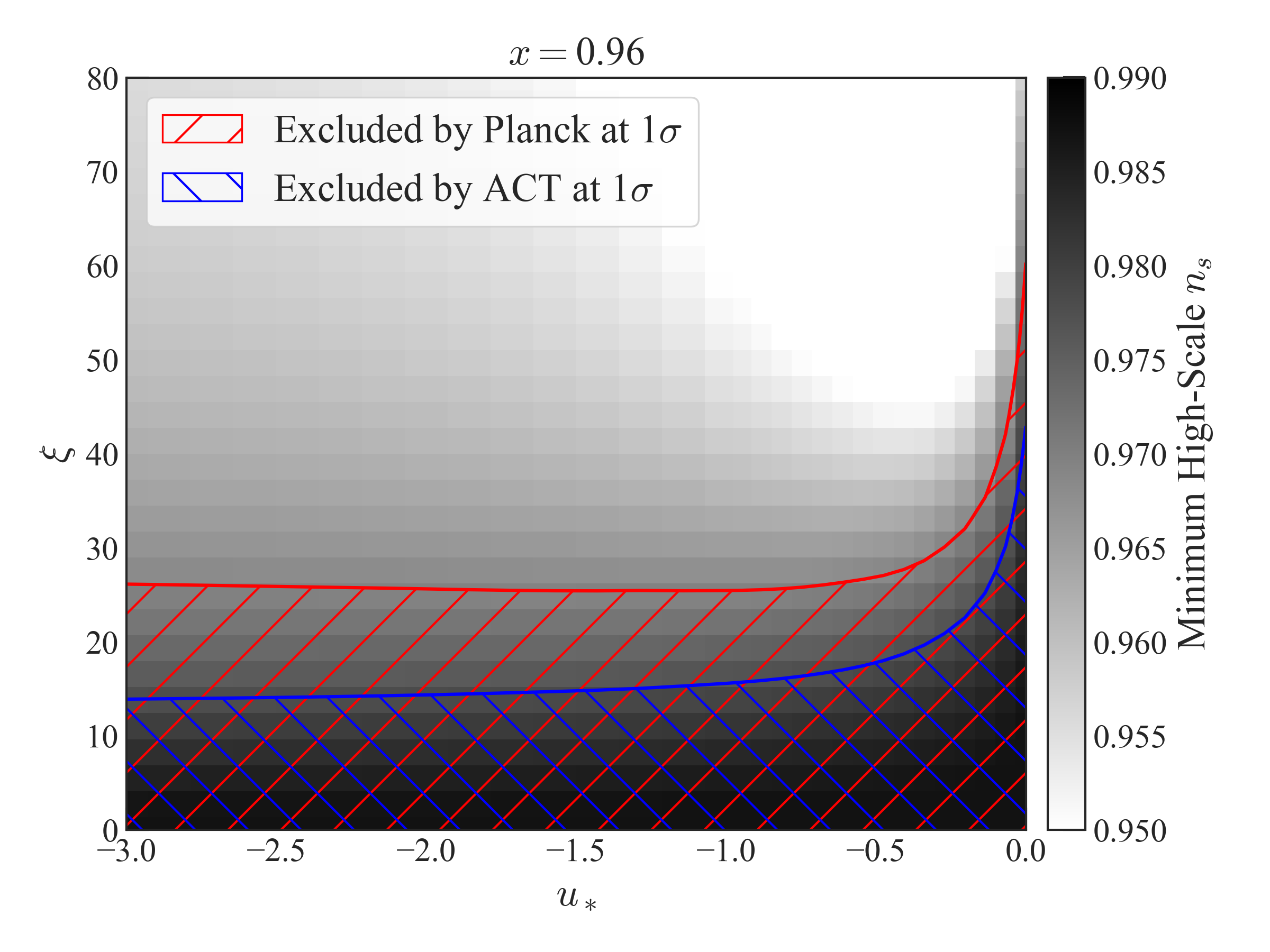}
    \caption{The minimum spectral scalar index within the high-scale regime in non-minimally coupled chain inflation as a function of $\xi$ and $u_*$ for $x=0.90$ (left) and $x=0.96$ (right). The shaded regions show the parameter space for which the minimum $n_s$ is above the $1\sigma$ upper bound of the \textit{Planck} (red) and ACT (blue) constraints. }
    \label{fig:ns_XiExclusion}
\end{figure}

For $u_*<0$, the pivot scale crosses the horizon when $\chi<0$, such that the lifetime of the vacua decreases as inflation proceeds up until $\chi\approx0$, where the lifetime is at its minimum, after which it progressively increases for $\chi>0$ as we discussed before. 
The dependence of the decay rate on $u$ leads to a reduced duration of the expansion in vacua with $|u|<|u_*|$; these vacua did not contribute in models with $u_*\geq0$. Note that, as the pivot scale is moved away from the origin towards more negative values of $u_*$, more $e$-folds accumulate close to $u_*$, such that the critical field value at which inflation ends rapidly approaches the origin ($|\chi_c|\ll1$).
As the total number of $e$-folds $\mathcal{N_*}$ is reduced, the inflationary scale $V_*$ decreases following Eq.~\eqref{eq:HighScale_Efolds_Redefined}. 
This relaxes the lower bound on $\xi$ for valid high-scale solutions, as seen in the left panel of Figure~\ref{fig:ns_XiExclusion}, where the boundary of the exclusion region for $\xi_{\rm min}$ is seen to be lowered when considering regions with $u_*<0$.

However, it turns out that one cannot indefinitely lower the inflationary scale by shifting $u_*$ towards larger negative values. This occurs due to the system approaching a universal regime where $u_*$ is removed from the constraints when it becomes large enough. This is clear to see in the three constraints in Eqs.~\eqref{eq:NMC_AmplitudeConstraint},~\eqref{eq:ns_Constraint_NMC} and~\eqref{eq:HighScale_Efolds_Redefined}, since in all cases one can approximate $u_*/(1-u_*)\approx-1$ and thus $u_*$ is completely removed when it becomes negative enough. This is precisely what we observe in Figure~\ref{fig:ns_XiExclusion}. There we see that, once $|u_*|$ becomes large enough (which practically corresponds to $|u_*|\gtrsim 0.5$ for $x=0.96$ and $|u_*|\gtrsim 1$ for $x=0.9)$, the boundary of the exclusion region approaches a constant value along the horizontal axis and thus becomes independent of $u_*$. 

Following a similar method, we find a relation between $n_s$ and $V_*$ that is analogous to the one found in the $u_*=0$ regime. Given the universal behavior of the equations in the large $|u_*|$ limit, the constraints can now be approximately (once again neglecting the radiation's contribution to the expansion rate) solved to give
\begin{equation}\label{eq:uStar_MinusInfty_V*}
    \begin{aligned}
        V_*^{1/4}=&10\sqrt{3} M_{\rm Pl}\left(\frac{1-n_s}{3.48\times10^4(1+S_2)}\right)^{1/4} \\
        &\times\exp\left[0.55\frac{1+S_2}{1-n_s} e^{-S_2/4} {}_1F_1\left(\frac{3}{4}, \frac{7}{4}, \frac{S_2}{4}\right) -62\right] \, .
    \end{aligned}
\end{equation}
The remainder of the argument is analogous to the one presented for $u_*=0$, with an identical conclusion that $S_2$, and thus also $\xi_{\min}$, increases with decreasing $n_s$, as this requires a more significant deviation from the tilted cosine potential at large inflationary scales. The effects of the NMC are contained within $S_2$, as seen in Eq.~\eqref{eq:S2_VStar_SZero_app}, such that we recover the result for the tilted cosine shown in Eq.~\eqref{eq:ns_InflationScale_Relation} when $S_2=0$.
\section{Constraining the Viable Range of the Euclidean Action}\label{sec:ActionBoundsAppendix}

{The tunneling description used in chain inflation is only reliable over a finite range of Euclidean bounce actions. If the action is too large, tunneling becomes too slow to satisfy the CMB amplitude and percolation requirements; if it is too small, the semiclassical expansion and perturbative control of the local potential become questionable. This appendix summarizes the resulting upper and lower bounds on $S_E$ and explains how they are used in the parameter-space analysis.}

We are interested in determining the viable range of the Euclidean bounce action at the pivot scale, which we denote as $S_{E,*}$. We start by considering the bounds on the expansion rate at the pivot scale $ H_*\sim{\sqrt{V_*}}/M_{\rm Pl}$ leading to
\begin{equation}
     10^{-23}\ \GeV\lesssim H_*\lesssim10^5 \ \GeV \, ,
\end{equation}
where we have used the CMB and BBN bounds on the inflationary scale $10\ \text{MeV}\lesssim V_*^{1/4}\lesssim10^{12}\ \GeV$ \cite{Winkler:2020ape}. By requiring the correct normalization of the CMB power spectrum given in Eq.~\eqref{eq:ScalarPowerSpectrumConstraint}, these translate into bounds on the transition rate at the pivot scale 
\begin{equation}
    10^{-19}\ \GeV\lesssim\Gamma_*^{1/4}\lesssim10^9 \ \GeV\, .
\end{equation}
We can convert this into an equivalent constraint on the Euclidean action by using Eq.~\eqref{eq:GeneralGamma} for the transition rate
\begin{equation}
    \Gamma_*^{1/4}=\frac{\Lambda^2}{f}\underbrace{(1-x^2)^{1/4}}_{\mathcal{O}(1)}\sqrt{\frac{S_E}{2\pi}}\exp\left[{-\frac{S_E-13.15+\frac{15.8}{x^{2.9}}}{4}}\right] \, 
\end{equation}
for $0.8<x<0.96$, as required by viable chain inflationary models with fast enough transitions and no presence of a tunneling catastrophe. This functional dependence on the action $S_E$ means that the lower bound on the transition rate will translate into an upper bound for the Euclidean action, which we calculate in below. 

\subsection{Upper Bound}
The perturbative unitarity of the theory \cite{Winkler:2020ape} requires $\frac{d^4V}{d\phi^4}\big\lvert_{\phi=\phi_{\rm min}}<8\pi$, which within the tilted cosine parametrization is equivalent to 
\begin{equation}
    \frac{\Lambda^4}{f^4}\sqrt{1-x^2}<8\pi
\end{equation}
or ${\Lambda}/{f}\lesssim\mathcal{O}(1)$.
Additionally, since $f\lesssim10^{10} \ \GeV$ \cite{Winkler:2020ape,Freese:2021noj}, we may generally constrain $\Lambda\lesssim10^{10} \ \GeV$. However, we are also specifically constrained by the slow variation of the model's parameters imposed by $n_s$, which we must satisfy while avoiding fine-tuning of $\dot \Gamma$ and $\dot H$. The slow variation of $\Gamma$ is ensured by the smallness of the coefficients $S_i$ in the expansion \eqref{eq:S_expansion}. At the level of the expansion rate, this is equivalent to requiring $\epsilon = |\dot H|/H^2<1$, leading to  
\begin{equation}
\frac{\Delta H}{H^2\Delta t}=\underbrace{\frac{1.4\Gamma^{1/4}}{H}}_{4\times 10^4}\frac{\Delta H}{H}\lesssim1 \,.
\end{equation}
We then write the variation of the Hubble parameter as 
\begin{equation}
    \Delta H=\frac{\Delta H^2}{2H}\approx\frac{\Delta V}{6H}\approx\frac{\mu^3\Delta\phi}{6H}=\frac{2\pi\mu^3f}{6H}\approx\frac{\mu^3f}{H}=\frac{x\Lambda^4}{H} \,,
\end{equation}
which turns the slow variation of the expansion rate into a constraint on the potential's parameters
\begin{equation}
    H^2\gtrsim4\times10^4  \ x\Lambda^4\Rightarrow\Lambda\lesssim \frac{V_*^{1/4}}{10}
\end{equation}
and thus in the case of the lowest allowed $V_*^{1/4}=10 \ \text{MeV}$ from BBN we get $\Lambda^2/f\lesssim\Lambda\lesssim10^{-3} \ \GeV$. Therefore 
\begin{equation}
    10^{-19} \lesssim\frac{\Gamma_*^{1/4}}{\GeV}\lesssim 10^{-3} \sqrt{\frac{S_E}{2\pi}}\exp\left[{-\frac{S_E-13.15+\frac{15.8}{x^{2.9}}}{4}}\right] \, ,
\end{equation}
from which we conclude that, by rearranging and taking the fourth power of both sides,
\begin{equation}
10^{-64}\lesssim\frac{S_E^2}{4\pi^2}\exp\left[-S_E+13.15-\frac{15.8}{x^{2.9}}\right] \, .
\end{equation}
Taking $4\pi^2\sim\mathcal{O}(10)$ and $\frac{15.8}{x^{2.9}}-13.15\sim\mathcal{O}(10)$ for any observationally compatible tunneling parameter $x$, we find for the logarithm of the above equation 
\begin{equation}
    -63\ln10\lesssim2\ln S_E-S_E-10 \quad\Rightarrow\quad S_E-2\ln S_E\lesssim135 \quad \Rightarrow \quad S_{E,*}\lesssim 145 \,.
\end{equation}
This establishes the approximate upper bound on the Euclidean action at the moment of pivot scale horizon crossing. This differs from the result of $S_{E,*}<120$ quoted in Ref. \cite{Freese:2023szd}, which may follow from the different condition applied to the slow variation of the expansion rate in Ref. \cite{Freese:2021noj}, there written as $f\mu^3\lesssim4.7\times 10^{-6}H^2\neq2.5\times10^{-5}H^2$, where the latter value is the one used in the derivation above. The differing factor of 10 in the limit for $\Lambda$ propagates to give $S_{E,*}\lesssim143$, which is only slightly closer to the quoted result of 120, the difference to which may then be due to the accumulation of ignored numerical factors in the intermediate calculations. A more careful calculation using Eq.~\eqref{eq:VStar_SZero_uStar0} for $V_*^{1/4}=10$ MeV and $n_s\sim0.970$ gives $S_E\approx130$, which is precisely in between both of the aforementioned estimates, thus providing a good reference value for the maximum value of $S_E$.

\subsection{Lower Bound}
We can find a lower bound on the Euclidean bounce action by considering its form in Eq.~\eqref{eq:ActionNumericalApproximation}, which depends both on the ratio $f^4/\Lambda^4$, constrained by the perturbative unitarity of the theory to be larger than $\sqrt{1-x^2}/8\pi$, and the tunneling parameter $x$, which is itself approximately limited to the range $x\in [0.8,0.96]$ for viable models of chain inflation. By minimizing the product of the corresponding quantities in the definition of $S_E$, these constraints combine into a lower bound on the Euclidean action
\begin{equation}
    S_{E}=\frac{f^4}{\Lambda^4}\mathcal{S}(x)\gtrsim3.5  \, .
\end{equation}

We can also prove that this lowest value of $S_E$ is associated with the largest allowed chain inflation energy scales. This is done by calculating $\Gamma$ for the requirements that are naturally picked out by the model's constraints for large energy scales, as seen in Figure 2 of Ref. \cite{Freese:2021noj}. This is achieved by taking the maximum value $f_{\rm max}=10^{10} \ \GeV$ and saturating the perturbative unitarity condition. The minimum value $S_E\approx3.5$ is obtained by also taking the fastest tunneling possible ($x=0.96$). All of these combine to give $\Gamma^{1/4}_*\approx10^9 \ \GeV$, which aligns perfectly with the maximum tunneling rate imposes by the highest allowed potential scale for general chain inflation ($V_*^{1/4}=10^{12} \ \GeV$). In contrast, if we take the minimum value of $S_E=30$ quoted in Ref. \cite{Freese:2023szd} with the same necessary assumptions for $f$ and $\Lambda$, we find that $S_E=30$ requires $x=0.83$, for which $\Gamma^{1/4}_*\approx10^6 \ \GeV$. This value is far below the transition rate one must have to satisfy the CMB normalization constraint for $V_*^{1/4}=10^{12} \ \GeV$, instead being associated with a scale $V_*^{1/4}\approx10^{10}\ \GeV$.

\section{Validating the \texorpdfstring{$\dot\Gamma/H\Gamma\lesssim1$}{GammaDot /H Gamma <1} Assumption}\label{app:SlowlyVaryingGamma}

{The GW estimates used in the main text rely on treating the transition rate as slowly varying over the timescale of an individual phase transition. In this appendix we check this assumption explicitly for the representative branches of the non-minimally coupled model. We show that the quantity $|\dot{\Gamma}/H\Gamma|$ remains small compared with the relevant inverse transition timescale, justifying the use of the standard bubble-collision formulae in the parameter range of interest.}

We can quantify the quality of our assumption to consider $\Gamma$ as slowly varying by focusing on transitions occurring around the pivot scale ($\chi\approx\chi_*$) and the end of the chain ($\chi\approx\chi_c$), since they describe the limiting cases of the model. 

To determine this value at the end of the chain, we calculate $|\dot \Gamma_c/H_c\Gamma_c|$ and determine its magnitude. This can be determined as
\begin{equation}
\Big\lvert\frac{\dot\Gamma}{H\Gamma}\Big\lvert_{\chi=\chi_c}\simeq\frac{\Delta\tilde\chi}{H_c\Delta t_c}|S_1+2S_2|\simeq\frac{1.4\Gamma_*^{1/4}}{H_*N_*}\frac{H_*}{H_c}e^{-(S_1+S_2)/4}|S_1+2S_2|
\end{equation}
{ We now analyze this expression for both $u_*=0$ and $u_*\rightarrow-\infty$ to determine the limiting cases for which we have the best analytical control. Using Eq.~\eqref{eq:ns_Effective} to write $\frac{1.4\Gamma_*^{1/4}}{H_*N_*}=\frac{12}{5}\frac{(1-n_s)}{(2-S_1)}$ and $H_c\approx\sqrt{\rho_{r,c}/3}$, we get}
\begin{equation}
    \frac{12}{5}\frac{(1-n_s)}{(2-S_1)}\left(\frac{V_*}{\rho_{r,c}}\right)^{1/2}\times\begin{cases}
    2e^{-S_2/4}S_2 \quad\text{for }u_*=0\\
       0 \quad\quad\quad\quad\ \ \text{for }u_*\rightarrow-\infty
    \end{cases}\,,
\end{equation}
{where we have used the fact that for $|u_*|\gg1$ one finds $S_1=-2S_2$. For $n_s=0.970$, taking $S_2\approx10$ for $u_*=0$, and using the result of $\rho_{r,c}=0.7\rho_{r,*}\approx0.07V_*$ from Appendix \ref{sec:RadiationAppendix}, we find}
\begin{equation}
     \Big\lvert\frac{\dot\Gamma}{H\Gamma}\Big\lvert_{\chi=\chi_c ,\ u_*=0}\approx0.7<\frac{\beta(\alpha)}{H} \, ,
\end{equation}
{which justifies our assumption of using the results for $\beta$ in the $\Gamma\approx\text{const}$ regime from Ref.~\cite{Freese:2022qrl}.

The argument is analogous for $\chi\approx\chi_*$, with }
\begin{equation}
\Big\lvert\frac{\dot\Gamma}{H\Gamma}\Big\lvert_{\chi=\chi_*}\simeq\frac{\Delta\tilde\chi}{H_*\Delta t_*}|S_1|\simeq \frac{12}{5}\frac{1-n_s}{2-S_1}|S_1|=
\begin{cases}
0 & \text{for } u_* = 0 \\
\frac{12}{5}(1-n_s)\frac{S_2}{1+S_2} & \text{for } u_* \rightarrow -\infty
\end{cases}\,.
\end{equation}
{By taking $n_s=0.970$ and $S_2\approx3$, as required by models with $|u_*|\gg1$ (see Figure~\ref{fig:ns_vs_V*_ZoomGrid}), we find}
\begin{equation}
     \Big\lvert\frac{\dot{\Gamma}}{H\Gamma}\Big\lvert_{\tilde{\chi}=0, \ u_* \rightarrow -\infty} \approx 0.054 \,,
\end{equation}
such that we can always work within an approximation of a slowly-varying transition rate $\Gamma$.

\section{Bubble Collisions and  Backreaction}
\label{app:backreaction}

{
First order phase transitions proceed through bubble nucleation. Eventually, bubble walls from neighboring bubbles must meet and either    collide with  or pass through each other.  Even though many treatments assume that bubble collisions lead to a thermal radiation bath, the true bubble collision dynamics  can be complicated and the post-collision state is uncertain and model-dependent.

The  treatment of the tunneling, including bubble nucleation and collisions, used throughout this work  assumed that the energy released in each phase transition can be described by a component with a radiation-like equation of state. If this assumption holds, as discussed in Section~\ref{subsec:Radiation},  at the pivot scale
  the CMB amplitude and the tracking solution imply
\beq
\rho_{r,*} \simeq \frac{1.4\Gamma_*^{1/4}}{4H_*}\Delta V \simeq 10^4\Delta V \, .
\eeq
The above equation arises due to the requirement of $\sim 10^4$ transitions per $e$-fold.
Although this component ($\rho_{r,*}$) remains subdominant relative to the total vacuum energy, its energy density is much larger than the energy released in a single transition. This raises
the question of whether the collision products modify the tunneling relevant for subsequent
tunneling events or allow the field to classically cross over the barriers. In particular, if they rapidly form a thermal bath that couples strongly
to the tunneling field, finite-temperature corrections could make thermally assisted
transitions important.

The identification of the collision debris with an equilibrium thermal bath is, however,
not guaranteed. The interval between successive transitions is of order
\beq
\Delta t\sim\Gamma^{-1/4}\ll H^{-1},
\eeq
and the energy released in a bubble collision may initially reside in scalar gradients,
coherent field oscillations, relativistic particles, or localized nonlinear field configurations (oscillons).
The subsequent thermalization rate depends on the microscopic interactions of the theory~\cite{AARTS2000403,PhysRevD.72.025014,PhysRevLett.90.121301}.
Indeed, numerical studies have found substantial non-equilibrium structure following
bubble collisions, including large-amplitude oscillations and oscillon-like configurations
that can delay the redistribution of energy into a homogeneous bath
\cite{Pirvu:2023plk, Cutting:2018tjt}.

Importantly, even in the absence of thermal equilibrium, the collision products may affect
subsequent tunneling. The magnitude of this backreaction is not fixed by the total energy density alone, but also by the characteristic frequencies, couplings, coherence properties, and which field directions are excited. Several possibilities may therefore reduce the effect of the existing radiation on the next tunneling event, and thus a proper analysis requires dedicated simulations and an exact description of the microphysics for a particular model. We briefly describe three possible mechanisms for suppressing post-collision back-reaction.
 First, the collision energy may be carried predominantly by  hard modes, which suppress back-reaction by producing a smaller field variance at fixed energy density.  Parametrically $\langle\delta\phi^2\rangle \sim \rho_{\delta\phi}/\omega_{\delta\phi}^2$, where $\rho_{\delta\phi}$ is the energy density of $\phi$ fluctuations after the bubble collision and $\omega_{\delta\phi}$ is their typical frequency, with $\omega_{\delta\phi}\simeq k$ for hard (high-momentum) modes.
Alternatively, a fraction of the collision energy may be transferred to additional degrees of freedom whose feedback on the tunneling coordinate is suppressed.
More generally, in genuinely multidimensional realizations of chain inflation, successive transitions may occur along different field directions, so that only a limited subset of the debris produced in previous transitions couples appreciably to the field responsible for subsequent tunneling. For example, in the original formulation of chain inflation  a series of fields with ``nearest-neighbor" couplings undergoes individual tunneling events in close succession~\cite{Freese:2004vs}. While this is by no means an exhaustive list of possible post-tunneling dynamics, it provides a sense of the inherent ambiguity and model-dependence.

A quantitative assessment requires simulations that follow   the evolution of the non-equilibrium collision debris through repeated nucleation, bubble expansion, collisions, and subsequent tunneling events. The real-time dynamics of bubble nucleation and collisions remains an active area of numerical research, including scalar-field simulations, coupled scalar--fluid systems, and, more recently, dynamical nucleation from ensembles of metastable field configurations together with studies of bubble velocities and oscillon precursors in $1+1$ dimensions~\cite{Child:2012qg,Giblin:2013kea,Hindmarsh:2013xza,Pirvu:2023plk}. Extending such calculations to repeated nucleation and collisions in $3+1$ dimensions, as required for chain inflation, remains an important open problem.
}

\end{appendices}

\bibliographystyle{JHEP}
\bibliography{References.bib}

\end{document}